\documentclass[11pt]{article}

\usepackage[margin=1in]{geometry}
\usepackage{lmodern}
\usepackage{microtype}
\usepackage{setspace}
\usepackage{amsmath, amssymb, mathtools, bm}
\usepackage{amsthm}
\numberwithin{equation}{section}

\usepackage{xcolor}
\usepackage{graphicx}
\usepackage{tikz}
\usetikzlibrary{arrows.meta,calc,fit,positioning}
\usepackage{subcaption}
\usepackage{array}
\usepackage{tabularx}
\usepackage{booktabs}
\usepackage{float}
\usepackage{algorithm}
\usepackage{algorithmic}

\definecolor{qnavy}{HTML}{17365D}
\definecolor{qnavysoft}{HTML}{355374}
\definecolor{qteal}{HTML}{1B7F83}
\definecolor{qtealdark}{HTML}{0B5D63}
\definecolor{qgold}{HTML}{B9831F}
\definecolor{qgolddark}{HTML}{8A610D}
\definecolor{qink}{HTML}{1F252B}
\definecolor{qmuted}{HTML}{667085}
\definecolor{qpanel}{HTML}{F7F9FC}
\definecolor{qtealfill}{HTML}{EFF7F6}
\definecolor{qgoldfill}{HTML}{FFF8EA}
\definecolor{qline}{HTML}{CFD8E3}

\usepackage[round,authoryear]{natbib}
\usepackage{hyperref}
\hypersetup{colorlinks=true,linkcolor=black,citecolor=black,urlcolor=black}

\newcommand{\R}{\mathbb{R}}
\newcommand{\E}{\mathbb{E}}
\newcommand{\Normal}{\mathcal{N}}
\newcommand{\TN}{\mathcal{N}^+}
\newcommand{\ind}{\mathbf{1}}
\newcommand{\diag}{\mathrm{diag}}
\newcommand{\vect}[1]{\bm{#1}}
\newcommand{\mat}[1]{\bm{#1}}
\DeclareMathOperator{\argmin}{arg\,min}

\DeclareMathOperator{\IG}{IG}
\DeclareMathOperator{\GIG}{GIG}

\newcommand{\Exp}{\mathrm{Exp}}
\newcommand{\Unif}{\mathrm{Unif}}
\newcommand{\AL}{\mathrm{AL}}
\newcommand{\exAL}{\mathrm{exAL}}
\newcommand{\aCRPS}{\mathrm{aCRPS}}
\newcommand{\RHS}{\ensuremath{\mathrm{RHS}}}
\newcommand{\pkg}[1]{\textsf{#1}}
\newcolumntype{Y}{>{\raggedright\arraybackslash}X}
\newcommand{\TableStyle}{%
  \footnotesize
  \setlength{\tabcolsep}{4.5pt}%
  \renewcommand{\arraystretch}{1.12}%
}

\newcommand{\GlofasApplicationCurrentForecastWindowFigure}{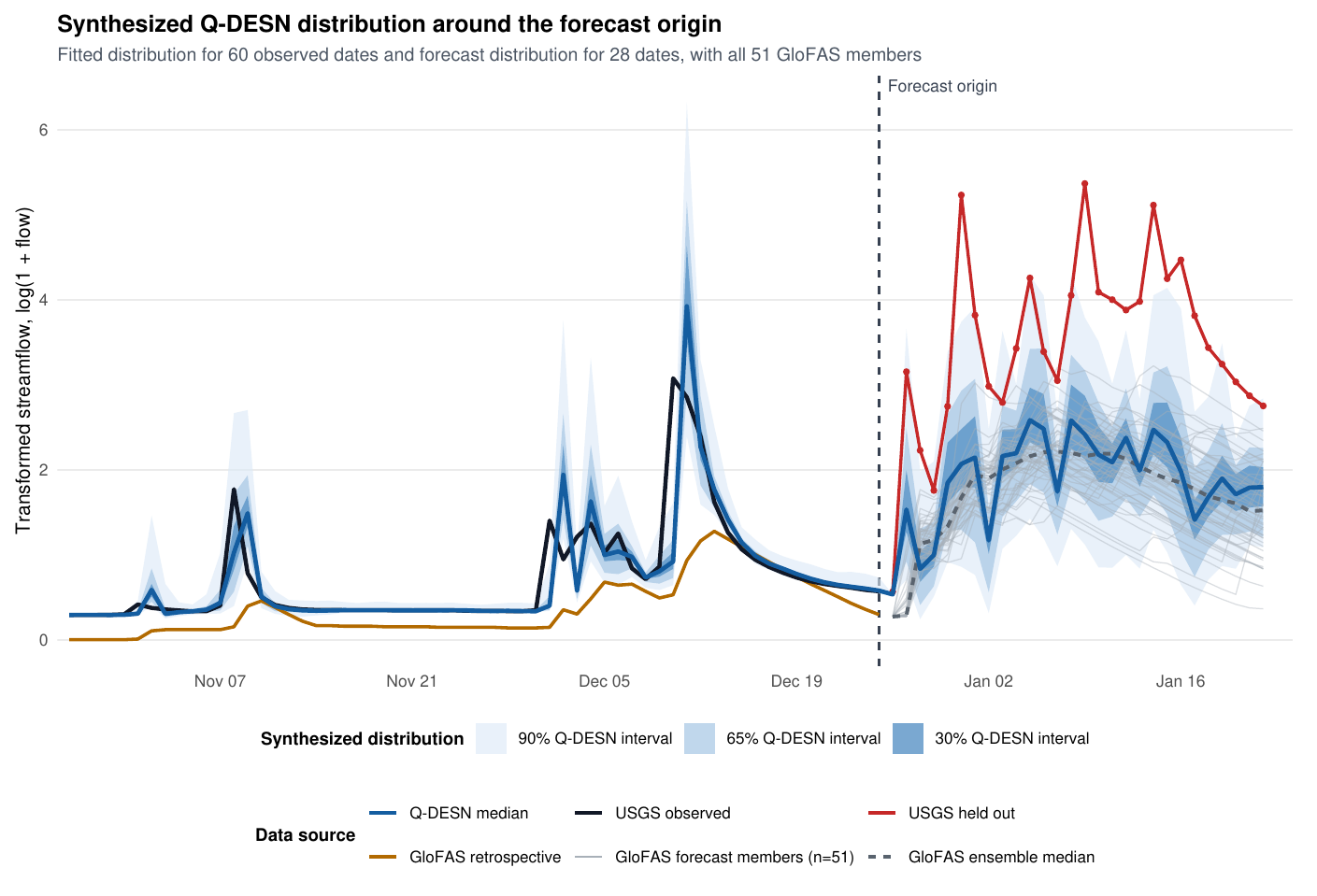}

\newcommand{\GlofasApplicationCurrentQdesnCheckLoss}{0.5654}
\newcommand{\GlofasApplicationCurrentRawCheckLoss}{0.7639}

\newcommand{\GlofasApplicationCurrentQdesnIntervalScore}{10.8032}
\newcommand{\GlofasApplicationCurrentRawIntervalScore}{25.2028}

\newcommand{\GlofasApplicationCurrentQdesnAcrps}{1.1319}
\newcommand{\GlofasApplicationCurrentRawAcrps}{1.4424}

\newcommand{\GlofasApplicationCurrentQdesnMeanCoverage}{0.357}
\newcommand{\GlofasApplicationCurrentRawMeanCoverage}{0.000}
\newcommand{\GlofasApplicationCurrentScoredHorizons}{28}
\newcommand{\GlofasApplicationCurrentOriginDate}{2022-12-25}

\newcommand{\GlofasApplicationCurrentSharedRhsTau}{0.1}
\newcommand{\GlofasApplicationCurrentDiscrepancyRhsTau}{0.001}

\newcommand{\GlofasApplicationCurrentObservedHistoryAcrps}{0.0445}

\newcommand{\GlofasApplicationCurrentObservedHistoryCoverage}{0.906}
\newcommand{\GlofasApplicationCurrentObservedHistoryDates}{12,495}

\newcommand{\PricefmFullRegionFolds}{114}
\newcommand{\PricefmFullRegions}{38}
\newcommand{\PricefmFullFolds}{3}
\newcommand{\PricefmFullQdesnWins}{66}
\newcommand{\PricefmFullQdesnClose}{12}
\newcommand{\PricefmFullPricefmWins}{36}
\newcommand{\PricefmFullQdesnWinRate}{57.9\%}
\newcommand{\PricefmFullMeanQdesnAql}{6.826}
\newcommand{\PricefmFullMeanPricefmAql}{7.039}
\newcommand{\PricefmFullMeanDeltaAql}{-0.213}
\newcommand{\PricefmFullMedianDeltaAql}{-0.083}
\newcommand{\PricefmFullGraphRows}{56}
\newcommand{\PricefmFullTargetOnlyRows}{58}
\newcommand{\PricefmFullHorizonRows}{72}
\newcommand{\PricefmFullPaperQuantiles}{0.10, 0.25, 0.45, 0.50, 0.55, 0.75, 0.90}

\newcommand{\PricefmAlignedQdesnAqcr}{2.64\%}
\newcommand{\PricefmAlignedPricefmAqcr}{0.00\%}
\newcommand{\PricefmAlignedQdesnMae}{16.728}
\newcommand{\PricefmAlignedPricefmMae}{17.274}
\newcommand{\PricefmAlignedQdesnRmse}{25.453}
\newcommand{\PricefmAlignedPricefmRmse}{26.336}
\newcommand{\PricefmAlignedAqlReduction}{3.0\%}
\newcommand{\PricefmAlignedMaeReduction}{3.2\%}
\newcommand{\PricefmAlignedRmseReduction}{3.4\%}

\newcommand{\Id}[1]{\mat I_{#1}}
\newcommand{\zeros}[1]{\vect 0_{#1}}
\newcommand{\ones}[1]{\vect 1_{#1}}

\theoremstyle{plain}

\theoremstyle{definition}

\theoremstyle{remark}

\title{\vspace{-0.6cm}Bayesian Quantile Deep Echo State Networks\\for Nonlinear Time Series}
\author{Antonio De Leon$^{1}$ \and Raquel Prado$^{1}$ \and Bruno Sans\'o$^{1}$}
\date{\small $^{1}$Department of Statistics, University of California, Santa Cruz \\
\vspace{0.25em}August 28, 2026}

\begin{document}
\maketitle
\vspace{-1.25em}

\begin{abstract}
\noindent
Conditional quantiles are central to asymmetric decision losses, tail-risk
assessment, and interval forecasts, but Bayesian quantile regression for
nonlinear time series can be difficult when the temporal feature vector is
high-dimensional. We develop the quantile deep echo state network (Q--DESN), a
Bayesian quantile regression model conditional on fixed features generated by a
deep echo state network. Conditional on a specified reservoir construction and
feature map, posterior uncertainty is assigned to the regression coefficients,
likelihood, and shrinkage parameters.
Single-level fits use asymmetric Laplace or quantile-fixed generalized
asymmetric Laplace working likelihoods with ridge or regularized-horseshoe
priors. Posterior computation uses Markov chain Monte Carlo when computationally
practical and a model-specific variational Bayes approximation for analyses
requiring repeated fitting. For quantile grids, we compare
independent level-wise regressions followed by monotone rearrangement with a
joint quantile-vector regression that shrinks adjacent quantile-specific
coefficient differences. Synthetic studies, a single-origin retrospective
Global Flood Awareness System (GloFAS) streamflow case, and a retrospective
PriceFM comparison identify
settings where this fixed-feature Bayesian regression improves finite-grid
quantile scores.
\end{abstract}

\bigskip

\section{Introduction}
\label{sec:intro}

Forecast evaluation often requires summaries beyond the conditional mean.
Conditional means are natural under squared-error loss, but asymmetric
decision losses, tail-risk assessment, and interval forecasts call for
conditional quantiles. Under asymmetric piecewise-linear loss, conditional
quantiles are optimal point forecasts, and pairs of quantiles define interval
forecasts
\citep{KoenkerBassett1978,Koenker2005,Gneiting2011}. Let
\(Y_{t+h}\) denote the future response at horizon \(h\), and let
\(\mathcal F_t\) be the sigma-field generated by the information available
at forecast origin \(t\), including past responses and covariates known at the
origin or specified through a scenario. With
\(F_{t,h}(u)=\Pr(Y_{t+h}\le u\mid\mathcal F_t)\), define the level-\(p_0\)
conditional quantile by the generalized inverse
\[
Q_{t,h}(p_0)=F_{t,h}^{-1}(p_0)
=\inf\{u:F_{t,h}(u)\ge p_0\},\qquad p_0\in(0,1).
\]
When the forecasting question concerns one probability level, a single fitted
conditional quantile is the relevant summary. When several levels are fitted,
the resulting finite grid can be rearranged into a monotone quantile curve.
Such a curve is a predictive summary over the chosen grid. A complete
predictive distribution additionally specifies the quantile function between
the fitted levels and its behavior in the two tails
\citep{GneitingRaftery2007,GneitingKatzfuss2014,BarlowBrunk1972IsotonicDual,
chernozhukov2010}.

For nonlinear dynamic data, quantile forecast accuracy also depends on how
temporal dependence is represented. An echo state network (ESN) represents
past responses and covariates through a recurrent nonlinear state equation.
In the construction used here, a specified reservoir realization maps the
observed history to reservoir states and hence to a fixed nonlinear design
matrix. The ESN literature calls the final regression layer the readout
\citep{Jaeger2001EchoState,LukoseviciusJaeger2009ReservoirComputing,Lukosevicius2012PracticalGuideESN}.
A deep echo state network (DESN) extends this idea by stacking reservoir
layers, following the deep ESN construction of
\citet{GallicchioMicheliPedrelli2018DeepESNDesign}. Q--DESN uses Bayesian
quantile regression with the resulting fixed design and conditions on the
reservoir realization throughout the analysis.

The quantile deep echo state network (Q--DESN) combines this fixed DESN
design with a quantile working likelihood. For a chosen level \(p_0\), the
linear predictor \(\mu_t=\vect x_t^\top\vect\beta\) is the likelihood location
and, under the parameterization used here, the fitted \(p_0\) conditional
quantile. The baseline likelihood is asymmetric Laplace (AL). The extended
asymmetric Laplace likelihood, denoted \(\exAL\), is the quantile-fixed
generalized asymmetric Laplace construction of \citet{yan2025new}; it allows
mode, skewness, and tail behavior to vary while retaining the fitted quantile
level. In model labels, the prefix ``ex'' indicates this extended working
likelihood; labels without the prefix use AL.

The AL or \(\exAL\) likelihood is used as a working likelihood in the Bayesian
quantile-regression sense. It defines a posterior distribution for
quantile-specific coefficients without asserting that either likelihood is the
data-generating error law
\citep{sriram2013theoretical,YangWangHe2016ALInference,JiLeeRabeHesketh2025ValidSE}.
Posterior intervals for coefficient-derived quantiles and forecast bands
computed from the fitted likelihood are therefore model-based summaries
conditional on the fixed DESN design and the chosen working likelihood.
Frequentist coverage of credible intervals under likelihood misspecification is
a separate question from empirical coverage of the forecast bands; the
simulations and retrospective applications assess the latter empirically.

High-dimensional regressions on DESN features require regularization. The
default coefficient prior is ridge, which gives dense Gaussian shrinkage. As an
adaptive alternative, the regularized horseshoe (RHS) introduces global-local
shrinkage and a slab component for large coefficients
\citep{CarvalhoPolsonScott2010HS,PiironenVehtari2017RHS}. Under the
product representation used here, the local, global, and slab scale updates
retain closed-form inverse-gamma forms
\citep{NishimuraSuchard2023SSS,MakalicSchmidt2016SimpleSampler}.
Markov chain Monte Carlo (MCMC) is used for posterior simulation when
computational cost permits. A mean-field variational Bayes (VB) approximation
is used in analyses requiring many repeated fits. For the non-conjugate
Q--DESN \(\exAL\) scale--asymmetry block, the VB approximation uses the
Laplace--Delta construction of \citet{wang2013nonconjugatevb}; the label
VB--LD refers to this Q--DESN approximation.

This article makes three principal contributions. First, it introduces
Q--DESN as Bayesian conditional-quantile regression on a fixed DESN feature
map, with AL or \(\exAL\) working likelihoods and ridge or
regularized-horseshoe regularization. Second, it develops two distinct
strategies for quantile grids: independent level-wise regressions followed
by a pre-specified monotone rearrangement, and a joint quantile-vector
regression that shares information through regularized-horseshoe shrinkage on
adjacent coefficient differences, borrowing from Quantile-Varying Parameter
ideas \citep{KohnsSzendrei2025QVP}. This adjacent-difference prior encourages
similar slopes at neighboring quantile levels; exact noncrossing still
requires an explicit constraint or monotone rearrangement. Third, the article
gives MCMC and model-specific VB--LD computation and evaluates the resulting
procedures in simulations and two retrospective applications. The applications
are interpreted under their stated covariates and forecast origins.

The empirical sections use Q--DESN in three ways. The single-quantile
simulation compares QDESN and exQDESN regressions with dynamic quantile linear
model (DQLM) and extended DQLM (exDQLM) baselines. The joint simulation isolates
multi-quantile coefficient sharing
against independent level-wise regressions under known true conditional
quantile paths. The GloFAS and PriceFM examples use independent Q--DESN
regressions in retrospective forecasting comparisons. Thus, evidence for the
joint quantile-vector prior comes from the synthetic multi-quantile study,
whereas the applications assess fixed-feature Bayesian quantile regression in
two domain-specific forecasting settings.

\subsection*{Related Work}

Bayesian quantile regression provides the likelihood-based foundation for
Q--DESN. Classical quantile regression targets conditional quantiles
directly \citep{KoenkerBassett1978,Koenker2005}. Bayesian implementations
often use the asymmetric Laplace likelihood because its mode aligns with
check-loss minimization and because its mixture representation supports
conditional posterior computation
\citep{yu2001bayesian,KozumiKobayashi2011Gibbs}. In this article the
asymmetric likelihood is a working likelihood: it defines posterior uncertainty
for quantile-regression parameters, while the frequentist behavior of
model-based credible intervals requires separate assessment
\citep{sriram2013theoretical,YangWangHe2016ALInference,JiLeeRabeHesketh2025ValidSE}.
The quantile-fixed generalized asymmetric Laplace distribution of
\citet{yan2025new} relaxes the AL error-shape restriction after the fitted
quantile level has been fixed; this is the \(\exAL\) working likelihood used
below.

Dynamic quantile linear models provide Bayesian state-space baselines for
time-varying conditional quantiles. The DQLM combines a linear state evolution
with Bayesian quantile regression
\citep{GoncalvesMigonBastos2020DQLM}. The extended DQLM (exDQLM) of
\citet{barata2021_flex_quantile} adds the corresponding asymmetric-likelihood
flexibility, and both models are implemented in version 1.1.1 of the
Comprehensive R Archive Network (CRAN) package \pkg{exdqlm}
\citep{exdqlmRPackage}. Q--DESN
represents temporal dependence through fixed nonlinear reservoir features. The DQLM and
exDQLM fits therefore serve as structured linear dynamic quantile comparators
under the same held-out forecast evaluation.

Reservoir computing and ESNs provide the nonlinear dynamic features used by
Q--DESN. ESNs were introduced by \citet{Jaeger2001EchoState}
and later reviewed within the broader reservoir-computing literature by
\citet{LukoseviciusJaeger2009ReservoirComputing}; practical guidance is
given by \citet{Lukosevicius2012PracticalGuideESN}. In a leaky-integrator
ESN, the leak rate balances persistence of the previous state with movement
toward the current input-driven update. Analyses of the echo state property
show that spectral-radius heuristics are useful but insufficient stability guarantees
\citep{JaegerLukoseviciusPopoviciSiewert2007LeakyESN,YildizJaegerKiebel2012ESP}.
Broader reservoir-computing surveys and time-series reviews discuss their
role in forecasting and applied dynamic modeling
\citep{ZhangVargas2023RCSurvey,YanEtAl2024ReservoirComputingOpportunities,CardosoEtAl2024ESNReview,SunEtAl2024SystematicESN}.
Deep ESNs, topology variants, and reservoir-parameter selection extend the
basic construction by stacking reservoirs, changing connectivity, or choosing
reservoir parameters
\citep{GallicchioMicheliPedrelli2018DeepESNDesign,KimKing2020DeepESN,MaatGianniotisProtopapas2018IJCNN,BaiEtAl2023DeepESNBO,ViehwegTeutschMader2025Topologies}.
Q--DESN uses these reservoir constructions as fixed nonlinear features for
Bayesian quantile regression.

A related reservoir-computing literature embeds ESN features in probabilistic
forecasting models. Deep ESNs with uncertainty quantification, probabilistic
reservoir methods for load forecasting, activation-level uncertainty
propagation, and Bayesian or state-space ESN formulations show that reservoir
features can support uncertainty-aware prediction
\citep{McDermottWikle2018DeepESNUQ,GuerraScardapaneBianchi2023PLFRC,GandhiLeePanTheodorou2018TanhUncertainty,SinghRaman2025ESNStateSpace}.
We focus on Bayesian inference for conditional quantiles with fixed DESN
features, asymmetric working likelihoods, and high-dimensional coefficient
regularization.

Quantile-oriented ESN methods provide the closest forecasting comparison.
\citet{LvZhaoLiuWang2016QRESNE} use quantile-regression ESN ensembles to
construct prediction intervals for blast-furnace gas generation, and
\citet{BonasWikleCastruccio2024CalibratedDeepESN} use penalized quantile
regression to calibrate DESN forecasts for quasi-periodic climate processes.
Q--DESN is complementary to these approaches. Conditional on the generated
reservoir, it specifies an \(\exAL\) regression whose posterior includes
coefficients, likelihood scale, asymmetry, and shrinkage parameters. The joint extension further
couples quantile-specific regression coefficients during inference, which can
reduce adjacent-level crossings before monotone rearrangement is applied.

Shrinkage priors enter because fixed reservoir designs can be
high-dimensional and correlated. Ridge shrinkage provides the
default dense Gaussian regularization. The regularized horseshoe, building
on the horseshoe prior and its slab-regularized extension, is used as an
adaptive global-local alternative
\citep{CarvalhoPolsonScott2010HS,PiironenVehtari2017RHS}. Under the product
representation used here, scale updates remain closed form
\citep{NishimuraSuchard2023SSS,MakalicSchmidt2016SimpleSampler}. These
priors regularize the coefficients. Feature-level interpretability requires
additional diagnostics.

Finally, a grid of quantile levels requires a distinction between monotone
rearrangement after fitting and joint modeling across levels. Separate Q--DESN
fits can cross because each level is estimated independently. Isotonic
regression can project the fitted grid onto the monotone cone
\citep{BarlowBrunk1972IsotonicDual}, and rearrangement methods provide a
general approach to noncrossing quantile and probability curves
\citep{chernozhukov2010}. These operations enforce monotonicity of the displayed quantile curve
after fitting. A joint posterior over quantile-indexed coefficients requires a
joint model across levels. Constrained quantile-regression methods
instead impose non-crossing restrictions during estimation
\citep{WuLiu2009StepwiseNCQR,BondellReichWang2010NoncrossingQR}. Recent
work also shows that non-crossing constraints can be viewed as fused
shrinkage across quantile-specific coefficients
\citep{SzendreiBhattacharjeeSchaffer2025FusedLASSO}. Bayesian joint
quantile models share information across levels through process priors,
score-likelihood constructions, or smoothness penalties
\citep{YangTokdar2017JointQuantilePlanes,WuNarisetty2021ScoreLikelihood,
WangCai2024CompositeBayesianNoncrossing}. The Quantile-Varying Parameter
construction of \citet{KohnsSzendrei2025QVP} is especially relevant here
because it places a state-space shrinkage prior on adjacent quantile-indexed
coefficient changes. The joint Q--DESN extension below adapts that idea to
fixed-feature regressions and assigns regularized-horseshoe priors to adjacent
slope differences. The resulting prior encourages neighboring quantile curves
to have similar slopes, which can reduce crossings. In the simulation study
below, the joint AL regression substantially reduces adjacent-level crossings
relative to independently estimated AL regressions; exact monotonicity still requires an
explicit ordered parameterization, projection, or rearrangement.

The remainder of the paper is organized as follows.
Section~\ref{sec:notation} introduces notation, deep echo state network
(DESN) dynamics, and the \(\exAL\) representation.
Section~\ref{sec:model} presents the Q--DESN model and prior specification,
including independent and joint multi-quantile extensions.
Section~\ref{sec:inference} describes posterior computation, while
Sections~\ref{sec:forecast}--\ref{sec:selection} describe multi-step quantile
summaries, finite-grid scoring, and reservoir diagnostics.
Section~\ref{sec:simulation} presents the simulation design, and
Sections~\ref{sec:data}--\ref{sec:pricefm-application} give streamflow and
electricity-price forecasting applications.
Section~\ref{sec:discussion} summarizes limitations and future work. The
supplement gives full conditional derivations, variational details, the
evidence lower bound,
monotone rearrangement details, and supplementary simulation and application
tables.


\section{Notation and Preliminaries}
\label{sec:notation}

\noindent
The notation separates deterministic reservoir quantities from stochastic
coefficient and likelihood quantities. Matrices and vectors are bold, as in
\(\mat W\) and \(\vect x\); scalar observations, latent states, and parameters
use standard lowercase notation unless they are named constants. The symbols
\(\IG\), \(\GIG\), and \(\TN\) denote inverse-gamma, generalized inverse
Gaussian, and positive-truncated Normal distributions. For a square matrix
\(\mat A\), write \(\rho(\mat A)=\max_i |\lambda_i(\mat A)|\), where
\(\{\lambda_i(\mat A)\}\) are the eigenvalues. Thus \(\rho(\cdot)\) denotes
spectral radius, while \(\rho_d\in(0,1)\) denotes the target spectral radius
for layer \(d\). Indices use \(d=1,\dots,D\) and \(t=1,\dots,T\). When
conditioning information is shown explicitly, \(Q_p(y\mid\cdot)\) denotes the
conditional \(p\)-quantile of the scalar response. The identity matrix, zero
vector, and one vector of dimension \(m\) are denoted by \(\Id{m}\),
\(\zeros{m}\), and \(\ones{m}\), respectively.

\subsection{Deep Echo State Network (DESN)}
\label{subsec:desn}

\noindent
The DESN defines a fixed feature map conditional on its specification and
reservoir realization. Posterior inference then concerns the quantile-regression
parameters. We state the dimensions before defining the recursion.
\medskip

\noindent\textbf{Notation and dimensions.}
The input vector is
\(\vect u_t=(y_{t-1},\dots,y_{t-m},\vect z_t^\top)^\top\in\R^{m+q}\), with
\(m\) response lags and \(q\) exogenous variables. For layer \(d\),
\(\mat W_d\in\R^{n_d\times n_d}\) is recurrent,
\(\mat W^{\mathrm{in}}_1\in\R^{n_1\times(m+q)}\) maps the original input to
the first layer, and
\(\mat W^{\mathrm{in}}_d\in\R^{n_d\times\tilde n_{d-1}}\) maps the reduced
state from layer \(d-1\) to layer \(d\) for \(d\ge2\). The reducer
\(\mat Q_{d-1}\in\R^{\tilde n_{d-1}\times n_{d-1}}\) maps
\(\vect h_{t,d-1}\in\R^{n_{d-1}}\) to
\(\tilde{\vect h}_{t,d-1}\in\R^{\tilde n_{d-1}}\). The regression dimension is
\(r=n_D+\sum_{d=1}^{D-1}\tilde n_d+(m+q)\), and
\(\vect\beta=(\beta_0,\beta_1,\dots,\beta_r)^\top\in\R^{r+1}\).

\medskip
\noindent\textbf{Random sparse weights and spectral scaling.}
In contrast to fully trained recurrent networks, reservoir and input
weights are drawn once and kept fixed throughout inference. Let \(\circ\)
denote the Hadamard product. Each recurrent matrix is
\(\mat W_d=\Phi^{(W)}_d\circ Z^{(W)}_d\), and each input matrix is generated
analogously as
\(\mat W^{\mathrm{in}}_d=\Phi^{(\mathrm{in})}_d\circ Z^{(\mathrm{in})}_d\).
Mask entries are Bernoulli with probabilities \(\pi^{(W)}_d\) and
\(\pi^{(\mathrm{in})}_d\); base entries are independent draws from
zero-mean finite-variance distributions \(p_W\) and \(p_{\mathrm{in}}\), such
as \(\Normal(0,\sigma^2)\) or \(\Unif[-a,a]\). Masks and base draws are
mutually independent and independent across entries and layers. The recurrent
matrices are then scaled to the target spectral radii,
\[
\text{Spectral-radius scaling:}\qquad
\bar{\mat W}_d \;=\; \frac{\rho_d}{\rho(\mat W_d)}\,\mat W_d,\qquad
\rho_d\in(0,1).
\]
If \(\rho(\mat W_d)=0\), the matrix is resampled. Input matrices
\(\mat W^{\mathrm{in}}_d\) are not spectrally scaled.

\medskip
\noindent\textbf{DESN recursion and regression features.}
With layer-specific leak rates \(\alpha_d\in[0,1]\), elementwise nonlinearity
\(f(\cdot)\), and lower-layer transformation \(k(\cdot)\), the fixed feature
map is
\begin{equation}
\label{eq:desn-hierarchy}
\begin{alignedat}{3}
\text{Layer 1 candidate state} &:\quad&
\vect\omega_{t,1}
&=f\!\left(\bar{\mat W}_1\vect h_{t-1,1}
    +\mat W^{\mathrm{in}}_1\vect u_t\right),\\
\text{Layer 1 update} &:\quad&
\vect h_{t,1}
&=(1-\alpha_1)\vect h_{t-1,1}+\alpha_1\vect\omega_{t,1},\\
\text{Reduced lower-layer state} &:\quad&
\tilde{\vect h}_{t,d-1}
&=\mat Q_{d-1}\vect h_{t,d-1}, \qquad&& d=2,\dots,D,\\
\text{Layer d candidate state} &:\quad&
\vect\omega_{t,d}
&=f\!\left(\bar{\mat W}_d\vect h_{t-1,d}
    +\mat W^{\mathrm{in}}_d\tilde{\vect h}_{t,d-1}\right),
    \qquad&& d=2,\dots,D,\\
\text{Layer d update} &:\quad&
\vect h_{t,d}
&=(1-\alpha_d)\vect h_{t-1,d}+\alpha_d\vect\omega_{t,d},
    \qquad&& d=2,\dots,D,\\
\text{Regression feature vector} &:\quad&
\tilde{\vect x}_t
&=\big[\vect h_{t,D}^\top,\ k(\tilde{\vect h}_{t,1})^\top,\dots,
    k(\tilde{\vect h}_{t,D-1})^\top,\ \vect u_t^\top\big]^\top,\\
\text{Intercept and linear predictor} &:\quad&
\vect x_t&=(1,\tilde{\vect x}_t^\top)^\top,\qquad
\mu_t=\vect x_t^\top\vect\beta .
\end{alignedat}
\end{equation}
Here \(k(\cdot)\) is an elementwise activation applied to reduced lower-layer
states, commonly \(k=\mathrm{id}\) or \(k=f\).

\medskip
\noindent
The input vector \(\vect u_t\) bundles response lags and exogenous covariates;
the intercept is appended only in \(\vect x_t\). The leaky update is a convex
combination of the previous state and the current nonlinear update: small
\(\alpha_d\) retains more memory, while large \(\alpha_d\) reacts more quickly
to the current input. The leak rate \(\alpha_d\) and target radius \(\rho_d\)
therefore control the empirical time scale and persistence of the deterministic
state recursion. The echo-state property must be assessed separately. Reducers
\(\mat Q_{d-1}\) may be fixed random projections, optionally orthonormalized,
or linear reducers such as principal component analysis (PCA) computed from the
fitting data.

The regression feature vector concatenates the top-layer state, transformed reduced
lower-layer states, and \(\vect u_t\), yielding
\(\tilde{\vect x}_t\in\R^r\). Appending an intercept gives
\(\vect x_t\in\R^{r+1}\), and the coefficient vector maps this feature vector to the
scalar location \(\mu_t\). States are initialized at
\(\vect h_{0,d}=\zeros{n_d}\), or from
\(\Normal(\zeros{n_d},\sigma_{h,d}^2\Id{n_d})\). A washout period based on
observed lagged responses is applied before the states enter the regression;
this discards early transient states that depend strongly on the arbitrary
initialization. Washout length and scaling choices can affect short-horizon
behavior. The resulting weight matrices remain fixed throughout posterior
inference.

\subsection{Extended Asymmetric Laplace Working Likelihood}
\label{subsec:exal}
The quantile-fixed generalized asymmetric Laplace distribution of
\citet{yan2025new} extends the asymmetric Laplace (AL) working likelihood
while keeping the target quantile fixed. We denote this likelihood by
\(\exAL\), for extended asymmetric Laplace. The standard AL likelihood is
recovered as a special case, but its skewness is
fixed once \(p_0\) is chosen. The \(\exAL\) asymmetry parameter allows mode,
skewness, and tail behavior to vary within a quantile-fixed likelihood. At
fixed level \(p_0\), \(\mu\) is the \(p_0\)th quantile, \(\sigma>0\) is a
scale parameter, and \(\gamma\) is an asymmetry parameter with bounded support
\((L,U)\). The endpoints \(L\) and \(U\) solve \(g(\gamma)=1-p_0\) and
\(g(\gamma)=p_0\), respectively, where
\(g(\gamma)=2\Phi(-|\gamma|)\exp(\gamma^2/2)\), and \(\Phi\) is the standard
Normal cumulative distribution function.

\medskip
\noindent
Posterior computation uses the stochastic representation of the \(\exAL\)
distribution. If
\[
y\sim\exAL_{p_0}(\mu,\sigma,\gamma),
\]
then there exist independent \(\epsilon\sim\Normal(0,1)\), \(z\sim\Exp(1)\),
and \(s\sim\TN(0,1)\) such that
\[
y
= \mu + \lambda(\gamma)\sigma s
+ A(\gamma)\sigma z
+ \{B(\gamma)\sigma^2 z\}^{1/2}\epsilon,
\]
where \(A(\gamma)=\{1-2p_\gamma\}/\{p_\gamma(1-p_\gamma)\}\),
\(B(\gamma)=2/\{p_\gamma(1-p_\gamma)\}\),
\(C(\gamma)=\{\ind\{\gamma>0\}-p_\gamma\}^{-1}\),
\(p_\gamma=\ind\{\gamma<0\}+\{p_0-\ind\{\gamma<0\}\}/g(\gamma)\), and
\(\lambda(\gamma)=C(\gamma)|\gamma|\).
Setting \(v=\sigma z\) gives
\(v\mid\sigma\sim\Exp(\text{rate}=1/\sigma)\) and the conditional Gaussian
form used for Q--DESN posterior computation:
\[
y \mid \mu,\sigma,\gamma,s,v
\sim \Normal\!\big(\mu+\lambda(\gamma)\sigma s + A(\gamma)v,\,
B(\gamma)\sigma v\big).
\]
All quantities \(g(\gamma)\), \(p_\gamma\), \(A(\gamma)\), \(B(\gamma)\),
\(C(\gamma)\), and \(\lambda(\gamma)\) depend on the target quantile
level \(p_0\); this dependence is suppressed for readability. Once
\((p_0,\gamma)\) is fixed, these are deterministic constants in the Gaussian
augmentation.

\section{Model Specification}
\label{sec:model}

\noindent
Fix a quantile level \(p_0\in(0,1)\). After reservoir construction and
washout, \(\mat X=(\vect x_1,\ldots,\vect x_T)^\top\) is treated as fixed,
and all posterior uncertainty is conditional on this design. The statistical
model is a Bayesian quantile regression for \(\vect y\) given \(\mat X\). The
observation model and priors define the posterior distribution; the Gaussian
augmentation below is a computational representation of the asymmetric
working likelihood. Dependence on \(p_0\) is suppressed when no ambiguity
arises.

\subsection{Quantile Deep Echo State Network (Q--DESN)}
\label{subsec:qdesn}

The Q--DESN is obtained by inserting the fixed DESN feature map into the
exAL working likelihood of Section~\ref{subsec:exal}. Let
\(\tilde{\vect x}_t\) be the feature vector in
Eq.~\eqref{eq:desn-hierarchy}, and set
\(\vect x_t=(1,\tilde{\vect x}_t^\top)^\top\). Conditional on the fixed design
matrix, the marginal working likelihood is
\[
y_t\mid\vect x_t,\vect\beta,\sigma,\gamma
\sim
\exAL_{p_0}(\vect x_t^\top\vect\beta,\sigma,\gamma),
\qquad t=1,\ldots,T.
\]
Thus the dynamic linear predictor \(\vect x_t^\top\vect\beta\) is the model's
conditional \(p_0\)-quantile, while \(\sigma\) and \(\gamma\) control the
scale and shape of the working error distribution. For posterior computation,
the same exAL mixture representation is applied at each time point:

\begin{equation}
\label{eq:qdesn-aug}
\begin{alignedat}{2}
\text{Quantile linear predictor} &:\quad&
\mu_t&=\vect x_t^\top\vect\beta,\qquad
\vect\beta=(\beta_0,\beta_1,\dots,\beta_r)^\top\in\R^{r+1},\\
\text{Gaussian conditional} &:\quad&
y_t \mid \vect\beta,\sigma,\gamma,s_t,v_t,\vect x_t
&\sim \Normal\!\Big(\mu_t+\lambda(\gamma)\sigma s_t + A(\gamma)v_t,\,
B(\gamma)\sigma v_t\Big),\\
\text{Positive-shift latent} &:\quad&
s_t&\sim \TN(0,1),\\
\text{Scale-mixture latent} &:\quad&
v_t\mid\sigma&\sim \Exp(\text{rate}=1/\sigma).
\end{alignedat}
\end{equation}
The display separates the statistical quantile model from the computational
augmentation. The first line defines the quantile path; the second gives the
conditional Gaussian observation equation; and the last two lines introduce
auxiliary variables. The constants \(A(\gamma)\), \(B(\gamma)\), and
\(\lambda(\gamma)\) are inherited from Section~\ref{subsec:exal}. The DESN
feature map is fixed before posterior inference, while \(\vect\beta\),
\(\sigma>0\), and \(\gamma\in(L,U)\) are posterior unknowns. Integrating out
\((s_t,v_t)\) recovers the marginal exAL likelihood. These variables expose
conditionally Gaussian updates while preserving the quantile target.
Complete-data joint distributions for the AL and exAL likelihoods with ridge
and \RHS{} priors are given in Section S4 of the supplement.

\subsection{Single-Level Prior Specification}
\label{subsec:priors}

The intercept is modeled separately from the slope coefficients:
\(\beta_0\sim\Normal(b_0,V_0)\), with \(b_0\in\R\) and \(V_0>0\). This
lets shrinkage act on the high-dimensional DESN features without forcing the
overall fitted quantile level toward zero.

\noindent\textbf{Ridge prior.}
For the non-intercept coefficients
\(\vect\beta_{\mathrm{slope}}=(\beta_1,\ldots,\beta_r)^\top\), the default ridge prior is
\begin{equation}
\vect\beta_{\mathrm{slope}}\mid\kappa_\beta^2
\sim \Normal(\zeros{r},\,\kappa_\beta^2\Id{r}),
\label{eq:ridge-prior-main}
\end{equation}
where \(\kappa_\beta^2>0\) is fixed or assigned a weakly informative scale
prior. This Gaussian prior gives dense baseline regularization.

\noindent\textbf{Adaptive shrinkage.}
The regularized horseshoe (\RHS{}) prior is the adaptive shrinkage
alternative. It starts from the ordinary horseshoe
\citep{CarvalhoPolsonScott2010HS}, adds slab regularization
\citep{PiironenVehtari2017RHS}, and uses the product representation of
\citet{NishimuraSuchard2023SSS} to retain closed-form scale updates. For
non-intercept coefficients \(j=1,\ldots,r\), the ordinary horseshoe is
\begin{equation}
\beta_j\mid\tau,\lambda_j\sim\Normal(0,\tau^2\lambda_j^2),\qquad
\lambda_j\sim C^+(0,1),\qquad
\tau\sim C^+(0,\tau_0),
\label{eq:hs-ordinary}
\end{equation}
where \(\tau\) is the global scale and \(\lambda_j\) is coefficient-specific.
This prior concentrates mass near zero while allowing some coefficients to
escape strong shrinkage. The regularized version introduces a slab scale
\(\zeta>0\) to control the largest prior variances and
sets
\begin{equation}
V_j(\lambda_j,\tau,\zeta)=
\left(\zeta^{-2}+\tau^{-2}\lambda_j^{-2}\right)^{-1}
=\frac{\zeta^2\tau^2\lambda_j^2}{\zeta^2+\tau^2\lambda_j^2},
\qquad j=1,\ldots,r,
\label{eq:rhs-var}
\end{equation}
with conditional prior
\begin{equation}
\beta_j\mid\lambda_j,\tau,\zeta\sim\Normal(0,V_j),\qquad j=1,\ldots,r.
\label{eq:beta-prior-main}
\end{equation}
This is the RHS coefficient prior used in the adaptive-shrinkage Q--DESN
fits. Equation~\eqref{eq:rhs-var} gives the interpretation: if
\(\tau^2\lambda_j^2\ll\zeta^2\), then
\(V_j\approx\tau^2\lambda_j^2\), recovering ordinary horseshoe behavior; if
\(\tau^2\lambda_j^2\gg\zeta^2\), then \(V_j\approx\zeta^2\), so the slab caps
the effective prior variance.

The regularized horseshoe can be written directly as in
\citet{PiironenVehtari2017RHS}. For Gibbs and coordinate-ascent variational
updates, however, the direct global-local block includes an extra
\(\{1+\tau^2\lambda_j^2/\zeta^2\}^{1/2}\) factor, so the ordinary horseshoe
updates for \((\tau,\lambda_j)\) are not preserved. We therefore use the
equivalent product representation of \citet{NishimuraSuchard2023SSS}, which
keeps the same conditional coefficient prior while restoring closed-form scale
updates:
\begin{equation}
p(\beta_j,\lambda_j\mid\tau,\zeta)
\propto
\exp\!\left(-\frac{\beta_j^2}{2\tau^2\lambda_j^2}\right)
\exp\!\left(-\frac{\beta_j^2}{2\zeta^2}\right)
\lambda_j^{-1}(1+\lambda_j^2)^{-1},
\label{eq:ns-joint-prior}
\end{equation}
where proportionality is with respect to \((\beta_j,\lambda_j)\). The
\((\tau,\zeta)\)-dependent Gaussian normalizing factors are retained in the
full-conditionals derivations so that the inverse-gamma shape parameters are
recovered correctly. Thus the RHS option gives adaptive global-local shrinkage
while retaining inverse-gamma updates for the local and global scales.

Closed-form scale updates use the inverse-gamma representation of half-Cauchy
priors \citep{MakalicSchmidt2016SimpleSampler}:
\begin{align}
\lambda_j^2\mid\nu_j&\sim\IG\!\left(\frac12,\frac{1}{\nu_j}\right),\qquad
\nu_j\sim\IG\!\left(\frac12,1\right),\quad j=1,\ldots,r,
\label{eq:hs-aux-local}\\
\tau^2\mid\xi&\sim\IG\!\left(\frac12,\frac{1}{\xi}\right),\qquad
\xi\sim\IG\!\left(\frac12,\frac{1}{\tau_0^2}\right).
\label{eq:hs-aux-global}
\end{align}
The slab scale is either fixed or assigned
\begin{equation}
\zeta^2\sim\IG(a_\zeta,b_\zeta).
\label{eq:zeta-prior}
\end{equation}
Under \eqref{eq:ns-joint-prior}, this prior remains compatible with
closed-form updating \citep{NishimuraSuchard2023SSS}.

The likelihood parameters are assigned
\(\sigma\sim\IG(a_\sigma,b_\sigma)\) and
\(\gamma\sim \pi_\gamma(\gamma)\,\ind\{L<\gamma<U\}\), with \((L,U)\)
determined by the quantile-fixing construction as described in
Section~\ref{subsec:exal}. For \RHS{} fits, collect
\(\Theta=(\vect\beta,\sigma,\gamma,\vect\lambda^2,\tau^2,\vect\nu,\xi,\zeta^2)\),
where \(\vect\lambda^2=(\lambda_1^2,\dots,\lambda_r^2)^\top\) and
\(\vect\nu=(\nu_1,\dots,\nu_r)^\top\). The entry \(\zeta^2\) is omitted when
the slab scale is fixed. Ridge fits replace the global-local shrinkage block
by \(\kappa_\beta^2\) when that scale is assigned a prior.

\subsection{Joint Quantile-Vector Regression with DESN Features}
\label{subsec:joint-qdesn}

The single-level model can be fit independently over a grid of quantile levels,
with separate regression coefficients at each level.
When several conditional quantiles are estimated together, let
\(0<p_1<\cdots<p_Q<1\) be the fixed quantile grid and let \(\mat Z\) denote
the \(T\times r\) non-intercept part of the fixed design matrix. For Q--DESN,
\(\mat Z\) is obtained from the DESN feature matrix after removing the
intercept column. The joint quantile-vector regression separates intercepts
from high-dimensional slopes:
\begin{equation}
Q_{p_q}(y_t\mid\mathcal F_t)
=\alpha_q+\tilde{\vect x}_t^\top\vect\beta_q,
\qquad q=1,\ldots,Q,\quad t=1,\ldots,T,
\label{eq:joint-qdesn-readout}
\end{equation}
where \(\alpha_q\in\R\) is the quantile-specific intercept and
\(\vect\beta_q\in\R^r\) is the slope vector at level \(p_q\). By default, the
regularized-horseshoe prior is assigned to the high-dimensional slope vectors; intercepts are either
weakly regularized independently or represented through ordered gaps when a
stronger monotonicity convention is desired.

This construction generalizes the single-level RHS Q--DESN by replacing the
single intercept-slope coefficient vector \((\beta_0,\vect\beta_{\mathrm{slope}})\) with
\(\{(\alpha_q,\vect\beta_q)\}_{q=1}^Q\). For \(Q=1\), the individual RHS
Q--DESN is recovered by taking \(\alpha_1\) as the single-level intercept,
identifying \(\vect\beta_1\) with the single-level non-intercept slope vector,
and fixing \(\vect\beta_{\mathrm{base}}=\zeros{r}\). The regularized-horseshoe
prior is then assigned directly to \(\Delta_1=\vect\beta_1\).

\noindent\textbf{Composite working likelihood.}
For each \(q\), the working likelihood uses the exAL family at the
corresponding target level,
\begin{equation}
y_t\mid\alpha_q,\vect\beta_q,\sigma_q,\gamma_q
\sim
\exAL_{p_q}\!\left(
\alpha_q+\tilde{\vect x}_t^\top\vect\beta_q,\sigma_q,\gamma_q
\right).
\label{eq:joint-qdesn-exal}
\end{equation}
The product of the \(Q\) quantile-specific contributions defines a composite
working likelihood for the repeated use of the same response series. The
resulting posterior jointly estimates the quantile-indexed regression
coefficients. The analyses use this untempered composite posterior.
Future-response simulation uses the forecast construction
in Section~\ref{subsec:forecast_hstep}.

\noindent\textbf{Shrinkage across adjacent quantile levels.}
For a multi-quantile grid, the joint prior adapts the Quantile-Varying
Parameter idea that slopes at neighboring quantile levels should be
similar unless the data support tail-specific differences
\citep{KohnsSzendrei2025QVP}. Let
\(\vect\beta_{\mathrm{base}}\in\R^r\) be a baseline slope vector and define
adjacent coefficient differences
\[
\Delta_{1,j}=\beta_{1,j}-\beta_{\mathrm{base},j},\qquad
\Delta_{q,j}=\beta_{q,j}-\beta_{q-1,j},\quad q=2,\ldots,Q .
\]
We assign an ordinary RHS prior to the baseline slope and RHS priors with
quantile-gap-specific global scales to the adjacent coefficient differences:
\begin{align}
\beta_{\mathrm{base},j}\mid\lambda^0_j,\tau^0,\zeta_0
&\sim \Normal(0,V^0_j),&
V^0_j
&=\frac{\zeta_0^2(\tau^0)^2(\lambda^0_j)^2}
{\zeta_0^2+(\tau^0)^2(\lambda^0_j)^2},
\label{eq:joint-baseline-rhs}\\
\Delta_{q,j}\mid\lambda^\Delta_{q,j},\tau^\Delta_q,\zeta_\Delta
&\sim \Normal(0,V^\Delta_{q,j}),&
V^\Delta_{q,j}
&=\frac{\zeta_\Delta^2(\tau^\Delta_q)^2(\lambda^\Delta_{q,j})^2}
{\zeta_\Delta^2+(\tau^\Delta_q)^2(\lambda^\Delta_{q,j})^2}.
\label{eq:joint-innovation-rhs}
\end{align}
This construction separates feature effects shared across quantile levels,
\(\vect\beta_{\mathrm{base}}\), from effects that vary across adjacent quantile
levels. It is related to interquantile shrinkage and composite-quantile
regularization in frequentist quantile regression
\citep{JiangBondellWang2014InterquantileShrinkage,ZouYuan2008CompositeQR},
but the shrinkage is placed on the quantile-indexed regression coefficients
conditional on the fixed feature design.

The next display rewrites the \(Q\) augmented likelihood contributions as one
stacked Gaussian regression conditional on the exAL latent variables and the
quantile-specific intercepts. This computational representation is algebraically
equivalent to the joint quantile model above.
Let
\(\vect\beta_{\mathrm{st}}=(\vect\beta_1^\top,\ldots,\vect\beta_Q^\top)^\top\),
\(\mat Z_{\mathrm{st}}=\Id{Q}\otimes\mat Z\), and
\(\vect\alpha_{\mathrm{st}}
=(\alpha_1\ones{T}^\top,\ldots,\alpha_Q\ones{T}^\top)^\top\). Under the
exAL augmentation, define
\[
d_{q,t}=\lambda(\gamma_q)\sigma_q s_{q,t}+A(\gamma_q)v_{q,t},\qquad
\omega_{q,t}=B(\gamma_q)\sigma_q v_{q,t},
\]
and stack these quantities into \(\vect d\) and diagonal
\(\mat\Omega=\diag(\omega_{q,t})\). The conditional Gaussian likelihood is
\[
\vect y_{\mathrm{st}}\mid\vect\beta_{\mathrm{st}},\vect\alpha,\vect d,\mat\Omega
\sim
\Normal\!\left(
\vect\alpha_{\mathrm{st}}+\mat Z_{\mathrm{st}}\vect\beta_{\mathrm{st}}+\vect d,\,
\mat\Omega
\right),
\qquad
\vect y_{\mathrm{st}}=\ones{Q}\otimes\vect y .
\]
Let \(\mat H\) be the block first-difference matrix such that
\[
\mat H\vect\beta_{\mathrm{st}}
=
(\vect\beta_1^\top,
(\vect\beta_2-\vect\beta_1)^\top,\ldots,
(\vect\beta_Q-\vect\beta_{Q-1})^\top)^\top .
\]
With
\(\tilde{\vect\beta}_{\mathrm{base}}
=(\vect\beta_{\mathrm{base}}^\top,\zeros{r}^\top,\ldots,\zeros{r}^\top)^\top\)
and diagonal \(\mat D_\Delta\) containing the RHS adjacent-difference variances
\(V^\Delta_{q,j}\), the conditional prior is proportional to
\[
\exp\left[
-\frac12
(\mat H\vect\beta_{\mathrm{st}}-\tilde{\vect\beta}_{\mathrm{base}})^\top
\mat D_\Delta^{-1}
(\mat H\vect\beta_{\mathrm{st}}-\tilde{\vect\beta}_{\mathrm{base}})
\right].
\]
Consequently, the Gaussian full conditional for the stacked vector of
quantile-specific slopes has
precision
\begin{equation}
\mat K_\beta
=
\mat Z_{\mathrm{st}}^\top\mat\Omega^{-1}\mat Z_{\mathrm{st}}
+\mat H^\top\mat D_\Delta^{-1}\mat H,
\label{eq:joint-kbeta-main}
\end{equation}
and mean
\begin{equation}
\vect m_\beta
=
\mat K_\beta^{-1}
\left[
\mat Z_{\mathrm{st}}^\top\mat\Omega^{-1}
(\vect y_{\mathrm{st}}-\vect\alpha_{\mathrm{st}}-\vect d)
+\mat H^\top\mat D_\Delta^{-1}\tilde{\vect\beta}_{\mathrm{base}}
\right].
\label{eq:joint-mbeta-main}
\end{equation}
The block-banded structure in \eqref{eq:joint-kbeta-main} permits sparse
Gaussian calculations for the joint model.

\noindent\textbf{Adjacent-difference shrinkage and quantile crossings.}
The non-crossing interpretation is local in adjacent quantile gaps. For
neighboring levels,
\[
Q_{p_q}(y_t\mid\mathcal F_t)-Q_{p_{q-1}}(y_t\mid\mathcal F_t)
=
(\alpha_q-\alpha_{q-1})
+\tilde{\vect x}_t^\top(\vect\beta_q-\vect\beta_{q-1}).
\]
On a bounded feature domain, for example after an explicit scaling or clipping
condition that puts each component of \(\tilde{\vect x}_t\) in \([-1,1]\), the
inequality
\(\alpha_q-\alpha_{q-1}\ge
\|\vect\beta_q-\vect\beta_{q-1}\|_1\) is sufficient for non-crossing. The
adjacent-difference RHS prior shrinks \(\vect\beta_q-\vect\beta_{q-1}\) toward
zero and can reduce adjacent-level crossings, although it does not enforce
monotonicity. Exact monotonicity requires an
explicit ordered-intercept parameterization, posterior rejection rule, isotonic
projection, or monotone rearrangement. The synthetic joint study below reports
crossings before rearrangement separately from the pre-specified monotone rearrangement used for
scored quantile grids.

\section{Posterior Inference}
\label{sec:inference}

\subsection{Single-Level Augmented Posterior}
\label{subsec:posterior-target}

For one probability level \(p_0\), the unknown quantities are the regression
coefficients and the working-likelihood and shrinkage parameters, conditional
on the fixed DESN design. Let \(\mat Z\) denote the
non-intercept design,
\(\alpha\) the intercept, and \(\vect\beta_1\) the non-intercept slope vector.
Let \(\Theta_\Delta\) collect the local, global, auxiliary, and slab scales for
the RHS prior when the adaptive prior is used. For the exAL working likelihood,
let \(\{v_t,s_t:t=1,\ldots,T\}\) be the scale-mixture and positive-shift
latent variables; for AL, the \(s_t\) and \(\gamma\) blocks are omitted. With
\(\vect d_1\), \(\mat\Omega_1\), and \(\omega_t\) collecting the corresponding
complete-data shifts and variances from the augmented likelihood, the
single-level complete-data posterior can be written, up to constants not
depending on unknowns, as
\begin{align}
&p(\alpha,\vect\beta_1,\sigma,\gamma,\{v_t,s_t\}_{t=1}^T,\Theta_\Delta
\mid\vect y,\mat Z)
\propto
\exp\left[
-\frac12
(\vect y-\alpha\ones{T}-\mat Z\vect\beta_1-\vect d_1)^\top
\mat\Omega_1^{-1}
(\vect y-\alpha\ones{T}-\mat Z\vect\beta_1-\vect d_1)
\right]
\nonumber\\
&\quad\times
\prod_{t=1}^T
\left\{
(\omega_t)^{-1/2}p(v_t\mid\sigma)p(s_t)
\right\}
p(\sigma)\pi_{\gamma}(\gamma)\ind\{L<\gamma<U\}
p(\alpha)\,p_\Delta^{(1)}(\vect\beta_1\mid\mat D_\Delta)p(\Theta_\Delta).
\label{eq:single-posterior-collapse-main}
\end{align}
Here
\[
p_\Delta^{(1)}(\vect\beta_1\mid\mat D_\Delta)
\propto
\exp\left(-\frac12\vect\beta_1^\top\mat D_\Delta^{-1}\vect\beta_1\right)
\]
is the ordinary RHS slope prior, with \(\Delta_1=\vect\beta_1\). The ridge
case replaces this factor by the Gaussian ridge prior. This display gives the
posterior distribution for the individual Q--DESN regression; the joint
quantile-vector construction below reuses the same augmented Gaussian
structure after stacking levels and replacing the direct slope prior by a
regularized-horseshoe prior on quantile-indexed slope increments.

\subsection{Joint Quantile-Vector Extension}
\label{subsec:joint-inference-main}

For a grid \(p_1<\cdots<p_Q\), let \(\Theta_\Delta\) collect the RHS scales for
adjacent coefficient differences and let \(\Theta_0\) collect the baseline-slope RHS
scales. Let \(\Theta_Q\) collect \(\vect\alpha\),
\(\vect\beta_{\mathrm{st}}\), the likelihood parameters,
\(\Theta_\Delta\), and, for \(Q>1\),
\((\vect\beta_{\mathrm{base}},\Theta_0)\). Let
\(\vect a_Q=\{v_{q,t},s_{q,t}:q=1,\ldots,Q;\,t=1,\ldots,T\}\) collect the
exAL auxiliary variables. With the stacked quantities
\(\vect y_{\mathrm{st}}\), \(\vect\alpha_{\mathrm{st}}\),
\(\mat Z_{\mathrm{st}}\), \(\vect d\), and \(\mat\Omega\) defined in
Section~\ref{subsec:joint-qdesn}, the joint complete-data posterior is
\begin{align}
&p(\Theta_Q,\vect a_Q\mid\vect y,\mat Z)
\propto
\exp\left[
-\frac{1}{2}
(\vect y_{\mathrm{st}}-\vect\alpha_{\mathrm{st}}
-\mat Z_{\mathrm{st}}\vect\beta_{\mathrm{st}}-\vect d)^\top
\mat\Omega^{-1}
(\vect y_{\mathrm{st}}-\vect\alpha_{\mathrm{st}}
-\mat Z_{\mathrm{st}}\vect\beta_{\mathrm{st}}-\vect d)
\right]
\nonumber\\
&\quad\times
\prod_{q=1}^Q\prod_{t=1}^T
\left\{
(\omega_{q,t})^{-1/2}
p(v_{q,t}\mid\sigma_q)p(s_{q,t})
\right\}
\prod_{q=1}^Q p(\sigma_q)\pi_{\gamma_q}(\gamma_q)\ind\{L_q<\gamma_q<U_q\}
\nonumber\\
&\quad\times
p(\vect\alpha)\,
p_\Delta^{(Q)}(\vect\beta_{\mathrm{st}}\mid\vect\beta_{\mathrm{base}},\mat D_\Delta)\,
p(\Theta_\Delta)\,\mathcal B_Q,
\label{eq:joint-posterior-target-main}
\end{align}
where
\[
\mathcal B_Q=
\begin{cases}
p(\vect\beta_{\mathrm{base}}\mid\Theta_0)p(\Theta_0), & Q>1,\\
1, & Q=1.
\end{cases}
\]
The Gaussian prior factor \(p_\Delta^{(Q)}\) is induced by the
first-difference matrix \(\mat H\) and difference covariance
\(\mat D_\Delta\) in \eqref{eq:joint-kbeta-main}--\eqref{eq:joint-mbeta-main}.
For \(Q>1\), it is the adjacent-difference prior centered on the baseline
slope, so the \(\vect\beta_{\mathrm{base}}\) argument is active. The AL
special case omits \(s_{q,t}\), \(\gamma_q\), and the positive-truncated
Normal factors. Setting \(Q=1\), taking \(\alpha_1=\alpha\),
\(\vect y_{\mathrm{st}}=\vect y\), \(\mat Z_{\mathrm{st}}=\mat Z\),
\(\vect\alpha_{\mathrm{st}}=\alpha\ones{T}\),
\(\vect\beta_{\mathrm{st}}=\vect\beta_1\), and dropping \(\mathcal B_Q\)
recovers the single-level posterior in
\eqref{eq:single-posterior-collapse-main}. The exAL
augmentation preserves the Gaussian slope-path update in
\eqref{eq:joint-kbeta-main}--\eqref{eq:joint-mbeta-main}, but the
scale-asymmetry blocks \((\sigma_q,\gamma_q)\) remain non-conjugate. We use two
computational approaches for this fixed-design posterior: an MCMC sampler with
separate exAL block updates and a mean-field variational Bayes approximation
with Laplace--Delta updates for the non-conjugate blocks.
In the algorithms below, GIG, positive-truncated Normal, Gaussian, and
inverse-gamma entries denote closed-form full conditionals for MCMC and
closed-form coordinate factors for VB. The exAL scale-asymmetry block requires
a non-conjugate update.

\subsection{Markov Chain Monte Carlo}

The augmented likelihood and shrinkage hierarchy give a partially Gibbs
sampler. Closed-form blocks are sampled from their named full conditionals;
the complete-data posterior densities, full conditionals, coordinate-ascent
variational inference (CAVI) factors, and evidence lower bound (ELBO) terms for
AL and exAL likelihoods under ridge and \RHS{} priors are given in Sections
S5--S9 of the supplement. Under the Nishimura--Suchard hierarchy, the
global-local scales have inverse-gamma updates. Each exAL block
\((\sigma_q,\gamma_q)\) remains non-conjugate and is updated with
Metropolis--Hastings or slice moves on transformed coordinates.
Algorithm~\ref{alg:mcmc} targets
\eqref{eq:joint-posterior-target-main}; with \(Q=1\), it uses the collapsed
posterior in \eqref{eq:single-posterior-collapse-main}. Ridge and AL variants
omit the corresponding shrinkage,
\(s_{q,t}\), and \(\gamma_q\) blocks. Diagnostics use trace plots, effective
sample size summaries, crossing diagnostics for multi-quantile grids, and
monitoring of the \((\sigma_q,\gamma_q)\) updates, including acceptance rates
when Metropolis--Hastings is used.

\begin{algorithm}[H]
\caption{MCMC sampler for the Q--DESN posterior. For \(Q=1\),
set \(\vect\beta_{\mathrm{base}}=\zeros{r}\), place the RHS prior on
\(\Delta_1=\vect\beta_1\), and omit baseline-slope blocks. AL and ridge
variants omit the corresponding \(s_{q,t}\), \(\gamma_q\), and shrinkage
blocks.}
\label{alg:mcmc}
\begin{algorithmic}[1]
\REQUIRE Data $\{y_t,\tilde{\vect x}_t\}_{t=1}^T$, quantile grid $p_1<\cdots<p_Q$, and priors
\REQUIRE MCMC iterations $N$ and convergence-monitoring settings
\STATE Initialize $\vect\alpha$, $\vect\beta_{\mathrm{st}}$, likelihood parameters, latent variables, and RHS scales
\IF{$Q>1$}
  \STATE Initialize $\vect\beta_{\mathrm{base}}$ and its baseline RHS scales
\ENDIF
\FOR{$i=1$ to $N$}
  \STATE \textbf{Latent scales:} for each \(q,t\), draw \(v_{q,t}\mid\cdot\) from the AL or exAL GIG full conditional
  \STATE \textbf{exAL shifts:} for exAL, draw \(s_{q,t}\mid\cdot\) from its positive-truncated Normal full conditional; omit this step for AL
  \STATE \textbf{Slope path:} compute \(\mat K_\beta\) and \(\vect m_\beta\) from \eqref{eq:joint-kbeta-main}--\eqref{eq:joint-mbeta-main}
  \STATE \textbf{Slope path:} draw \(\vect\beta_{\mathrm{st}}\mid\cdot\sim\Normal(\vect m_\beta,\mat K_\beta^{-1})\)
  \IF{$Q>1$}
    \STATE \textbf{Baseline:} draw \(\vect\beta_{\mathrm{base}}\) from its Gaussian conditional and draw its RHS scales from inverse-gamma full conditionals
  \ENDIF
  \STATE \textbf{Intercepts:} draw \(\vect\alpha\) from its Gaussian conditional, or update the ordered-gap block when exact intercept monotonicity is imposed
  \STATE \textbf{\RHS{} scales:} draw adjacent-difference local, global, auxiliary, and slab scales from inverse-gamma full conditionals; for \(Q=1\), this is the ordinary single-level RHS block for \(\vect\beta_1\)
  \STATE \textbf{AL scale:} for AL, draw each \(\sigma_q\mid\cdot\) from its inverse-gamma full conditional and omit \(\gamma_q\)
  \STATE \textbf{exAL scale--asymmetry:} for exAL, update each \((\sigma_q,\gamma_q)\) on transformed coordinates targeting its non-conjugate conditional kernel
  \STATE Retain post-burn-in draws and compute convergence and quantile-crossing diagnostics
\ENDFOR
\end{algorithmic}
\end{algorithm}

\subsection{Variational Approximation}
\label{sec:vb}

For analyses requiring repeated model fitting, we use a mean-field variational
approximation to the same augmented posterior. The
Gaussian, GIG, positive-truncated Normal, inverse-gamma, and RHS scale factors
retain the closed-form coordinate structure implied by
\eqref{eq:joint-posterior-target-main}. The exAL scale-asymmetry block
\((\sigma_q,\gamma_q)\) is nonconjugate and is approximated with the
Laplace--Delta treatment for non-conjugate variational inference of
\citet{wang2013nonconjugatevb}, adapted to Q--DESN as detailed in
the supplement. We denote this model-specific approximation by VB--LD. MCMC
remains the primary posterior simulation method for the MCMC results reported
below.

Equations~\eqref{eq:joint-posterior-target-main} and
\eqref{eq:single-posterior-collapse-main} define the posterior distributions
used for inference, and Algorithm~\ref{alg:mcmc} implements the corresponding
MCMC updates. The supplement provides the single-level full conditionals,
variational factors, ELBO, joint-model calculations, and quantile-crossing
diagnostics. For a single fitted quantile, Algorithm~\ref{alg:mcmc} uses
\(Q=1\) and the collapsed RHS prior. For multiple fitted levels, the same
sampler uses the baseline and adjacent-difference blocks. The supplement also
gives the monotone-rearrangement details.

\section{Multi-Step Quantile Forecasting and Scoring}
\label{sec:forecast}

For a forecast origin \(T\), a quantile path is the sequence of fitted
conditional quantiles over horizons \(h=1,\ldots,H\). A single-level Q--DESN
fit supplies this path for one probability level \(p_0\). Independent fits at
\(K\) levels \(p_1<\cdots<p_K\) supply one path at each level, and their raw
quantile estimates may cross. We obtain ordered quantiles for display and
scoring with the weighted isotonic projection
\begin{equation}
\vect q_{T,h}^*
=
\argmin_{z_1\le\cdots\le z_K}
\sum_{k=1}^K w_k(\widehat q_{T,h,k}-z_k)^2,
\qquad w_k>0,
\label{eq:monotone-reporting-main}
\end{equation}
where \(\vect q_{T,h}^*=(z_1,\ldots,z_K)^\top\) at the minimizer. The
reported analyses use \(w_k=1\). When forecasts between fitted levels are
needed, linear interpolation defines the quantile curve over \([p_1,p_K]\);
any extension into the two tails is specified separately for the application.

\subsection{Multi-Step Quantile Forecast Summaries}
\label{subsec:forecast_hstep}

Forecast summaries are computed conditional on the fitted quantile-regression
coefficients, the fixed reservoir construction, and the covariates available
or specified at the forecast origin. Before forecasting, the reservoir state is
obtained by running Eq.~\eqref{eq:desn-hierarchy} through the observed record
with observed response lags. After the forecast origin, future response lags
are unknown; if they are needed for recursive multi-step forecasts, they must
be generated within each Monte Carlo trajectory or supplied by a stated
scenario.

For a single-level fit at probability \(p_0\), let
\(\theta_{p_0}^{(m)}
=(\vect\beta_{p_0}^{(m)},\sigma_{p_0}^{(m)},\gamma_{p_0}^{(m)})\) denote a
posterior or variational draw. Conditional on the draw-specific future design
row, the fitted quantile location is
\[
\mu_{p_0,T+h}^{(m)}
=\vect x_{T+h}^{(m,p_0)\top}\vect\beta_{p_0}^{(m)}
\]
for horizon \(h\). These \(\mu_{p_0,T+h}^{(m)}\) values are draws of the
fitted \(p_0\)-level quantile location. If a scalar future response path is
needed for recursive lags, the fitted AL or \(\exAL\) working likelihood can be used to generate
\[
y_{T+h}^{(m,p_0)}
\sim
\exAL_{p_0}\!\left(
\mu_{p_0,T+h}^{(m)},
\sigma_{p_0}^{(m)},\gamma_{p_0}^{(m)}
\right).
\]
The resulting response draws are simulations from the fitted working
likelihood. Summaries of \(\mu_{p_0,T+h}^{(m)}\) describe posterior or
variational uncertainty in the fitted quantile location. A closed-form marginal
quantile of \(Y_{T+h}\mid\mathcal F_T\) requires an explicitly stated
marginalization or simulation rule.

For the joint quantile-vector regression, a draw contains the coefficients and
likelihood parameters at each quantile level. The same future design vector
then returns the raw fitted quantile vector
\[
\mu_{j,T+h}^{(m)}
=
\alpha_j^{(m)}
+\tilde{\vect x}_{T+h}^{(m)\top}\vect\beta_j^{(m)},
\qquad j=1,\ldots,Q.
\]
Let
\(\vect q_{T+h}^{(m)}
= (\mu_{1,T+h}^{(m)},\ldots,\mu_{Q,T+h}^{(m)})^\top\). This vector
estimates the conditional quantile path over the fitted grid.

If exact monotonicity is required before scoring or display, let
\(\mathcal M\) denote the stated monotone rearrangement, such as isotonic
projection or rearrangement
\citep{BarlowBrunk1972IsotonicDual,chernozhukov2010}, and set
\[
\vect q_{T+h}^{*,(m)}=\mathcal M\{\vect q_{T+h}^{(m)}\}.
\]
The ordered values define a draw-specific quantile curve on the fitted
interval \([p_1,p_Q]\) by linear interpolation. The reported joint study
evaluates this finite grid using fixed forecast-design rows derived from the
simulated series. Recursive response simulation over \((0,1)\) would
additionally require a specified tail extension.

For forecast evaluation, let \(0<p_1<\cdots<p_K<1\), with \(K\geq2\), and
write \(\mathcal P_K=\{p_1,\ldots,p_K\}\) for the probability levels used to
compute the score. The evaluation levels may differ from the fitted levels
when an estimated quantile function is available at other probabilities. For
one forecast origin and horizon, let
\(q_{T,h,k}^*\) be the reported conditional \(p_k\)-quantile of
\(Y_{T+h}\mid\mathcal F_T\), after monotone rearrangement when it is applied,
and let \(y\) be the realized response. The check loss is
\begin{equation}
\rho_p(e)=e\{p-\ind(e<0)\}.
\label{eq:check-loss-main}
\end{equation}
For a predictive distribution \(F\) with finite first moment and quantile
function \(F^{-1}\), the continuous ranked probability score (CRPS) has the
integrated check-loss representation \citep{GneitingRaftery2007}:
\begin{equation}
\operatorname{CRPS}(F,y)
=2\int_0^1 \rho_p\!\left\{y-F^{-1}(p)\right\}\,dp.
\label{eq:crps-integrated-check-main}
\end{equation}
When quantile forecasts are evaluated on \(\mathcal P_K\), we use the
finite-grid trapezoidal approximation over \([p_1,p_K]\),
\begin{equation}
\aCRPS_{\mathcal P_K}(y;\vect q_{T,h}^*)
=2\sum_{k=1}^K \omega_k
  \rho_{p_k}\!\left(y-q_{T,h,k}^*\right),
\label{eq:acrps-main}
\end{equation}
where
\begin{equation}
\begin{aligned}
\omega_1&=\frac{p_2-p_1}{2},
&\omega_k&=\frac{p_{k+1}-p_{k-1}}{2},\quad 2\le k\le K-1,
&\omega_K&=\frac{p_K-p_{K-1}}{2}.
\end{aligned}
\label{eq:acrps-weights-main}
\end{equation}
These weights satisfy \(\sum_{k=1}^K\omega_k=p_K-p_1\), so \(\aCRPS\)
retains the scale of integrated check loss over the covered probability range.
The joint simulation and GloFAS analyses use seven common evaluation levels
with endpoints \(p_1=0.05\) and \(p_K=0.95\), so their trapezoidal weights sum
to \(0.90\). For fitted quantile regressions, the evaluation levels coincide
with the fitted grid.

In simulations, the known conditional response distribution permits a
distribution-averaged version of the same finite-grid score. Let
\(\mathcal G\) index the evaluated origin--horizon pairs, let \(F_{0,r}\)
denote the true conditional distribution for pair \(r\), and let
\(q_{r,k}^{*,(b)}\) be the monotonically rearranged \(p_k\)-quantile from
posterior draw \(b\). We call the resulting quantity the
data-generating-process (DGP)-integrated finite-grid quantile score and define
it by
\begin{equation}
S_{\mathrm{DGP}}^{(b)}
=\frac{2}{|\mathcal G|}
\sum_{r\in\mathcal G}\sum_{k=1}^K \omega_k
\E_{Y_r\sim F_{0,r}}
\left[
\rho_{p_k}\!\left\{Y_r-q_{r,k}^{*,(b)}\right\}
\right].
\label{eq:dgp-integrated-score-main}
\end{equation}
The posterior draws produce a posterior distribution for
\(S_{\mathrm{DGP}}\). Its mean and equal-tailed 95\% credible interval are
the primary summaries in the joint simulation. The expectation averages over
the known conditional response distribution at the held-out design points,
while conditioning on the simulated fitting series and selected model
specification. The interval therefore excludes variation across newly
generated fitting series and model selection. The score of the
posterior-mean quantile curve and scores against realized outcomes provide
additional diagnostics. In the applications, where \(F_{0,r}\) is unknown,
forecast evaluation uses the realized \(\aCRPS\) in
Eq.~\eqref{eq:acrps-main}. GloFAS applies monotone rearrangement to its
variational posterior-mean quantile estimates before scoring.

\section{Model Diagnostics and Reservoir Specification}
\label{sec:selection}

Let
\[
\mathcal R =
\{D,\{n_d\},\{\tilde n_d\},m,T_0,\{\alpha_d\},\{\rho_d\},
\{\pi^{(W)}_d\},\{\pi^{(\mathrm{in})}_d\},f,k,\text{random-number seed}\}
\]
denote the depth, layer sizes, reduction sizes, lag memory, washout, leak
rates, spectral radii, sparsity probabilities, activations, and random seed
used to construct Eq.~\eqref{eq:desn-hierarchy}. Once \(\mathcal R\) and the
preprocessing convention are fixed, posterior inference is conditional on the
resulting design matrix \(\mat X\). Reservoir specification is therefore part
of the empirical design: the Bayesian analysis estimates regression,
likelihood, and shrinkage parameters conditional on the specified fixed feature
map. Integrating over reservoir specifications would require an additional
hierarchical model.

The ESN literature treats these choices as application-dependent design parameters.
Reservoir size controls the number and diversity of state variables, while leak rates and spectral radii jointly
affect memory and stability, and input scaling determines how strongly the
reservoir is driven into nonlinear regimes
\citep{JaegerLukoseviciusPopoviciSiewert2007LeakyESN,Lukosevicius2012PracticalGuideESN}.
For deep reservoirs, stacking layers can create hierarchical temporal
representations, with additional depth justified by held-out forecast criteria and
state diagnostics
\citep{GallicchioMicheliPedrelli2018DeepESNDesign}. The condition
\(\rho_d<1\) is a useful conservative design choice. For a driven nonlinear
reservoir, empirical state diagnostics are still needed to assess stability
\citep{YildizJaegerKiebel2012ESP}.
Several ESN modification papers identify diagnostic concerns that are visible
before forecast scoring: regression coefficients can reveal predictively
relevant reservoir units, state feedback changes the effective recurrent
dynamics through an external feedback channel, redundant state trajectories
can degrade regression conditioning, and
random reservoir realizations can induce substantial seed-to-seed variability
\citep{FanWangJin2017PNR,EhlersNurdinSoh2025StateFeedback,SaadatFarshadEliasiShokoohiMehr2025OnlineESN,WuFokoueKudithipudi2018StatisticalChallengesESN}.
In Q--DESN these results motivate diagnostics and component-wise sensitivity
analyses. After the feature map is fixed, posterior inference is for
the quantile-regression parameters conditional on that feature map.

The empirical design fixes the fitting, model-selection, and held-out evaluation
period, forecast origins, horizons, quantile levels, input transformations, and
scoring rules before feature maps are compared. We then assess
input alignment, feature dimensions, washout handling, finite state values,
empirical spectral radii, near-constant states or states concentrated near the
activation bounds,
and conditioning of the fixed-feature regression. State-matrix diagnostics include
near-zero-variance columns, pairwise state correlations, singular-value or
covariance spectra, effective rank, and a condition-number or eigenvalue-spread
summary. These diagnostics distinguish a larger reservoir that adds
nonredundant temporal features from one that mainly adds collinear columns to
the fixed design matrix.

The relevant evaluation metric depends on the statistical question. In
simulations with a known conditional quantile path, quantile-path RMSE, mean
absolute error, bias, and correlation assess recovery of the true path. In both
simulations and applications, forecast scores must use common forecast origins
and horizons. For one fitted quantile level, check loss is the proper score for
the reported quantile. Across several levels, \(\aCRPS\) approximates
integrated check loss over the evaluation grid, while interval score, empirical
coverage, and interval length describe selected central intervals. Calibration
of posterior credible bands under a misspecified working likelihood requires
separate assessment.

Reservoir specifications vary depth, width, lag length, washout, leak rate,
spectral radius, sparsity, input scaling, and activation. Specifications are
compared using fitting-period criteria and state diagnostics, including
comparisons across reservoir realizations when computationally feasible. The
selected specification is held fixed for evaluation. Comparisons between VB
and MCMC and across fixed feature maps are reported as sensitivity analyses.

Each empirical comparison is interpreted conditional on its specification grid,
feature-construction defaults, state-quality summaries, forecast origins,
horizons, scoring metrics, reservoir seeds, and final \(\mathcal R\). Claims
about a reservoir specification are therefore conditional on the corresponding
evaluation design. Evidence from additional seeds, forecast origins, and scoring
criteria is needed before generalizing beyond the reported comparison.
State-feedback and principal-neuron reinforcement methods define alternative
feature maps for future study.

\section{Simulation Studies}
\label{sec:simulation}

\noindent
The simulation study addresses two complementary questions. The first asks
whether a single-quantile Q--DESN regression can recover and forecast a
known dynamic quantile path under a 500-observation benchmark following the
family-based comparison strategy of \citet{yan2025new}. The target quantile
level is fixed in advance, the data-generating mechanism is known, and
competing models are compared under matched error families, fitting-sample sizes,
and quantile levels. The second asks whether a joint multi-quantile regression can
recover several known conditional quantile paths under posterior simulation and
reduce adjacent-level crossings before monotone rearrangement. The supplement
reports the corresponding VB analyses and additional forecast comparisons.

In contrast to the linear-regression setting of \citet{yan2025new}, the true
conditional quantile paths, referred to as oracle paths in this simulation, are
dynamic. Q--DESN represents those paths through
high-dimensional deterministic reservoir features. The design therefore
separates two questions: how accurately a fitted model recovers the known
conditional quantile path, and how well it forecasts held-out dynamic blocks.
We write \(p\) for the single target level in the first study and \(\tau\) for
levels in the multi-quantile grid.

The main-text tables report MCMC summaries under fixed evaluation designs:
data-generating mechanisms, rolling-origin definitions, fitting samples,
held-out periods, scoring rules, and model classes are specified before the
comparisons are formed. The supplement gives corresponding approximate VB summaries and additional
diagnostic summaries for the same simulation settings. The single-quantile and
multi-quantile studies use separate fixed designs. Their results address
distinct comparisons and provide complementary evidence.

\subsection{Single-Quantile Dynamic Data-Generating Processes}

For each error family \(f\) and target quantile level
\(p\in\{0.05,0.25,0.50\}\), the synthetic series is generated as
\[
y_{t,p,f}=\mu_{t,f}+\varepsilon_{t,p,f},\qquad
\varepsilon_{t,p,f}=e_{t,f}-H_f^{-1}(p),
\]
where \(H_f\) is the distribution function of the raw error \(e_{t,f}\).
The shift by \(H_f^{-1}(p)\) makes the \(p\)-quantile of
\(\varepsilon_{t,p,f}\) equal to zero. Hence
\[
Q_p(y_{t,p,f}\mid \vect\theta_{t,f})=\mu_{t,f}.
\]
The true conditional quantile path is generated by a six-dimensional dynamic linear
state with local level, local slope, and two harmonic pairs:
\[
\mu_{t,f}=\vect F^\top \vect\theta_{t,f},\qquad
\vect F=(1,0,1,0,1,0)^\top,
\]
with local linear trend and two sine--cosine harmonic pairs. The seasonal
period is \(90\), the harmonics are \(1\) and \(2\), and the state
innovation covariance \(\mat W_\theta\) has diagonal square roots
\[
\sqrt{\diag(\mat W_\theta)}
=(0.005,\ 0.00002,\ 0.004,\ 0.004,\ 0.003,\ 0.003).
\]
The initial covariance is \(\mat C_0=0.01\Id{6}\), and each simulated
trajectory is initialized deterministically at the family-specific state in
Table~\ref{tab:simulation-dgp}. For a fixed family, the same latent path
\(\{\mu_{t,f}\}\) is used across the three values of \(p\); only the centering
constant \(H_f^{-1}(p)\) changes.

\begin{table}[!htbp]
\centering
\small
\begin{tabular}{@{}>{\raggedright\arraybackslash}p{0.15\textwidth}>{\raggedright\arraybackslash}p{0.29\textwidth}>{\centering\arraybackslash}p{0.09\textwidth}>{\centering\arraybackslash}p{0.09\textwidth}>{\raggedright\arraybackslash}p{0.25\textwidth}@{}}
\toprule
Family & Raw error distribution \(H_f\) & Initial level & Initial slope & Seasonal settings \\
\midrule
Gaussian & \(\Normal(0,10^2)\) & 40 & 0.012 &
harmonics \((24,0.35)\), \((8,-0.80)\) \\
Laplace & Laplace with scale \(10\) & 35 & 0.011 &
harmonics \((28,-0.15)\), \((10,0.75)\) \\
Gaussian mixture &
\(0.1\Normal(0,0.5^2)+0.9\Normal(1,15^2)\) & 45 & 0.014 &
harmonics \((32,0.85)\), \((12,-1.25)\) \\
\bottomrule
\end{tabular}
\caption{Single-quantile dynamic data-generating mechanisms. Raw errors are shifted by
\(H_f^{-1}(p)\) so that the \(p\)-quantile of the observation error is zero;
the same latent dynamic quantile path is then shared across target levels
within each family. Seasonal entries report amplitude and phase for the two
harmonic pairs in the initial state.}
\label{tab:simulation-dgp}
\end{table}

\subsection{Single-Quantile Fit and Forecast Evaluation}

Each synthetic series is first run for \(T_{\mathrm{warm}}=2000\) warmup
observations, which are discarded before analysis. The next
\(T_{\mathrm{main}}=10000\) observations form the retained series. The first
9000 support model specification and fitting, and the final 1000 are held out for
forecast evaluation. The reported single-quantile comparison uses
\(T_{\mathrm{fit}}=500\), namely observations 8501--9000 immediately preceding
the fitting origin. Forecast
scores are computed from rolling origins within the held-out period; origins
are spaced 30 observations apart, and each origin is scored for horizons up to
30 steps ahead. The same evaluation design is used for all model specifications and
score calculations.

\subsection{Single-Quantile Competing Methods}

For each family and target quantile level in the single-quantile simulation
comparison, the reported Q--DESN rows vary the working likelihood and
inference method while using regularized-horseshoe (RHS) coefficient shrinkage.
Entries labeled Q--DESN AL--RHS use the AL working likelihood, whereas entries
labeled Q--DESN exAL--RHS use the exAL working likelihood. The main comparison
focuses on regularized-horseshoe fits. For each displayed metric, the reservoir
specification and realization were selected before held-out evaluation and then
held fixed.
Dynamic quantile linear model (DQLM) and extended dynamic quantile linear
model (exDQLM) baselines are fit to the same family, quantile level,
fitting-sample size, and forecast-evaluation design using version 1.1.1 of the
CRAN package \pkg{exdqlm} \citep{exdqlmRPackage}. The exDQLM calculations use its
partially collapsed MCMC update and structured
\(q(\gamma)q(\sigma\mid\gamma)\) variational approximation for the
scale--asymmetry block.
For rolling-origin MCMC evaluation, the initial parameter posterior is retained
while the filtering distribution is updated sequentially using the responses
available at each origin. In the exDQLM analysis, the observation-error mean
and variance used in this update are averaged over paired retained draws of the
scale and asymmetry parameters. The updated state is then propagated over the
requested forecast horizons.

The primary comparison is a fixed-design benchmark against structured dynamic
quantile linear models under linear dynamic mechanisms that favor those
baselines. Repeated data-generating realizations, broader reservoir-parameter comparisons, and
non-Bayesian ESN regression baselines remain topics for future sensitivity
analyses.

\subsection{Single-Quantile Criteria for Comparison}

Let \(q_{p,t}^{(b)}\) be retained draw \(b\) of the conditional \(p\)-quantile
at fitting index \(t\), and let \(q_{p,r}^{(b)}\) be the corresponding forecast
draw for rolling-origin pair \(r\). Write \(\mathcal I\) for the 500 fitting
indices and \(\mathcal G\) for the 1000 scored origin--horizon pairs. Because
the oracle path \(\mu\) is known, the draw-wise criteria are
\begin{align}
R_{\mathrm{fit}}^{(b)}
  &=\left\{\frac{1}{|\mathcal I|}\sum_{t\in\mathcal I}
    \bigl(q_{p,t}^{(b)}-\mu_t\bigr)^2\right\}^{1/2}, \\
A_{\mathrm{fore}}^{(b)}
  &=\frac{1}{|\mathcal G|}\sum_{r\in\mathcal G}
    \left|q_{p,r}^{(b)}-\mu_r\right|, \\
C_{\mathrm{fore}}^{(b)}
  &=\frac{1}{|\mathcal G|}\sum_{r\in\mathcal G}
    \rho_p\!\left(y_r-q_{p,r}^{(b)}\right),
\end{align}
where \(\rho_p\) is the check loss in Eq.~\eqref{eq:check-loss-main}, and
\(\mu_r\) and \(y_r\) denote the oracle quantile and realized outcome for pair
\(r\). Thus fit RMSE and forecast MAE measure recovery of the oracle quantile
path, whereas forecast check loss is a proper score against realized
observations.

For each criterion \(M^{(b)}\), the tables report the posterior mean
\(B^{-1}\sum_{b=1}^B M^{(b)}\) and the empirical 0.025 and 0.975 posterior
quantiles. The VB panels use draws from the variational approximation and are
therefore approximate. Because the criteria are nonlinear functions of the
quantile path, these posterior means generally differ from applying a criterion
once to a posterior point path.

For each model, error family, quantile level, and criterion, the reservoir
specification was selected in a separate calibration analysis before the
posterior metric distribution was evaluated. Consequently, different criteria
within a model may correspond to different reservoir specifications. The
intervals condition on the selected specification, simulated series,
rolling-origin design, and reservoir realization; repeated-simulation,
reservoir-selection, and model-selection uncertainty are excluded.

\subsection{Single-Quantile Fit and Forecast Results}

\begin{figure}[!htbp]
\centering
\includegraphics[width=0.98\textwidth]{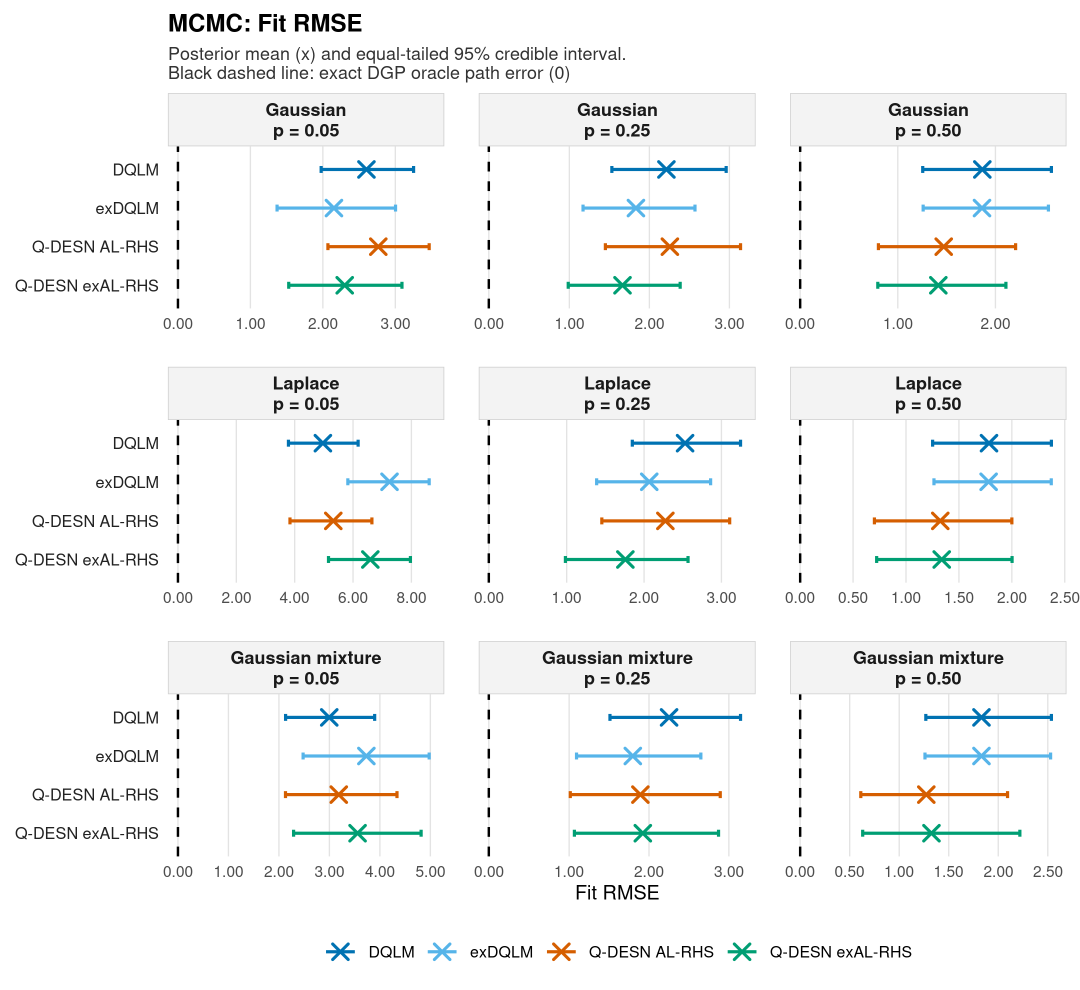}
\caption{MCMC posterior uncertainty for fit RMSE in the single-quantile simulation study. Horizontal segments show equal-tailed 95\% posterior intervals; crosses mark posterior means. Each panel uses its own horizontal scale, and lower values are better. The black dashed line marks the exact DGP oracle value of zero for this conditional-quantile path-error criterion. Intervals condition on the simulated data, evaluation design, and case-specific model specification.}
\label{fig:simulation-500obs-mcmc-fit-rmse-intervals}
\end{figure}

\begin{figure}[!htbp]
\centering
\includegraphics[width=0.98\textwidth]{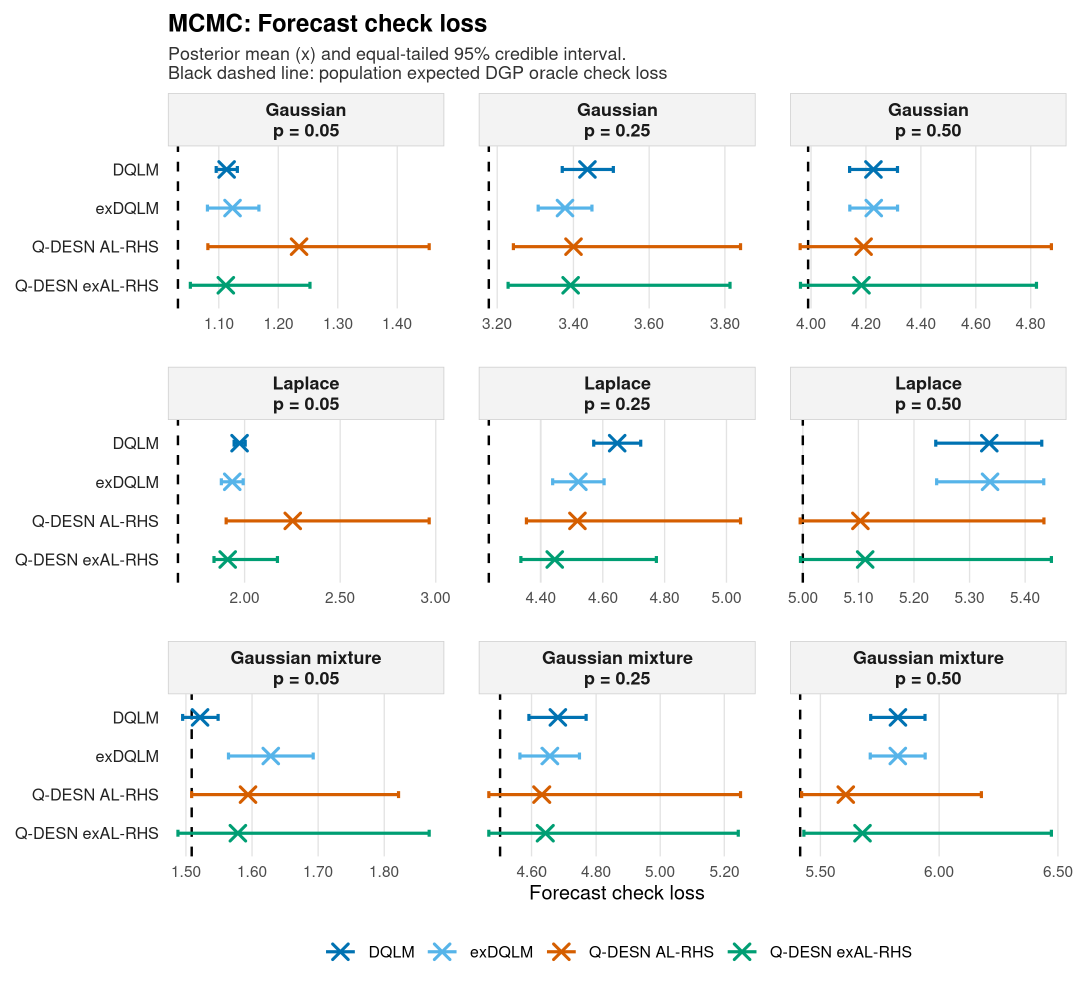}
\caption{MCMC posterior uncertainty for forecast check loss in the single-quantile simulation study. Horizontal segments show equal-tailed 95\% posterior intervals; crosses mark posterior means. Each panel uses its own horizontal scale, and lower values are better. The black dashed line marks the population expected check loss at the true conditional quantile. Because the intervals condition on one simulated series, finite-sample check-loss summaries may cross this population reference. Intervals condition on the simulated data, evaluation design, and case-specific model specification.}
\label{fig:simulation-500obs-mcmc-forecast-check-loss-intervals}
\end{figure}

The main comparison uses the shorter fitting design: 500 observations before
the fitting origin, followed by 1000 held-out observations for forecast
scoring. Forecast scores use rolling origins spaced 30 observations apart and
horizons up to 30 steps ahead, at the three target levels
\(p=0.05,0.25,0.50\). Figures~\ref{fig:simulation-500obs-mcmc-fit-rmse-intervals}--%
\ref{fig:simulation-500obs-mcmc-forecast-check-loss-intervals} summarize the
MCMC posterior metric distributions; the supplement gives the corresponding
numerical tables.

Each criterion is recomputed for every retained conditional-quantile draw. Table entries report the posterior mean and equal-tailed 95\% credible interval of the resulting draw-wise criterion. Fit RMSE measures recovery of the true conditional quantile over the 500-observation training sample; forecast MAE and check loss average the 1,000 rolling-origin and horizon evaluations used in the point comparison. The intervals condition on the simulated data, evaluation design, and criterion-specific fitted model. Variational Bayes intervals are approximate.

The exDQLM entries use exdqlm 1.1.1; the DQLM and both Q--DESN entries use their stated specifications. In rolling-origin exDQLM MCMC evaluation, the initial parameter posterior is retained while the filtering distribution is updated sequentially with responses available at each origin. The observation-error mean and variance used in this update are averaged over paired retained draws of the scale and asymmetry parameters before the updated state is propagated over the requested forecast horizons. Point estimates and posterior intervals use distinct estimators: point estimates summarize a single fixed path for VB and chain-level fixed paths for MCMC, whereas interval centers are posterior means of draw-wise criteria. The two summaries are therefore interpreted separately.

5 of the 108 displayed MCMC summaries are marked by daggers to indicate diagnostic cautions; the supplement provides details.

The displayed bands are posterior distributions of the draw-wise aggregate criteria. For Q--DESN and exQ--DESN, each metric draw preserves a common posterior draw of the quantile-regression coefficients across forecast origins and horizons. For DQLM and exDQLM, the origin- and horizon-specific posterior quantile marginals are combined using a pre-specified product coupling. The resulting intervals quantify model-specific posterior uncertainty for the reported criteria, conditional on the simulated series, selected specification, and evaluation design.

Posterior means were numerically unchanged across coupling assumptions, whereas interval widths varied; we therefore report the product-coupling intervals with this sensitivity analysis.

All displayed Q--DESN and exQ--DESN values use feature scaling and other
preprocessing quantities estimated from observations available through the
fitting origin. The DQLM and exDQLM baselines use the same simulated
trajectories, fitting sample, forecast origins, horizons, and scoring rules. The
Q--DESN exAL--RHS MCMC rows use the \(v\)-augmented transition described in the
supplement, which separates the scale update from the shape update and samples
the shape on the logit transform of its bounded support.

Across the 27 family--quantile--criterion comparisons in the MCMC panels, Q--DESN exAL--RHS has the lowest posterior mean in 10 comparisons and Q--DESN AL--RHS in 8; DQLM has the lowest mean in 4 and exDQLM in 5. The two Q--DESN variants account for 18 of the 27 lowest posterior means.

For all 27 comparisons, the equal-tailed intervals of the two lowest posterior means overlap. Boldface identifies the lowest posterior mean under each pre-specified case- and criterion-specific model specification. The interval overlap limits conclusions about differences between the two lowest-scoring methods.

\begin{figure}[!htbp]
\centering
\includegraphics[width=0.98\textwidth]{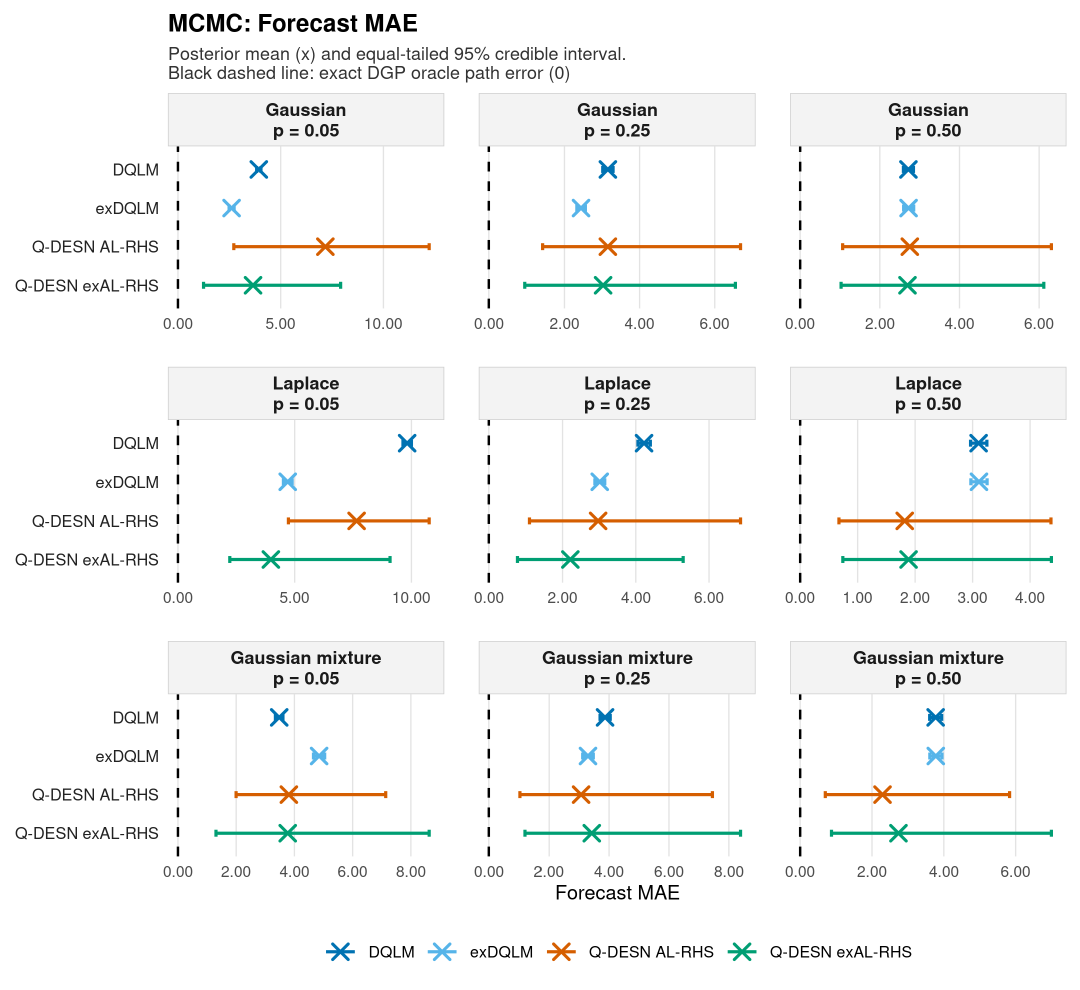}
\caption{MCMC posterior uncertainty for forecast MAE in the single-quantile simulation study. Horizontal segments show equal-tailed 95\% posterior intervals; crosses mark posterior means. Each panel uses its own horizontal scale, and lower values are better. The black dashed line marks the exact DGP oracle value of zero for this conditional-quantile path-error criterion. Intervals condition on the simulated data, evaluation design, and case-specific model specification.}
\label{fig:simulation-500obs-mcmc-forecast-mae-intervals}
\end{figure}

The supplement gives the complete numerical MCMC tables and approximate VB
intervals, together with a five-chain point-path sensitivity analysis for the
Gaussian family at \(p=0.25\).

\subsection{Joint Multi-Quantile Study with a Pre-Specified Feature Design}

The second simulation study assesses whether estimating a quantile vector
jointly improves forecast accuracy or coherence relative to fitting the same
levels independently. Each regression is fit at
\[
\tau\in\{0.05,0.10,0.25,0.50,0.75,0.90,0.95\}.
\]
The comparison uses eight synthetic mechanisms, a 500-observation fitting
sample, and a pre-specified feature and prior specification for each
mechanism--model combination. Three mechanisms use Gaussian, Laplace, and
Gaussian-mixture innovations. The remaining five introduce asymmetric tails,
persistent heavy tails, Student-\(t\) errors, a regime shift, and nonlinear
dynamics represented by reservoir features. We compare joint and independent
Q--DESN regressions under the \(\AL\) and \(\exAL\) working likelihoods, all
with regularized-horseshoe shrinkage, giving 32 comparisons.

The held-out analysis is sequentially conditional. Regression coefficients
and reservoir weights remain fixed, while newly observed response lags become
available as evaluation advances through the held-out period. Each
mechanism--model comparison contains 990 aligned origin--horizon evaluations.
The four model classes use the same evaluation indices and known conditional
response distributions. Variational approximations provide initial values for
the MCMC analysis. Every \(\exAL\) fit uses the scale-collapsed update for
\((\gamma,\sigma)\) described in the supplement.

The primary criterion is \(S_{\mathrm{DGP}}\) in
Eq.~\eqref{eq:dgp-integrated-score-main}, computed for every retained MCMC
draw after monotone rearrangement. Joint fits preserve posterior dependence
across quantile levels within each draw. Draws from independently fitted
levels are combined using a pre-specified, chain-balanced product coupling
detailed in the supplement. The product of the level-specific working
likelihoods serves as a composite objective for the quantile-indexed
coefficients; predictive evaluation is based on the resulting quantile vector.
Forecast MAE and RMSE compare the estimated paths with the known conditional
quantiles and are reported as recovery diagnostics.

\begin{table}[!htbp]
\centering
\scriptsize
\resizebox{\textwidth}{!}{%
\begin{tabular}{@{}>{\raggedright\arraybackslash}p{0.23\textwidth}rrrr@{}}
\toprule
Simulation setting & \shortstack{Joint Q--DESN\\\(\AL\)--\(\RHS\)} & \shortstack{Independent Q--DESN\\\(\AL\)--\(\RHS\)} & \shortstack{Joint exQDESN\\\(\exAL\)--\(\RHS\)} & \shortstack{Independent exQDESN\\\(\exAL\)--\(\RHS\)} \\
\midrule
Asymmetric-Laplace tail & 0.3170 [0.3147, 0.3213] & 0.3167 [0.3149, 0.3196] & \textbf{0.3163 [0.3143, 0.3200]} & 0.3168 [0.3147, 0.3205] \\
Gaussian-mixture innovations & 0.3866 [0.3843, 0.3897] & \textbf{0.3865 [0.3844, 0.3895]} & 0.3870 [0.3841, 0.3911] & 0.3872 [0.3842, 0.3917] \\
Laplace innovations & \textbf{0.3220 [0.3204, 0.3252]} & 0.3228 [0.3210, 0.3253] & 0.3220 [0.3201, 0.3254] & 0.3231 [0.3206, 0.3282] \\
Nonlinear reservoir dynamics & \textbf{0.4273 [0.4247, 0.4305]} & 0.4285 [0.4254, 0.4323] & 0.4293 [0.4258, 0.4342] & 0.4302 [0.4262, 0.4356] \\
Gaussian innovations & \textbf{0.3101 [0.3080, 0.3135]} & 0.3108 [0.3084, 0.3144] & 0.3119 [0.3084, 0.3179] & 0.3120 [0.3087, 0.3173] \\
Persistent heavy tails & 0.3644 [0.3614, 0.3685] & 0.3636 [0.3607, 0.3675] & 0.3643 [0.3608, 0.3698] & \textbf{0.3629 [0.3601, 0.3677]} \\
Regime shift & 0.4369 [0.4264, 0.4587] & \textbf{0.4311 [0.4263, 0.4417]} & 0.4394 [0.4269, 0.4639] & 0.4315 [0.4264, 0.4426] \\
Student-\(t\) location--scale & \textbf{0.3362 [0.3343, 0.3395]} & 0.3368 [0.3349, 0.3398] & 0.3377 [0.3346, 0.3441] & 0.3372 [0.3348, 0.3413] \\
\bottomrule
\end{tabular}
}%
\caption{Posterior data-generating-process-integrated finite-grid quantile scores for the joint multi-quantile simulation. Entries are posterior means with equal-tailed 95\% credible intervals. The score averages the seven-level integrated check loss over the known conditional response distribution and the held-out forecast design. Lower values are better; boldface identifies the numerical minimum, based on the unrounded posterior means, within each simulation setting. Because every joint-minus-independent posterior contrast interval under the pre-specified pairing includes zero, the bolded comparisons are descriptive.}
\label{tab:joint-qdesn-dgp-integrated-score}
\end{table}

\begin{figure}[!htbp]
\centering
\includegraphics[width=0.94\textwidth]{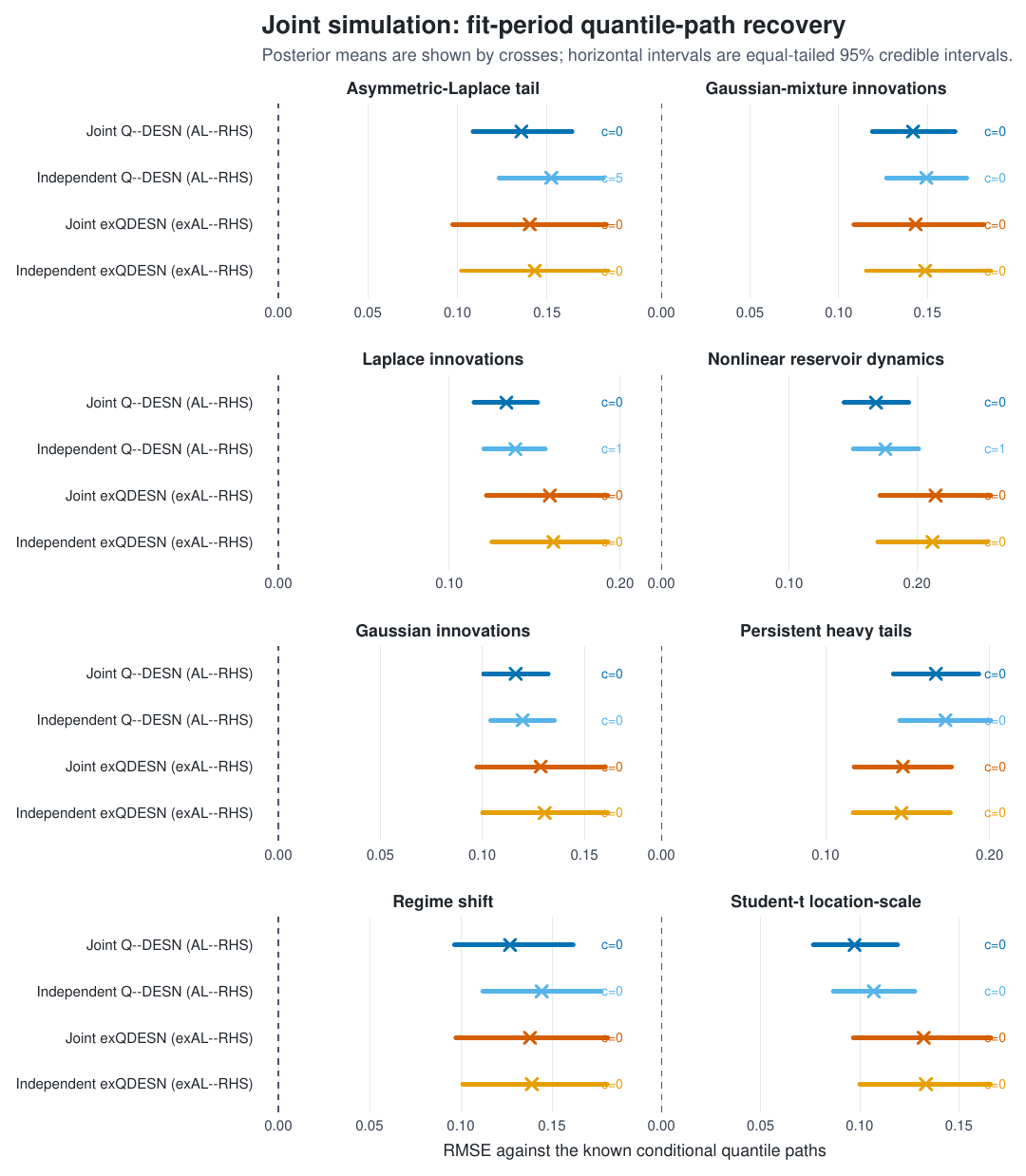}
\caption{Posterior fit-period recovery of the known conditional quantile paths in the joint multi-quantile simulation. Each panel uses the same four model classes and the same seven quantile levels. Crosses show posterior means and horizontal bars show equal-tailed 95\% credible intervals computed from 8,000 retained MCMC draws per comparison after monotone rearrangement. Lower RMSE is better. The labels at the right of each panel report raw adjacent-level crossing counts in the posterior-mean fitted grid before monotone rearrangement; all corresponding counts after rearrangement are zero. Panel-specific x-scales are used for readability.}
\label{fig:joint-qdesn-phase181-fit-rmse-intervals}
\end{figure}

\begin{figure}[!htbp]
\centering
\includegraphics[width=0.94\textwidth]{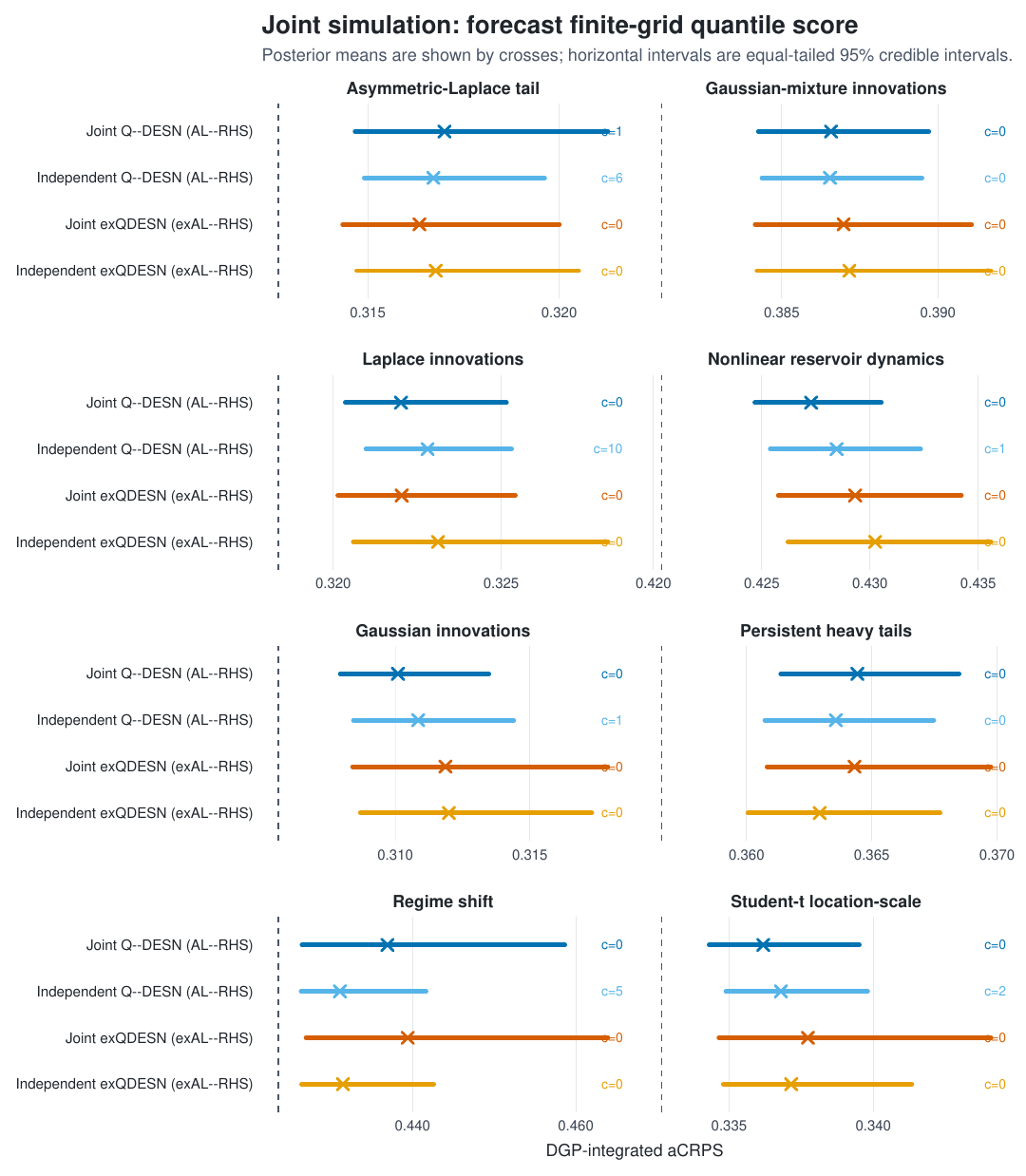}
\caption{Posterior DGP-integrated finite-grid quantile scores for the joint multi-quantile simulation. Each panel uses the same four model classes, the same held-out forecast design, and the same seven evaluation levels as Table~\ref{tab:joint-qdesn-dgp-integrated-score}. Crosses show posterior means and horizontal bars show equal-tailed 95\% credible intervals from 8,000 retained MCMC draws per comparison. Lower \(\aCRPS\) is better. Dashed vertical lines mark the oracle finite-grid score in each simulation setting. The labels at the right of each panel report raw adjacent-level crossing counts in the posterior-mean forecast grid before monotone rearrangement; all corresponding counts after rearrangement are zero. The numerical rankings remain descriptive because all paired joint-minus-independent contrast intervals include zero. Panel-specific x-scales are used for readability.}
\label{fig:joint-qdesn-phase181-forecast-acrps-intervals}
\end{figure}

Figures~\ref{fig:joint-qdesn-phase181-fit-rmse-intervals} and
\ref{fig:joint-qdesn-phase181-forecast-acrps-intervals} give the corresponding
posterior uncertainty summaries for fit-period path recovery and forecast
finite-grid scoring, using the same model classes and simulation settings as
Table~\ref{tab:joint-qdesn-dgp-integrated-score}.

Table~\ref{tab:joint-qdesn-dgp-integrated-score} gives the posterior mean and
equal-tailed 95\% credible interval of the DGP-integrated score. Joint methods
have the smallest numerical mean in five of eight settings: the
asymmetric-Laplace-tail, Laplace-innovation, nonlinear-dynamics,
Gaussian-innovation, and Student-\(t\) location--scale settings. Independent
methods have the smallest mean for Gaussian-mixture innovations, persistent
heavy tails, and the regime shift. Within likelihood family, the joint mean is
lower in nine of the 16 comparisons. All 16 joint-minus-independent posterior
contrast intervals under the pre-specified pairing include zero, so the
numerical rankings are interpreted descriptively.

The posterior-mean forecast grids show one adjacent-level crossing
for joint Q--DESN under \(\AL\), compared with 25 for the corresponding
independent regressions. Neither \(\exAL\) method crosses on these
posterior-mean grids, and monotone rearrangement yields ordered quantiles in all 32
comparisons. Draw-level crossing frequencies and MCMC diagnostics provide a
broader assessment in the supplement. MCMC diagnostics for the posterior
score meet the pre-specified numerical thresholds in 21 of 32 comparisons;
the remaining 11 are interpreted cautiously. A separate 50-replicate variational analysis
is retained as a sensitivity study of quantile-path recovery.

\section{Application: GloFAS Retrospective Streamflow Quantile Forecasting}
\label{sec:data}

\noindent\textbf{Retrospective case study.}
The empirical application considers a medium-range retrospective streamflow
quantile-forecasting case using the Global Flood Awareness System (GloFAS) and
a reference gauge record. GloFAS is a global hydrological forecast and monitoring system
jointly developed by the European Commission and the European Centre for
Medium-Range Weather Forecasts (ECMWF) and operated within the Copernicus
Emergency Management Service
\citep{AlfieriEtAl2013GloFAS,CEMS2026GloFASDocumentation}. Its operational
forecast and reforecast products provide ensemble river-discharge information
for flood early warning and forecast verification
\citep{HarriganEtAl2023GloFASForecasts}. Previous calibration work has focused
on hydrological model parameters using daily streamflow observations
\citep{HirpaEtAl2018GloFASCalibration}. Here Q--DESN provides a statistical
quantile-regression correction at one forecast origin. The covariate set
includes retrospectively observed future weather information.

\subsection{Data and Forecast-Origin Design}

Let \(T\) denote a forecast origin and let \(H\) be the issued forecast
horizon. The historical reference record is \(y_t\), the transformed United
States Geological Survey (USGS) streamflow through time \(T\). The retrospective GloFAS product over the same
period is \(g_t^{\mathrm{ret}}\), and
\(g_{T+h,m}^{\mathrm{ens}}\) denotes GloFAS ensemble member \(m\) issued at
origin \(T\) for horizon \(h=1,\ldots,H\). The response values
\(y_{T+1},\ldots,y_{T+H}\) are used only for forecast evaluation.

\noindent\textbf{Retrospective information set.}
The reported analysis uses the forecast origin
\GlofasApplicationCurrentOriginDate{} and scores seven fitted quantile levels
over \GlofasApplicationCurrentScoredHorizons{} issued target dates. The
application is a single-origin retrospective case. Its response
values are held out for scoring, and the covariate specification combines
observed history and Global Ensemble Forecast System (GEFS) forecasts with
realized-future contributions to the precipitation and soil-moisture
covariates. Consequently, the results describe retrospective
performance under this information set; evaluation using only forecast-origin
information remains future work.

\subsection{Application Model}

\noindent\textbf{Persistence-anchored discrepancy.}
For \(t\le T\), the application uses two deterministic Q--DESN feature maps.
The reference block \(\vect x_t^{\mathrm{ref}}\) is driven by the transformed
reference history and the same covariate specification. The discrepancy block
\(\vect x_t^{\mathrm{disc}}\) is driven by the retrospective discrepancy
\(g_t^{\mathrm{ret}}-y_t\), using the same preprocessing, lag,
reservoir, reducer, and washout conventions. For \(t>T\), the
persistence-plus-adjustment specification anchors the future discrepancy
feature at the last historical
retrospective discrepancy and the discrepancy regression models an adjustment
around that persistence baseline. The empirical issued-ensemble quantile is
retained as the unadjusted GloFAS benchmark in the forecast comparison; the
future ESN feature map for the discrepancy is driven by the
persistence-anchored discrepancy state.

For a fixed quantile level \(p_0\), the reference-process quantile is
represented by one Q--DESN regression and the GloFAS discrepancy by a second
regression,
\[
q^{\mathrm{ref}}_{p_0,t}=\vect x_t^{\mathrm{ref}\top}\vect\beta_{p_0},
\qquad
d^{\mathrm{glo}}_{p_0,t}=\vect x_t^{\mathrm{disc}\top}\vect\alpha_{p_0},
\qquad
q^{\mathrm{glo}}_{p_0,t}=q^{\mathrm{ref}}_{p_0,t}+d^{\mathrm{glo}}_{p_0,t}.
\]
The path \(q^{\mathrm{ref}}_{p_0,t}\) is the reference-process quantile of interest,
while \(d^{\mathrm{glo}}_{p_0,t}\) is the quantile-specific discrepancy between
GloFAS and the reference process. The retrospective and ensemble GloFAS
products inform the forecast-system quantile path \(q^{\mathrm{glo}}_{p_0,t}\);
the scored reference quantity remains \(q^{\mathrm{ref}}_{p_0,T+h}\).

At forecast origin \(T\), the issued GloFAS ensemble provides \(M\)
river-discharge forecasts at each covered target date. The
application summary is a VB quantile-regression summary of the reference
process and the persistence-anchored discrepancy adjustment. For variational
draw or summary index \(s=1,\ldots,S\), we compute the reference quantile and
the resulting forecast-system quantile:
\[
q^{\mathrm{ref},(s)}_{p_0,T+h}
=\vect x_{T+h}^{\mathrm{ref},(s)\top}\vect\beta_{p_0}^{(s)},
\qquad
q^{\mathrm{glo},(s)}_{p_0,T+h}
=q^{\mathrm{ref},(s)}_{p_0,T+h}
+d_T^{\mathrm{ret}}
+\vect x_{T+h}^{\mathrm{disc},(s)\top}\vect\alpha_{p_0}^{(s)}.
\]
Here \(d_T^{\mathrm{ret}}\) is the last historical retrospective discrepancy
on the transformed scale. Posterior means, medians, intervals, and monotone
multi-quantile summaries are computed after this quantile construction.

The working-likelihood specification is
\begin{align*}
y_t\mid q^{\mathrm{ref}}_{p_0,t},\sigma_{\mathrm{ref}},\gamma_{\mathrm{ref}}
&\sim \exAL_{p_0}(q^{\mathrm{ref}}_{p_0,t},\sigma_{\mathrm{ref}},\gamma_{\mathrm{ref}}),\\
g_t^{\mathrm{ret}}\mid q^{\mathrm{glo}}_{p_0,t},\sigma_{\mathrm{glo}},\gamma_{\mathrm{glo}}
&\sim \exAL_{p_0}(q^{\mathrm{glo}}_{p_0,t},\sigma_{\mathrm{glo}},\gamma_{\mathrm{glo}}),\\
g_{T+h,m}^{\mathrm{ens}}\mid q^{\mathrm{glo}}_{p_0,T+h},\sigma_{\mathrm{glo}},\gamma_{\mathrm{glo}}
&\sim \exAL_{p_0}(q^{\mathrm{glo}}_{p_0,T+h},\sigma_{\mathrm{glo}},\gamma_{\mathrm{glo}}),
\quad h=1,\ldots,H,\quad m=1,\ldots,M.
\end{align*}
This display gives the general exAL form. The AL specialization used by the
reported AL analysis omits \(s\)-latent variables and \(\gamma\) blocks,
and replaces the exAL constants by the corresponding AL constants at the fitted
level.
The issued ensemble members enter through a working likelihood that treats
them as conditionally independent and exchangeable given the forecast-system
quantile path and component-specific likelihood parameters. ECMWF ensemble
forecasts use perturbed initial states and model perturbations to produce
distinct forecast alternatives
\citep{ECMWF2026ENSGeneration}; the conditional-independence factorization is
therefore a modeling approximation for the Q--DESN regression. In exAL variants,
the default application model shares \(\sigma_{\mathrm{glo}}\) and
\(\gamma_{\mathrm{glo}}\) between retrospective and issued-ensemble GloFAS
observations. Under the AL specialization, the scale
\(\sigma_{\mathrm{glo}}\) is shared.

The seven quantile levels \(0.05\), \(0.15\), \(0.35\), \(0.50\), \(0.65\),
\(0.80\), and \(0.95\) are fitted independently using AL Q--DESN regressions
and VB. Their
variational quantile summaries are monotonically rearranged to obtain the
displayed intervals.

Separate fixed DESN feature maps are used for the reference process and
discrepancy, with their reservoir specifications reported in the supplement.
Forecast-period discrepancy states follow the persistence-plus-adjustment
specification described above.

The two coefficient blocks use separate regularized-horseshoe global scales,
\(\tau_{0,\mathrm{ref}}=\GlofasApplicationCurrentSharedRhsTau{}\) and
\(\tau_{0,\mathrm{disc}}=\GlofasApplicationCurrentDiscrepancyRhsTau{}\),
respectively.

\subsection{Retrospective Application Results}

The reported fit was evaluated over
\GlofasApplicationCurrentObservedHistoryDates{} observed dates through the
forecast origin. After monotone rearrangement, its observational-window
\(\aCRPS\) was \(\GlofasApplicationCurrentObservedHistoryAcrps{}\), and
empirical 90\% interval coverage was
\(\GlofasApplicationCurrentObservedHistoryCoverage{}\). These fitting-period
scores were used for model selection, with the forecast period reserved for
evaluation. The fitted \(p_0=0.95\) quantile occasionally takes extreme values
in the observational period, so upper-tail forecasts are interpreted
cautiously.

\noindent\textbf{Forecast-period scores.}
Table~\ref{tab:glofas-application-score} reports the forecast scores for the
reported retrospective analysis. Relative to raw GloFAS, Q--DESN reduces
the mean check loss from \(\GlofasApplicationCurrentRawCheckLoss{}\) to
\(\GlofasApplicationCurrentQdesnCheckLoss{}\), the interval score from
\(\GlofasApplicationCurrentRawIntervalScore{}\) to
\(\GlofasApplicationCurrentQdesnIntervalScore{}\), and \(\aCRPS\) from
\(\GlofasApplicationCurrentRawAcrps{}\) to
\(\GlofasApplicationCurrentQdesnAcrps{}\). Mean empirical interval coverage
increases from \(\GlofasApplicationCurrentRawMeanCoverage{}\) to
\(\GlofasApplicationCurrentQdesnMeanCoverage{}\), but the Q--DESN value remains
far below the nominal 90\% level. These summaries describe one specified
forecast origin under a blended retrospective information set. Multi-origin
evaluation with forecast-origin covariates is needed to assess real-time
performance and interval calibration.

\begin{table}[!htbp]
\centering
\caption{Forecast scoring for the GloFAS retrospective analysis.
Scores are computed on the transformed streamflow scale for the
forecast origin \GlofasApplicationCurrentOriginDate{}. Lower check loss,
interval score, and \(\aCRPS\) are better; \(\aCRPS\) is computed from
Eq.~\eqref{eq:acrps-main} at seven common evaluation levels. Mean
coverage reports empirical coverage for the nominal 90\% interval and
is interpreted descriptively. The final column gives the relative reduction in
mean check loss.}
\label{tab:glofas-application-score}
\small
\begin{tabular}{lrrrrrr}
\toprule
Model & Horizons & Check & Interval & aCRPS & Coverage & Check reduction \\
\midrule
Q--DESN correction & 28 & 0.5654 & 10.8032 & 1.1319 & 0.357 & 26.0\% \\
Raw GloFAS & 28 & 0.7639 & 25.2028 & 1.4424 & 0.000 & Reference \\
\bottomrule
\end{tabular}

\end{table}

\begin{figure}[H]
\centering
\includegraphics[width=0.94\textwidth]{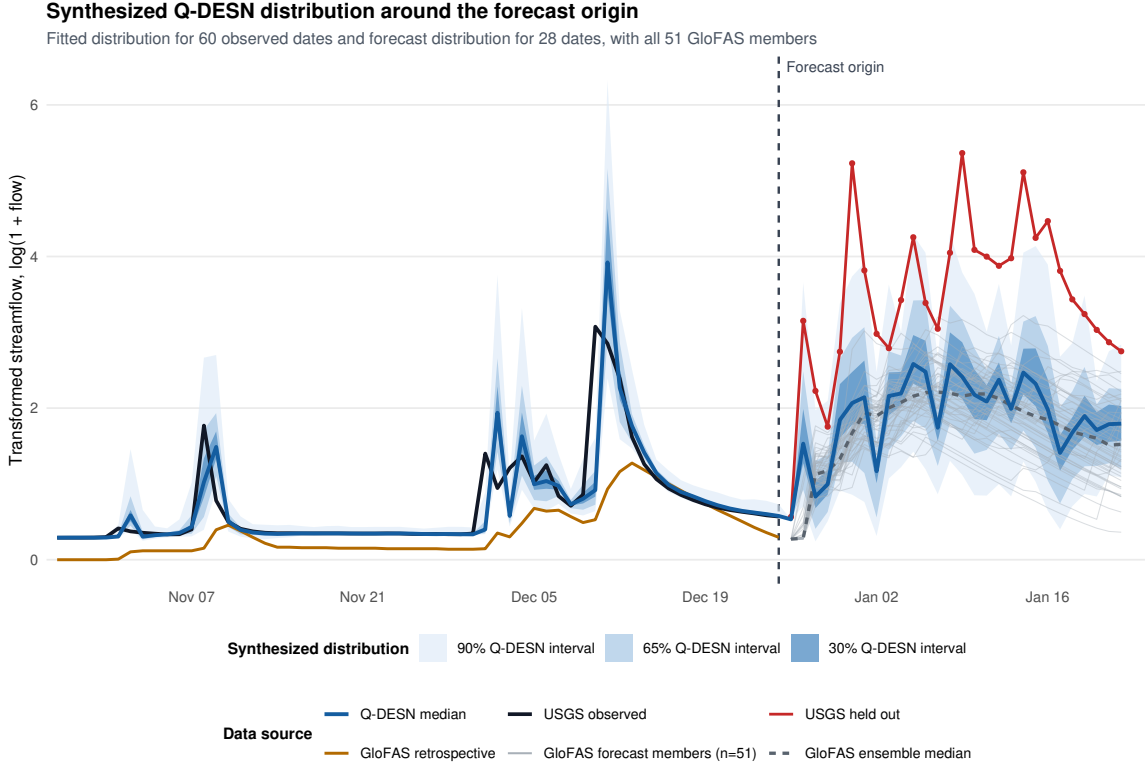}
\caption{Q--DESN quantile intervals for the GloFAS retrospective analysis over
the final 60 observed dates and the following 28-day forecast period. The
nested blue bands are the 90\%, 65\%, and 30\% quantile intervals, and the blue
path is the median. The vertical dashed line marks the forecast origin. Black
and red paths denote observed and held-out USGS streamflow, respectively;
orange denotes retrospective GloFAS, and thin gray paths show all 51 issued
GloFAS ensemble members, with their median dashed. Forecast scores use the
held-out measurements in Table~\ref{tab:glofas-application-score}.}
\label{fig:glofas-application-forecast-period}
\end{figure}

\section{Application: PriceFM Retrospective Electricity-Price Comparison}
\label{sec:pricefm-application}

As a second empirical case, we consider a retrospective comparison of
probabilistic day-ahead electricity-price forecasts using version~4 of the
PriceFM benchmark
\citep{YuEtAl2026PriceFM}. The current benchmark release spans 1~January~2022
through 1~January~2026.
Forecasts are evaluated in euros per megawatt-hour (EUR/MWh) using average
quantile loss (AQL) on the PriceFM quantile grid at matched held-out timestamps.
The released data cover European bidding regions at 15-minute resolution and
include day-ahead price, load, solar, and wind forecasts. PriceFM is a
pretrained neural forecasting model that incorporates region-level covariates
and restricts interregional dependence according to sparse links in the
European transmission network. The
comparison contains 114 region--fold evaluations against matched PriceFM
predictions.

\subsection{Comparison Design}

The comparison contains \PricefmFullRegionFolds{} region--fold
evaluations, spanning \PricefmFullRegions{} European bidding regions and
\PricefmFullFolds{} rolling folds. Each evaluation is scored on the same
fold-aligned held-out timestamps used by the PriceFM predictions. Each evaluation
uses the PriceFM paper quantile levels
\PricefmFullPaperQuantiles. Throughout this section, \(\Delta\) AQL denotes
Q--DESN AQL minus PriceFM AQL. Negative values indicate lower Q--DESN AQL;
positive values indicate lower PriceFM AQL.

\noindent\textbf{Retrospective covariates.}
We compare Q--DESN forecasts with the released PriceFM forecasts at matched
region--fold timestamps. Across the comparison, \PricefmFullTargetOnlyRows{}
evaluations use own-region predictors and \PricefmFullGraphRows{} evaluations
also include neighboring-region summaries. Both predictor sets use retrospectively
observed own-region load, solar, and wind lead covariates in the comparison;
evaluations with neighborhood summaries add retrospectively observed neighboring-region lead
summaries. The score comparison is therefore retrospective. A PriceFM study of the
joint quantile-vector RHS prior would require the same fold-aligned design with
that joint regression fitted directly.

PriceFM reports two model-comparison designs. Its Table~II evaluates an
all-region fitted setting, whereas its Table~III evaluates leave-one-region-out
generalization. Because the Q--DESN specifications here are fitted with
observations from each target region, Table~II is the relevant
paper-level analogue. Table~II is therefore the relevant published reference
for the present region-specific fits.

\subsection{Results Across Regions and Folds}

Across the aligned benchmark, Q--DESN has lower AQL in
\PricefmFullQdesnWins{} of \PricefmFullRegionFolds{} region/fold evaluations
(\PricefmFullQdesnWinRate{}), \PricefmFullQdesnClose{} evaluations fall within
the prespecified practical-equivalence margin, and PriceFM has lower
AQL in \PricefmFullPricefmWins{} evaluations. The mean Q--DESN AQL is
\PricefmFullMeanQdesnAql{}, compared with mean PriceFM AQL
\PricefmFullMeanPricefmAql{}, giving mean \(\Delta\) AQL
\PricefmFullMeanDeltaAql{} and median \(\Delta\) AQL
\PricefmFullMedianDeltaAql{}. The aggregate difference is therefore modest and
heterogeneous. The fold-level supplementary summary shows that the mean
\(\Delta\) AQL is most negative in fold 2, modestly negative in fold 1, and
essentially neutral in fold 3. On the same 114-comparison panel, the
average quantile crossing rate (AQCR), averaged without region or fold weights,
across the region--fold
evaluations is \PricefmAlignedQdesnAqcr{} for Q--DESN and
\PricefmAlignedPricefmAqcr{} for the fixed PriceFM forecasts from the public release. Mean MAE is
\PricefmAlignedQdesnMae{} versus \PricefmAlignedPricefmMae{}, and mean RMSE is
\PricefmAlignedQdesnRmse{} versus \PricefmAlignedPricefmRmse{}, respectively.
Thus Q--DESN reduces aligned AQL, MAE, and RMSE by approximately
\PricefmAlignedAqlReduction{}, \PricefmAlignedMaeReduction{}, and
\PricefmAlignedRmseReduction{}, while retaining a nonzero quantile-crossing
rate.

\begin{table}[!htbp]
\centering
\caption{PriceFM-aligned retrospective model comparison. The first panel reports direct
unweighted means from the common 38-region, three-fold comparison; boldface marks
the better of those two directly aligned rows. The second panel reproduces the
Table~II values from PriceFM version~4
\citep{YuEtAl2026PriceFM}. The published PriceFM values use different fitted
models and aggregation and are included for context. AQL, MAE, and RMSE are in
EUR/MWh; AQCR is in percent. Rank is reproduced only for the paper-reported
panel; dashes denote quantities without a cross-panel rank. Lower is better.
The comparison uses retrospectively observed own-region covariate leads in all
reported rows and additional retrospectively observed neighboring-region leads
when neighborhood summaries are included.}
\label{tab:pricefm-paper-aligned-main-comparison}
\begingroup
\TableStyle
\scriptsize
\setlength{\tabcolsep}{5pt}
\begin{tabular}{@{}lrrrrr@{}}
\toprule
& \multicolumn{2}{c}{Probabilistic} & \multicolumn{2}{c}{Pointwise} & \\
\cmidrule(lr){2-3}\cmidrule(lr){4-5}
Model & AQL & AQCR (\%) & MAE & RMSE & Rank \\
\midrule
\multicolumn{6}{@{}l}{\textit{Direct fold-aligned comparison (this article; 114 region/fold evaluations)}} \\
Reported Q--DESN & \textbf{6.83} & 2.64 & \textbf{16.73} & \textbf{25.45} & -- \\
Fixed PriceFM predictions & 7.04 & \textbf{0.00} & 17.27 & 26.34 & -- \\
\addlinespace[3pt]
\multicolumn{6}{@{}l}{\textit{PriceFM v4 Table II (paper-reported all-region fitted evaluation)}} \\
Na\"ive$^{1}$ & 15.29 & 0.00 & 22.06 & 34.68 & 11 \\
Na\"ive$^{2}$ & 15.35 & 0.00 & 23.31 & 34.31 & 13 \\
Na\"ive$^{3}$ & 15.46 & 0.00 & 22.64 & 32.61 & 12 \\
\addlinespace[1pt]
FEDFormer & 8.22 & 15.33 & 20.15 & 31.75 & 8 \\
PatchTST & 8.06 & 18.21 & 20.20 & 31.59 & 8 \\
iTransformer & 8.24 & 13.96 & 21.03 & 32.11 & 9 \\
TimesNet & 7.98 & 13.42 & 19.48 & 30.94 & 7 \\
TimeXer & 8.30 & 14.77 & 21.94 & 31.88 & 10 \\
\addlinespace[1pt]
GraphConv & 6.61 & 6.88 & 16.81 & 25.97 & 4 \\
GraphAttn & 7.13 & 10.33 & 17.00 & 26.11 & 6 \\
GraphSAGE & 6.78 & 6.01 & 17.56 & 26.03 & 5 \\
GraphDiffusion & 6.69 & 5.72 & 16.44 & 25.93 & 2 \\
GraphARMA & 6.72 & 6.03 & 16.56 & 25.84 & 3 \\
\addlinespace[1pt]
PriceFM & \textbf{5.80} & \textbf{0.00} & \textbf{14.28} & \textbf{22.39} & \textbf{1} \\
\bottomrule
\end{tabular}
\endgroup

\end{table}

The matched PriceFM forecasts are the appropriate distribution-level comparator
because they share the Q--DESN region--fold timestamps. Their AQL differs from
the PriceFM paper's optimized Table~II AQL,
so the two PriceFM rows are kept separate. Numerically, the
aligned Q--DESN AQL lies within the range reported for the graph baselines in
PriceFM Table~II, but differences in fitted models and aggregation preclude a
cross-panel rank claim. The supplement reports fold-level counts of comparisons
favoring Q--DESN, falling within the equivalence margin, or favoring PriceFM.

The predictor set also matters. Evaluations with neighborhood summaries have the
most negative average \(\Delta\) AQL, whereas own-region-only evaluations are
closer to neutral.
Because predictor sets and fitted model specifications vary across region--fold
evaluations, this comparison is descriptive. It summarizes where the reported
Q--DESN analyses have the lowest scores on the aligned benchmark. Results by
predictor set and forecast horizon are reported in the supplement.

\subsection{Results by Forecast Horizon}

Horizon-specific results are available for \PricefmFullHorizonRows{} of the
\PricefmFullRegionFolds{} region/fold evaluations. Negative \(\Delta\) values
indicate lower Q--DESN AQL. Because the horizon summaries cover a retained
subset of the aligned panel, they are interpreted descriptively.

In this retrospective comparison, Q--DESN has lower aggregate AQL, MAE, and
RMSE than the matched PriceFM forecasts, with substantial variation across
folds and regions. The optimized PriceFM Table~II result provides a separate
published reference. Paired uncertainty analyses and direct evaluation of the
joint quantile model remain topics for future study.

\section{Discussion}
\label{sec:discussion}

Q--DESN treats the fixed DESN reservoir as a nonlinear feature map and places
Bayesian quantile inference on the regression coefficients. Posterior
simulation and variational inference concern the regression, likelihood, and
shrinkage parameters conditional on the fixed reservoir features. The empirical
results evaluate this conditional Bayesian quantile-regression analysis under
specified reservoir constructions.

The AL and exAL families are used as working likelihoods for conditional
quantiles. They provide posterior distributions and conditionally tractable
augmented updates. The
simulation and application sections report fit recovery, forecast scoring,
empirical predictive coverage, monotonicity, and convergence diagnostics to
assess the model-based summaries under possible misspecification. Sandwich
corrections and generalized-Bayes calibration for composite quantile
likelihoods remain important topics for further study.

The multi-quantile construction makes the same distinction. Independent
single-level Q--DESN fits followed by isotonic regression or rearrangement
produce monotone quantile curves from level-wise regressions. The joint
quantile-vector extension shares information across quantile levels through RHS
shrinkage on adjacent coefficient differences. This shrinkage can reduce
crossings but does not enforce monotonicity. Ordered parameterizations or
monotone rearrangement can be used when monotonicity is required.

The single-quantile dynamic simulations assess recovery and held-out
forecasting under known data-generating mechanisms. The joint multi-quantile
study assesses quantile-path recovery and crossings under known conditional
quantile paths. The GloFAS application studies one
retrospective streamflow forecast origin with independent Q--DESN AL--VB fits,
blended future weather covariates, monotone reporting, and undercoverage of the
nominal 90\% interval. The PriceFM comparison uses matched PriceFM forecasts with
retrospectively observed lead covariates. Broader
rankings of Q--DESN against dynamic quantile or foundation-model alternatives
require evaluations across additional data sets, forecast origins, and
reservoir specifications.

Several extensions are natural. On the statistical side, future work should
study calibrated composite-likelihood scaling, hard or probabilistic
non-crossing constraints, richer covariance structures for quantile
differences, and sensitivity to reservoir scaling on bounded feature domains.
On the computational side, the joint model calls for sparse precision solvers,
targeted covariance extraction for VB, and diagnostics comparing VB with
MCMC on small joint problems. On the application side, multi-origin operational
GloFAS studies with covariates known at the forecast origin, paired PriceFM
uncertainty analyses, and explicitly joint PriceFM analyses are needed before the joint
quantile-vector prior can be interpreted in an application-level benchmark.

Q--DESN combines fixed nonlinear recurrent features with Bayesian regression
for dynamic conditional quantiles. The results demonstrate conditional-quantile
estimation under the specified simulation designs and retrospective
applications and identify the settings requiring broader validation.

\bibliographystyle{apalike}
\bibliography{refs}

\end{document}


\maketitle

\begin{abstract}
\noindent
Conditional on the fixed nonlinear features generated by a deep echo state
network (DESN), the quantile deep echo state network (Q--DESN) is a
high-dimensional Bayesian quantile regression model. We use the asymmetric
Laplace (AL) working likelihood and, under the label exAL, the quantile-fixed
generalized asymmetric Laplace distribution of \citet{yan2025new}. The
posterior includes regression coefficients, likelihood scale and asymmetry
parameters, auxiliary mixture variables, and shrinkage parameters. This
supplement derives the augmented posterior under Gaussian ridge and
Nishimura--Suchard regularized-horseshoe priors, the full conditional
distributions for Markov chain Monte Carlo (MCMC), and the coordinate updates
and evidence lower bound for variational Bayes with the Laplace--Delta
approximation (VB--LD). It also develops the joint multi-quantile regression
with regularized-horseshoe shrinkage on adjacent coefficient differences and
provides additional simulation and application results.
\end{abstract}

\section{Purpose and Relation to the Main Article}
\label{sec:supp_relation_main}

The main article introduces Q--DESN, defines its fixed deep reservoir feature
map, and presents the simulation design. Once
the reservoir weights, input weights, reducers, leak rates, and washout
convention are fixed, the likelihood depends on the data only through a
deterministic design matrix
\(\mat X=[\vect x_1^\top;\ldots;\vect x_T^\top]\). Thus the
Q--DESN regression is a static Bayesian quantile regression problem with a
high-dimensional nonlinear design. Consequently, the static AL and exAL
regression derivations used by \pkg{exdqlm}
\citep{barata2021_flex_quantile,exdqlmRPackage} can be adapted to the
Q--DESN regression by replacing ordinary covariate rows with rows of the fixed
DESN design matrix.

This supplement derives the augmented posterior, full conditional
distributions, mean-field coordinate updates, and evidence lower bound. Gaussian
ridge is the baseline coefficient prior, and RHS denotes the regularized
horseshoe. The software label \texttt{rhs\_ns} identifies the
Nishimura--Suchard product representation used to obtain closed-form scale
updates. For exAL, Taylor and Laplace approximations are evaluated separately
for \(\gamma<0\) and \(\gamma>0\); the point \(\gamma=0\) is handled under the
nested AL model. Subsequent sections cover reservoir specification, the
fixed-design likelihood and prior, MCMC and variational computation, the joint
quantile-vector extension, monotone rearrangement, supplementary empirical
results, and the GloFAS forecast-discrepancy model.

\section{Reservoir Specification and Diagnostics}
\label{sec:supp_reservoir_diagnostics}

Table~\ref{tab:supp-reservoir-selection} summarizes the reservoir-design
quantities and their associated diagnostics. Posterior inference conditions on
the resulting feature map and quantifies uncertainty in the regression,
likelihood, and shrinkage parameters.

\begin{table}[!htbp]
\centering
\begingroup
\TableStyle
\begin{tabularx}{\textwidth}{@{}>{\raggedright\arraybackslash}p{0.16\textwidth}
>{\raggedright\arraybackslash}p{0.25\textwidth}
>{\raggedright\arraybackslash}p{0.28\textwidth}
>{\raggedright\arraybackslash}X@{}}
\toprule
Component & Role in the fixed feature map & Design consideration & Potential concern \\
\midrule
\(D\) & Number of reservoir layers and hierarchy of time scales &
Evaluate a small prespecified set and retain additional depth only when
held-out forecast criteria improve consistently across reservoir realizations. &
Unnecessary depth can add numerically unstable or redundant design columns. \\
\(\{n_d\},\{\tilde n_d\}\) & Number of reservoir states and lower-layer feature
dimension & Treat as the main complexity control, with reduction sizes chosen
as part of the same feature map. &
A large design relative to the sample size can yield weakly identified
coefficients or a poorly conditioned regression; redundant states can reduce
the effective rank of \(\mat X\). \\
\(m\) & Explicit lag memory entering the reservoir input & Choose a
domain-informed grid and require all lags to be available at the forecast
origin. & Excessive lags increase dimension and can incorporate observations
unavailable at the forecast origin if alignment is incorrect. \\
\(\{\alpha_d\},\{\rho_d\}\) & Reservoir time scale, persistence, and stability &
Compare jointly because both affect memory. Near-one radii require empirical
stability checks. & States concentrated near the activation bounds,
near-constant states, or sensitivity to initial conditions. \\
\(\{\pi^{(W)}_d\},\{\pi^{(\mathrm{in})}_d\}\) and input scaling & Recurrent
connectivity and input-induced variation & Fix from a small set of literature-guided
defaults; broaden the comparison when diagnostics show weak state variation or
concentration near the activation bounds. & Low connectivity can produce weakly
varying states; high connectivity or input scaling can concentrate states near
the activation bounds. \\
\(f,k,T_0\) & Nonlinearity, transformation of reduced lower-layer states, and initial-condition
washout & Fix before feature-map comparison, with \(T_0\) chosen relative to the
memory scale. & Short washout can leave initial-condition artifacts. \\
Random reservoir realization & Realization of the random feature map & Assess
variation across repeated reservoir realizations. &
Held-out scores can vary materially across realizations. \\
\bottomrule
\end{tabularx}
\endgroup
\caption{Reservoir specification and diagnostics for the fixed
Q--DESN feature map.}
\label{tab:supp-reservoir-selection}
\end{table}

\section{Distributional Conventions}
\label{sec:dist_conventions}

\noindent\textbf{Shared parameterizations.}
The posterior kernels below use the following distributional
parameterizations. These conventions are used in the MCMC,
coordinate-ascent variational inference (CAVI), and evidence lower bound (ELBO)
derivations below; in particular, all inverse-gamma distributions use a rate in
the reciprocal scale.

The inverse-gamma (IG) parameterization is
\[
p(x\mid a,b)=\frac{b^a}{\Gamma(a)}x^{-a-1}\exp(-b/x),\qquad x>0,
\]
and the generalized inverse Gaussian (GIG) parameterization
\[
p(x\mid \lambda,\chi,\psi)
\propto x^{\lambda-1}\exp\left\{-\frac12(\chi/x+\psi x)\right\},
\qquad x>0.
\]
The notation \(\TN(m,V)\) denotes a Normal distribution with mean \(m\) and
variance \(V\), truncated to \((0,\infty)\). The functions \(\Phi\) and
\(\phi\) denote the standard Normal distribution and density functions.

\noindent\textbf{AL constants.}
For a fixed target quantile \(p_0\in(0,1)\), the AL constants are
\[
A_0=\frac{1-2p_0}{p_0(1-p_0)},\qquad
B_0=\frac{2}{p_0(1-p_0)}.
\]
\noindent\textbf{exAL branch constants.}
The notation \(\exAL\) denotes the quantile-fixed generalized asymmetric
Laplace distribution of \citet{yan2025new}. This convention emphasizes the
role of the likelihood as an extension of AL while preserving the
quantile-fixed parameterization that identifies \(\mu\) as the target
conditional quantile. The exAL likelihood
introduces an asymmetry parameter \(\gamma\in(L,U)\), where \(L\) and
\(U\) are determined by the quantile-identifying construction. Define
\[
g(\gamma)=2\Phi(-|\gamma|)\exp(\gamma^2/2),\qquad
p_\gamma=\ind\{\gamma<0\}
+\frac{p_0-\ind\{\gamma<0\}}{g(\gamma)}.
\]
Then
\[
A_\gamma=\frac{1-2p_\gamma}{p_\gamma(1-p_\gamma)},\quad
B_\gamma=\frac{2}{p_\gamma(1-p_\gamma)},\quad
C_\gamma=\{\ind\{\gamma>0\}-p_\gamma\}^{-1},\quad
D_\gamma=C_\gamma|\gamma|.
\]
The main article denotes the same exAL shift coefficient by
\(\lambda(\gamma)=C(\gamma)|\gamma|\); the supplement uses \(D_\gamma\)
to reduce visual clutter in posterior kernels.

\noindent\textbf{AL nesting and branchwise regularity.}
When \(\gamma=0\) and the \(s_t\) latents are omitted, the exAL hierarchy
reduces to the AL working likelihood with constants \(A_0,B_0\).
This nesting is exact. The surrounding \(p_\gamma\), \(A_\gamma\),
\(B_\gamma\), and \(D_\gamma\) maps are branchwise smooth for
\(\gamma<0\) and \(\gamma>0\), but the absolute value and indicator branches
make generic two-sided derivative statements at \(\gamma=0\) invalid except in
special symmetric cases. In particular, the one-sided derivatives of
\(p_\gamma\) at zero are proportional to \(1-p_0\) from the left and to
\(p_0\) from the right. At \(\gamma=0\), calculations use the nested AL model;
for \(\gamma\ne0\), numerical derivatives are evaluated within the
corresponding sign region.

\section{DESN Feature Map and Fixed-Design Quantile Regression}
\label{sec:desn_features}

The Q--DESN regression inherits the feature map defined in the main article. The
reservoir weights, input weights, reducers, leak rates, and washout convention
are fixed before posterior inference. Posterior inference conditions on the
design matrix \(\mat X\) constructed from these features. Scalar responses are denoted by
lowercase \(y_t\), both in likelihood statements and posterior kernels.

Let \(\vect u_t=(y_{t-1},\ldots,y_{t-m},\vect z_t^\top)^\top\) collect response
lags and exogenous covariates. For layers \(d=1,\ldots,D\), Q--DESN draws sparse
reservoir matrices \(\mat W_d\), input matrices \(\mat W^{\mathrm{in}}_d\), and
optional reducers \(\mat Q_d\), \(d=1,\ldots,D-1\), once and keeps them fixed.
With layer-specific leak rates \(\alpha_d\), activation \(f\), and spectral
radius targets \(\rho_d\),
the layer states satisfy
\begin{align}
\vect h_{t,1}
&=(1-\alpha_1)\vect h_{t-1,1}
+\alpha_1 f\{\bar{\mat W}_1\vect h_{t-1,1}+\mat W^{\mathrm{in}}_1\vect u_t\},
\label{eq:supp_desn_layer1}\\
\vect h_{t,d}
&=(1-\alpha_d)\vect h_{t-1,d}
+\alpha_d f\{\bar{\mat W}_d\vect h_{t-1,d}
+\mat W^{\mathrm{in}}_d\tilde{\vect h}_{t,d-1}\},\qquad d\ge2,
\label{eq:supp_desn_layerd}\\
\tilde{\vect h}_{t,d-1}&=\mat Q_{d-1}\vect h_{t,d-1},
\end{align}
where \(\bar{\mat W}_d=\rho_d\mat W_d/\rho(\mat W_d)\). The fixed feature
vector is
\begin{equation}
\tilde{\vect x}_t
=\{ \vect h_{t,D}^\top,\ k(\tilde{\vect h}_{t,1})^\top,\ldots,
k(\tilde{\vect h}_{t,D-1})^\top,\ \vect u_t^\top\}^\top,
\qquad
\vect x_t=(1,\tilde{\vect x}_t^\top)^\top.
\label{eq:supp_feature_stack}
\end{equation}
After a fixed washout period, the reservoir quantities in
\eqref{eq:supp_feature_stack} are treated as deterministic. Let
\(\vect\beta=(\beta_0,\beta_1,\ldots,\beta_r)^\top\) and
\[
\mat X=[\vect x_1^\top;\ldots;\vect x_T^\top]\in\R^{T\times(r+1)},\qquad
\mu_t=\vect x_t^\top\vect\beta .
\]
All posterior inference is conditional on \(\mat X\).

\section{Q--DESN Working Likelihoods, Priors, and Augmented Posterior}
\label{sec:model_joint}

The augmented posterior densities below are used by both computational
schemes. The AL and exAL likelihoods are working likelihoods for a
fixed target quantile, while the auxiliary variables provide equivalent
mixture representations for computation.

\subsection{AL and exAL Augmented Likelihoods}

The exAL Q--DESN observation hierarchy is
\begin{align}
v_t\mid\sigma &\sim \Exp(\text{rate}=1/\sigma),&
s_t&\sim\TN(0,1),\\
y_t\mid\vect\beta,\sigma,\gamma,v_t,s_t,\mat X
&\sim
\Normal\left(
\vect x_t^\top\vect\beta+\sigma D_\gamma s_t+A_\gamma v_t,
\sigma B_\gamma v_t
\right).
\label{eq:supp_exal_hierarchy}
\end{align}
The marginal \(p_0\)-quantile of \(y_t\mid\vect x_t\) is
\(\vect x_t^\top\vect\beta\). The AL special case is
\begin{align}
v_t\mid\sigma &\sim \Exp(\text{rate}=1/\sigma),\\
y_t\mid\vect\beta,\sigma,v_t,\mat X
&\sim
\Normal\left(
\vect x_t^\top\vect\beta+A_0v_t,
\sigma B_0v_t
\right).
\label{eq:supp_al_hierarchy}
\end{align}

\subsection{Coefficient Priors}
\label{sec:priors}

The coefficient prior regularizes the high-dimensional regression after the
reservoir features have been fixed. The ridge prior is the default Gaussian
baseline. The regularized-horseshoe prior is the adaptive global-local shrinkage
alternative; its product representation keeps the local, global, and slab scale
updates in closed form.

For the ridge coefficient prior, let
\begin{equation}
\vect\beta\sim\Normal(\vect b_\beta,\mat V_\beta),
\qquad
\mat P_\beta=\mat V_\beta^{-1}.
\label{eq:supp_ridge_prior}
\end{equation}
Applications usually keep the intercept weakly regularized and use a
diagonal slope penalty, while \eqref{eq:supp_ridge_prior} allows a general
positive definite prior covariance.

For the regularized-horseshoe prior, the intercept is independent,
\(\beta_0\sim\Normal(b_0,V_0)\), and the non-intercept coefficients are
represented through the
Nishimura--Suchard-style regularized horseshoe construction
\citep{CarvalhoPolsonScott2010HS,PiironenVehtari2017RHS,NishimuraSuchard2023SSS}.
For \(j=1,\ldots,r\),
\begin{equation}
p_{\mathrm{ns}}(\beta_j,\lambda_j\mid\tau,\zeta)
\propto
\exp\left(-\frac{\beta_j^2}{2\tau^2\lambda_j^2}\right)
\exp\left(-\frac{\beta_j^2}{2\zeta^2}\right)
\lambda_j^{-1}(1+\lambda_j^2)^{-1}.
\label{eq:supp_ns_joint}
\end{equation}
Equation~\eqref{eq:supp_ns_joint} is the marginal joint description after the
half-Cauchy auxiliary variable for the local scale has been integrated out. The
MCMC and VB derivations use the equivalent augmented product
\[
\Normal(\beta_j\mid0,\tau^2\lambda_j^2)\,
\Normal(0\mid\beta_j,\zeta^2)
\]
together with inverse-gamma auxiliaries for \(\lambda_j^2\), \(\nu_j\),
\(\tau^2\), and \(\xi\). This augmented product representation is the one used
in the augmented posterior densities below.
Equivalently, the conditional coefficient law is Gaussian with effective
variance
\begin{equation}
\beta_j\mid\lambda_j,\tau,\zeta
\sim
\Normal(0,V_j),\qquad
V_j=\left(\zeta^{-2}+\tau^{-2}\lambda_j^{-2}\right)^{-1}
=\frac{\zeta^2\tau^2\lambda_j^2}{\zeta^2+\tau^2\lambda_j^2}.
\label{eq:supp_rhs_cond_beta}
\end{equation}
For inverse-gamma scale updates, use the half-Cauchy augmentation
\citep{MakalicSchmidt2016SimpleSampler}:
\begin{align}
\lambda_j^2\mid\nu_j&\sim\IG(1/2,1/\nu_j),&
\nu_j&\sim\IG(1/2,1),\\
\tau^2\mid\xi&\sim\IG(1/2,1/\xi),&
\xi&\sim\IG(1/2,1/\tau_0^2),
\label{eq:supp_hs_aux}\\
\zeta^2&\sim\IG(a_\zeta,b_\zeta),
\label{eq:supp_zeta_prior}
\end{align}
where the last line is omitted when \(\zeta^2\) is fixed.

\subsection{Augmented Posterior Densities}

\noindent\textbf{Augmented-posterior convention.}
The augmented posterior distributions below include each AL or exAL mixture variable
once. The inverse-gamma variables in this representation augment only
the shrinkage scales; the likelihood augmentation is provided by the AL or exAL
mixture variables.

Let \(\vect v=(v_1,\ldots,v_T)^\top\) and
\(\vect s=(s_1,\ldots,s_T)^\top\). Under exAL and ridge,
\begin{align}
p(\vect y,\vect v,\vect s,\vect\beta,\sigma,\gamma\mid\mat X)
&=
\left\{\prod_{t=1}^T
p(y_t\mid\vect\beta,\sigma,\gamma,v_t,s_t,\vect x_t)
p(v_t\mid\sigma)p(s_t)\right\}
p(\vect\beta)p(\sigma)\pi_\gamma(\gamma)\ind\{L<\gamma<U\}.
\label{eq:supp_joint_exal_ridge}
\end{align}
\noindent\textbf{RHS target.}
Under exAL and the regularized-horseshoe prior, with
\(\Theta_{\mathrm{rhs}}=(\vect\lambda^2,\vect\nu,\tau^2,\xi,\zeta^2)\),
\begin{align}
&p(\vect y,\vect v,\vect s,\vect\beta,\sigma,\gamma,\Theta_{\mathrm{rhs}}\mid\mat X)
\nonumber\\
&\quad=
\left\{\prod_{t=1}^T
p(y_t\mid\vect\beta,\sigma,\gamma,v_t,s_t,\vect x_t)
p(v_t\mid\sigma)p(s_t)\right\}
p(\beta_0)
\prod_{j=1}^r
\left\{\Normal(\beta_j\mid0,\tau^2\lambda_j^2)
\Normal(0\mid\beta_j,\zeta^2)\right\}
\nonumber\\
&\qquad\times
\prod_{j=1}^r p(\lambda_j^2\mid\nu_j)p(\nu_j)
p(\tau^2\mid\xi)p(\xi)p(\zeta^2)
p(\sigma)\pi_\gamma(\gamma)\ind\{L<\gamma<U\}.
\label{eq:supp_joint_exal_rhs}
\end{align}
\noindent\textbf{AL reductions.}
The AL joint distributions are obtained by dropping \(\vect s\) and \(\gamma\),
replacing \(A_\gamma,B_\gamma\) by \(A_0,B_0\), and using
\eqref{eq:supp_al_hierarchy}. Thus, for example, the AL target with a ridge
prior is
\begin{align}
p(\vect y,\vect v,\vect\beta,\sigma\mid\mat X)
&=
\left\{\prod_{t=1}^T
p(y_t\mid\vect\beta,\sigma,v_t,\vect x_t)p(v_t\mid\sigma)\right\}
p(\vect\beta)p(\sigma).
\label{eq:supp_joint_al_ridge}
\end{align}
The AL target with the regularized-horseshoe prior is, analogously,
\begin{align}
&p(\vect y,\vect v,\vect\beta,\sigma,\Theta_{\mathrm{rhs}}\mid\mat X)
\nonumber\\
&\quad=
\left\{\prod_{t=1}^T
p(y_t\mid\vect\beta,\sigma,v_t,\vect x_t)p(v_t\mid\sigma)\right\}
p(\beta_0)
\prod_{j=1}^r
\left\{\Normal(\beta_j\mid0,\tau^2\lambda_j^2)
\Normal(0\mid\beta_j,\zeta^2)\right\}
\nonumber\\
&\qquad\times
\prod_{j=1}^r p(\lambda_j^2\mid\nu_j)p(\nu_j)
p(\tau^2\mid\xi)p(\xi)p(\zeta^2)p(\sigma).
\label{eq:supp_joint_al_rhs}
\end{align}
The posterior distribution is proportional to the corresponding joint density,
viewed as a function of unknown parameters and latent variables conditional on
\((\vect y,\mat X)\).

\section{MCMC Full Conditionals}
\label{sec:mcmc}

Conditional on the reservoir features, the ridge and Nishimura--Suchard
regularized-horseshoe samplers share the same likelihood updates; they differ
in the coefficient-prior precision and shrinkage-scale updates. The Gaussian,
GIG, truncated-normal, inverse-gamma, and conditional Gibbs steps follow from
the augmented posterior. The joint exAL \((\sigma,\gamma)\) block requires a
nonconjugate update.

\subsection{Coefficient Block}

For exAL, define
\begin{equation}
y_t^\star=y_t-\sigma D_\gamma s_t-A_\gamma v_t,\qquad
w_t=(\sigma B_\gamma v_t)^{-1},\qquad
\mat\Omega=\diag(w_1,\ldots,w_T).
\end{equation}
Let \(\vect y^\star=(y_1^\star,\ldots,y_T^\star)^\top\). For a ridge prior,
set \(\mat P=\mat P_\beta\) and \(\vect b=\vect b_\beta\). For
the regularized-horseshoe prior, set
\begin{equation}
\mat P=\diag\left(V_0^{-1},
\tau^{-2}\lambda_1^{-2}+\zeta^{-2},\ldots,
\tau^{-2}\lambda_r^{-2}+\zeta^{-2}\right),
\qquad
\vect b=(b_0,0,\ldots,0)^\top.
\label{eq:supp_prior_precision_mcmc}
\end{equation}
Then
\begin{align}
\mat\Sigma_{\beta\mid\cdot}
&=(\mat X^\top\mat\Omega\mat X+\mat P)^{-1},\\
\vect m_{\beta\mid\cdot}
&=\mat\Sigma_{\beta\mid\cdot}
(\mat X^\top\mat\Omega\vect y^\star+\mat P\vect b),\\
\vect\beta\mid\cdot
&\sim
\Normal(\vect m_{\beta\mid\cdot},\mat\Sigma_{\beta\mid\cdot}).
\label{eq:supp_beta_mcmc}
\end{align}
For AL, use the same update with
\[
y_t^\star=y_t-A_0v_t,\qquad
w_t=(\sigma B_0v_t)^{-1},\qquad
\mat\Omega=\diag(w_1,\ldots,w_T).
\]

\subsection{exAL Latent-Variable Blocks}

For \(v_t\), define
\[
\delta_t=y_t-\vect x_t^\top\vect\beta-\sigma D_\gamma s_t,\qquad
\chi_t=\frac{\delta_t^2}{\sigma B_\gamma},\qquad
\psi_t=\frac{A_\gamma^2}{\sigma B_\gamma}+\frac{2}{\sigma}.
\]
Then
\begin{equation}
v_t\mid\cdot\sim\GIG(1/2,\chi_t,\psi_t).
\label{eq:supp_v_mcmc}
\end{equation}
For \(s_t\), define \(\mu_{0t}=y_t-\vect x_t^\top\vect\beta-A_\gamma v_t\).
Then
\begin{align}
a_{s,t}&=1+\frac{\sigma D_\gamma^2}{B_\gamma v_t},&
m_{s,t}&=\frac{D_\gamma\mu_{0t}}{B_\gamma v_t+\sigma D_\gamma^2},\\
s_t\mid\cdot&\sim\TN(m_{s,t},a_{s,t}^{-1}).
\label{eq:supp_s_mcmc}
\end{align}

\subsection{AL Latent-Variable and Scale Blocks}

For AL,
\begin{equation}
v_t\mid\cdot
\sim
\GIG\left(
1/2,
\frac{(y_t-\vect x_t^\top\vect\beta)^2}{\sigma B_0},
\frac{A_0^2}{\sigma B_0}+\frac{2}{\sigma}
\right).
\label{eq:supp_al_v_mcmc}
\end{equation}
With \(\sigma\sim\IG(a_\sigma,b_\sigma)\),
\begin{align}
\sigma\mid\cdot
&\sim
\IG\left(
a_\sigma+\frac{3T}{2},
b_\sigma+\sum_{t=1}^T v_t
+\frac12\sum_{t=1}^T
\frac{(y_t-\vect x_t^\top\vect\beta-A_0v_t)^2}{B_0v_t}
\right).
\label{eq:supp_al_sigma_mcmc}
\end{align}

\subsection{exAL Scale and Asymmetry Blocks}

The pair \((\sigma,\gamma)\) is non-conjugate when updated jointly. The exAL
coefficient map is smooth within each sign branch for \(\gamma\), and the AL
model is exactly nested at \(\gamma=0\). Because \(p_\gamma\) and the associated
coefficients are defined separately for \(\gamma<0\) and \(\gamma>0\), generic
two-sided differentiability at zero need not hold for nonmedian quantiles.
The joint log-kernel is
\begin{align}
\ell_{\sigma,\gamma}
&=
-\frac12\sum_{t=1}^T\log(\sigma B_\gamma v_t)
-\frac12\sum_{t=1}^T
\frac{\{y_t-\vect x_t^\top\vect\beta-\sigma D_\gamma s_t-A_\gamma v_t\}^2}
{\sigma B_\gamma v_t}
\nonumber\\
&\quad
-T\log\sigma-\sum_{t=1}^T\frac{v_t}{\sigma}
-(a_\sigma+1)\log\sigma-\frac{b_\sigma}{\sigma}
+\log\pi_\gamma(\gamma),
\label{eq:supp_sigmagam_joint_kernel}
\end{align}
for \(L<\gamma<U\), with value \(-\infty\) outside this interval.
This kernel can be sampled by random-walk Metropolis, slice, or a
Laplace-informed proposal on transformed coordinates. Conditional on
\(\gamma\), the Gibbs update for \(\sigma\) is
\begin{align}
\sigma\mid\gamma,\cdot
&\sim
\GIG\left(
-a_\sigma-\frac{3T}{2},
2b_\sigma+2\sum_{t=1}^T v_t
+\sum_{t=1}^T\frac{(y_t-\vect x_t^\top\vect\beta-A_\gamma v_t)^2}
{B_\gamma v_t},
\sum_{t=1}^T\frac{D_\gamma^2s_t^2}{B_\gamma v_t}
\right).
\label{eq:supp_exal_sigma_mcmc}
\end{align}
Conditioning on \(\sigma\), the one-dimensional \(\gamma\) kernel is
\begin{align}
\pi(\gamma\mid\cdot)
&\propto
\pi_\gamma(\gamma)\ind\{L<\gamma<U\}
B_\gamma^{-T/2}
\exp\left[
\sum_{t=1}^T
\frac{D_\gamma s_t\{y_t-\vect x_t^\top\vect\beta-A_\gamma v_t\}}
{B_\gamma v_t}
\right.
\nonumber\\
&\qquad\left.
-\frac{1}{2\sigma}\sum_{t=1}^T
\frac{(y_t-\vect x_t^\top\vect\beta-A_\gamma v_t)^2}{B_\gamma v_t}
-\frac{\sigma}{2}\sum_{t=1}^T
\frac{D_\gamma^2s_t^2}{B_\gamma v_t}
\right].
\label{eq:supp_gamma_mcmc}
\end{align}

\subsection{Regularized-Horseshoe Shrinkage Blocks}
\label{sec:rhs_ns}

Conditioning on \(\vect\beta\), the regularized-horseshoe scale block has
inverse-gamma full conditionals:
\begin{align}
\lambda_j^2\mid\cdot
&\sim
\IG\left(1,\frac{1}{\nu_j}+\frac{\beta_j^2}{2\tau^2}\right),\\
\nu_j\mid\cdot
&\sim
\IG\left(1,1+\frac{1}{\lambda_j^2}\right),\\
\tau^2\mid\cdot
&\sim
\IG\left(\frac{r+1}{2},
\frac{1}{\xi}+\frac12\sum_{j=1}^r\frac{\beta_j^2}{\lambda_j^2}\right),\\
\xi\mid\cdot
&\sim
\IG\left(1,\frac{1}{\tau_0^2}+\frac{1}{\tau^2}\right),\\
\zeta^2\mid\cdot
&\sim
\IG\left(a_\zeta+\frac{r}{2},
b_\zeta+\frac12\sum_{j=1}^r\beta_j^2\right),
\label{eq:supp_rhs_mcmc}
\end{align}
with the final line omitted when \(\zeta^2\) is fixed. The relevant
distinction is that the Nishimura--Suchard joint construction preserves the
ordinary global-local full conditional block given \(\vect\beta\), while the
direct Piironen--Vehtari joint prior changes these inverse-gamma updates.

\begin{suppalgorithm}{Single-level Q--DESN MCMC sampler}
\label{alg:qdesn_mcmc}
\emph{Input:} response vector \(\vect y\), fixed DESN design matrix \(\mat X\),
target quantile \(p_0\), likelihood choice AL or exAL, coefficient prior ridge
or regularized horseshoe, and MCMC controls.
\begin{enumerate}[label=(\arabic*)]
\item Initialize \(\vect\beta,\sigma,\vect v\), and, for exAL,
\(\gamma,\vect s\). Initialize regularized-horseshoe scales when used.
\item For each MCMC iteration:
  \begin{enumerate}[label=(\alph*)]
  \item draw each \(v_t\mid\cdot\) from the GIG full conditional in
  \eqref{eq:supp_v_mcmc} for exAL or \eqref{eq:supp_al_v_mcmc} for AL;
  \item for exAL, draw each \(s_t\mid\cdot\) from the positive-truncated
  Normal full conditional in \eqref{eq:supp_s_mcmc};
  \item draw \(\vect\beta\mid\cdot\) from the Gaussian full conditional in
  \eqref{eq:supp_beta_mcmc};
  \item for the regularized-horseshoe prior, draw
  \((\lambda_j^2,\nu_j)_{j=1}^r,\tau^2,\xi,\zeta^2\) by
  \eqref{eq:supp_rhs_mcmc};
  \item for AL, draw \(\sigma\mid\cdot\) from
  \eqref{eq:supp_al_sigma_mcmc};
  \item for exAL, update \((\sigma,\gamma)\) on transformed coordinates
  targeting \eqref{eq:supp_sigmagam_joint_kernel}, or use the conditional
  updates in \eqref{eq:supp_exal_sigma_mcmc}--\eqref{eq:supp_gamma_mcmc};
  \item store retained draws and diagnostics after burn-in, with thinning only
  if used.
  \end{enumerate}
\end{enumerate}
\end{suppalgorithm}

\section{Variational Bayes and Laplace--Delta Approximation}
\label{sec:cavi}

The selected simulation and application analyses use VB--LD for
computationally efficient approximate inference. The approximation is
mean-field over the conditionally conjugate blocks and uses the Laplace--Delta
method of \citet{wang2013nonconjugatevb} for the nonconjugate exAL
scale--asymmetry block \((\sigma,\gamma)\). This section
gives the generic fixed-design updates; application-specific GloFAS factors
are given separately in Section~\ref{subsec:glofas_vb}.

For a single target level \(p_0\), use the mean-field family
\begin{equation}
q=
q_\beta(\vect\beta)\,q_\Theta(\Theta_\beta)\,
q_{\sigma,\gamma}(\sigma,\gamma)
\prod_{t=1}^T q_v(v_t)q_s(s_t),
\label{eq:supp_generic_vb_family}
\end{equation}
where \(q_s\) and the \(\gamma\) component are omitted for AL, and
\(q_\Theta\) is omitted under a fixed ridge prior. The coefficient factor is
Gaussian. Conditional on the latent variables, both AL and exAL have the form
\[
y_t=\vect x_t^\top\vect\beta+d_t+\epsilon_t,\qquad
\epsilon_t\mid\omega_t\sim\Normal(0,\omega_t),
\]
with \(d_t=A_0v_t\) and \(\omega_t=B_0\sigma v_t\) for AL, and
\[
d_t=\lambda(\gamma)\sigma s_t+A(\gamma)v_t,\qquad
\omega_t=B(\gamma)\sigma v_t
\]
for exAL. Define the variational canonical moments
\[
\bar w_t=\E_q(\omega_t^{-1}),\qquad
\bar r_t=\E_q\{\omega_t^{-1}(y_t-d_t)\}.
\]
If the prior contributes Gaussian canonical parameters
\((\bar{\mat P}_\beta,\bar{\vect h}_\beta)\), then
\begin{align}
q_\beta(\vect\beta)&=\Normal(\vect m_\beta,\mat\Sigma_\beta),\\
\mat\Sigma_\beta&=
\left(\mat X^\top\bar{\mat W}\mat X+\bar{\mat P}_\beta\right)^{-1},\\
\vect m_\beta&=
\mat\Sigma_\beta\left(\mat X^\top\bar{\vect r}
+\bar{\vect h}_\beta\right),
\label{eq:supp_generic_vb_beta}
\end{align}
where \(\bar{\mat W}=\diag(\bar w_1,\ldots,\bar w_T)\) and
\(\bar{\vect r}=(\bar r_1,\ldots,\bar r_T)^\top\). For a fixed ridge prior,
\(\bar{\mat P}_\beta=\diag(V_0^{-1},\kappa_\beta^{-2}\ones{r})\) and
\(\bar{\vect h}_\beta=(V_0^{-1}b_0,0,\ldots,0)^\top\). Under
the regularized-horseshoe prior, the intercept block is unchanged and the slope precision
uses
\[
\bar P_{\beta,j}
=
\E_q(\tau^{-2})\E_q(\lambda_j^{-2})+\E_q(\zeta^{-2}),
\qquad j=1,\ldots,r .
\]

For AL, the latent and scale updates are conjugate after taking expectations
over \(q_\beta\):
\begin{align}
q_v(v_t)
&=
\GIG\left(
\frac12,\
B_0^{-1}\E_q(\sigma^{-1})\overline e_t^{\,2},\
\E_q(\sigma^{-1})\left(\frac{A_0^2}{B_0}+2\right)
\right),\\
q_\sigma(\sigma)
&=\IG(a_\sigma^\star,b_\sigma^\star),
\label{eq:supp_generic_vb_al_latents}
\end{align}
where
\(\overline e_t^{\,2}
=\{y_t-\vect x_t^\top\vect m_\beta\}^2
+\vect x_t^\top\mat\Sigma_\beta\vect x_t\), and
\[
a_\sigma^\star=a_\sigma+\frac{3T}{2},\qquad
b_\sigma^\star=b_\sigma+\sum_{t=1}^T\E_q(v_t)
+\frac{1}{2B_0}\sum_{t=1}^T
\left[
\overline e_t^{\,2}\E_q(v_t^{-1})
-2A_0\{y_t-\vect x_t^\top\vect m_\beta\}
+A_0^2\E_q(v_t)
\right].
\]
For exAL, \(q_v(v_t)\) and \(q_s(s_t)\) have the same GIG and
positive-truncated Normal forms as the MCMC blocks after replacing products by
expectations under the other variational factors. The smooth expectations that
depend on \((\sigma,\gamma)\), such as
\(\E\{(\sigma B_\gamma)^{-1}\}\),
\(\E(D_\gamma/B_\gamma)\), and
\(\E\{A_\gamma^2/(\sigma B_\gamma)+2/\sigma\}\), are evaluated by the
Laplace--Delta approximation below.

The regularized-horseshoe CAVI updates are the MCMC inverse-gamma scale updates with
coefficient squares replaced by second moments
\(\overline{\beta_j^2}=m_{\beta,j}^2+(\mat\Sigma_\beta)_{jj}\):
\begin{align}
q(\lambda_j^2)
&=\IG\left(1,\ \E_q(\nu_j^{-1})
+\frac12\overline{\beta_j^2}\E_q(\tau^{-2})\right),\\
q(\nu_j)
&=\IG\left(1,\ 1+\E_q(\lambda_j^{-2})\right),\\
q(\tau^2)
&=\IG\left(\frac{r+1}{2},\
\E_q(\xi^{-1})+\frac12\sum_{j=1}^r
\overline{\beta_j^2}\E_q(\lambda_j^{-2})\right),\\
q(\xi)
&=\IG\left(1,\ \tau_0^{-2}+\E_q(\tau^{-2})\right),\\
q(\zeta^2)
&=\IG\left(a_\zeta+\frac r2,\
b_\zeta+\frac12\sum_{j=1}^r\overline{\beta_j^2}\right).
\label{eq:supp_generic_vb_rhs}
\end{align}
The slab update is omitted when \(\zeta\) is fixed.

For the exAL scale-asymmetry factor, map unconstrained coordinates
\(\vect\eta\) to \((\sigma,\gamma)\), for example by
\(\sigma=\exp(\eta_1)\) and a smooth transformation from \(\eta_2\) to
\((L,U)\). Let
\[
\ell_{\mathrm{LD}}(\vect\eta)
=
\E_{q_{-(\sigma,\gamma)}}\{\log p(\vect y,\vect v,\vect s
\mid\vect\beta,\sigma,\gamma,\mat X)\}
+\log p(\sigma)+\log\pi_\gamma(\gamma)+\log|J(\vect\eta)|.
\]
At the mode \(\hat{\vect\eta}\), with negative Hessian
\(\mat K_\eta=-\nabla^2\ell_{\mathrm{LD}}(\hat{\vect\eta})\), set
\[
q_{\sigma,\gamma}(\vect\eta)
\approx
\Normal(\hat{\vect\eta},\mat K_\eta^{-1}).
\]
For any branchwise smooth function \(h(\sigma,\gamma)\), expectations used by
the coordinate updates are evaluated by
\begin{equation}
\E_q\{h(\sigma,\gamma)\}
\approx
h\{T(\hat{\vect\eta})\}
+\frac12\tr\left[
\nabla^2_\eta h\{T(\hat{\vect\eta})\}\mat K_\eta^{-1}
\right],
\label{eq:supp_generic_delta}
\end{equation}
with the first term alone giving the Laplace-mode approximation when the delta
term is omitted.

\begin{suppalgorithm}{Single-level Q--DESN VB--LD coordinate updates}
\label{alg:qdesn_vb_ld}
\emph{Input:} response vector \(\vect y\), fixed DESN design matrix \(\mat X\),
target quantile \(p_0\), likelihood choice AL or exAL, coefficient prior ridge
or regularized horseshoe, and convergence controls.
\begin{enumerate}[label=(\arabic*)]
\item Initialize \(q_\beta\), latent-variable factors, likelihood scale
factors, and, when used, RHS and exAL scale-asymmetry factors.
\item Repeat until the ELBO or chosen parameter-change criterion has converged:
  \begin{enumerate}[label=(\alph*)]
  \item set \(q_\beta(\vect\beta)\) to the Gaussian factor in
  \eqref{eq:supp_generic_vb_beta};
  \item for AL, set \(q_v(v_t)\) and \(q_\sigma(\sigma)\) by
  \eqref{eq:supp_generic_vb_al_latents};
  \item for exAL, set \(q_v(v_t)\) and \(q_s(s_t)\) to the GIG and
  positive-truncated Normal factors implied by the corresponding MCMC blocks,
  replacing products by expectations under the other factors;
  \item for the regularized-horseshoe prior, set the inverse-gamma scale factors by
  \eqref{eq:supp_generic_vb_rhs};
  \item for exAL, set \(q_{\sigma,\gamma}\) by the Laplace--Delta Gaussian
  approximation on transformed coordinates and recompute branchwise smooth expectations
  using \eqref{eq:supp_generic_delta};
  \item monitor the ELBO in \eqref{eq:supp_generic_elbo}--\eqref{eq:supp_generic_elbo_decomp},
  interpreted as a Laplace--Delta approximation when the exAL block is active.
  \end{enumerate}
\item Return variational factors, fitted quantile summaries, ELBO history, and
convergence diagnostics.
\end{enumerate}
\end{suppalgorithm}

\section{ELBO Monitoring}
\label{sec:elbo}

\noindent\textbf{Monitored objective.}
The fixed-design VB approximation is assessed using the evidence lower bound
\[
\elbo(q)
=
\E_q\{\log p(\vect y,\vect v,\vect s,\vect\beta,\Theta_\beta,\sigma,\gamma
\mid\mat X)\}+H(q),
\label{eq:supp_generic_elbo}
\]
with the \(s\)-latent and \(\gamma\) terms omitted for AL. Expanding the exAL
case gives
\begin{align}
\elbo(q)
&=
\E_q\{\log p(\vect y\mid\vect\beta,\sigma,\gamma,\vect v,\vect s,\mat X)\}
+\E_q\{\log p(\vect v\mid\sigma)\}
+\E_q\{\log p(\vect s)\}
\nonumber\\
&\quad
+\E_q\{\log p(\vect\beta,\Theta_\beta)\}
+\E_q\{\log p(\sigma)\}
+\E_q\{\log\pi_\gamma(\gamma)\}
+H(q).
\label{eq:supp_generic_elbo_decomp}
\end{align}
\noindent\textbf{Approximation status.}
For AL, replace the exAL likelihood term by the AL likelihood term and omit
\(\E_q\{\log p(\vect s)\}\) and
\(\E_q\{\log\pi_\gamma(\gamma)\}\). The likelihood component is evaluated from
the conditional Gaussian density using the same expectations
\(\bar w_t\), \(\bar r_t\), \(\E_q(\log\sigma)\),
\(\E_q(\log B_\gamma)\), \(\E_q(\log v_t)\), and latent second moments that
enter the coordinate updates. The prior component includes the Gaussian
coefficient prior and, when the regularized-horseshoe prior is active, the inverse-gamma
local, global, auxiliary, and slab terms. The entropy component sums the
Gaussian coefficient entropy, GIG latent entropies, truncated-normal latent
entropies, inverse-gamma scale entropies, and the Gaussian entropy of the
Laplace--Delta factor for \((\sigma,\gamma)\). When the Laplace--Delta block is
active, \eqref{eq:supp_generic_elbo_decomp} approximates the mean-field ELBO.
Convergence of this quantity assesses numerical stability of the VB--LD
calculation; posterior accuracy is evaluated separately through comparison with
MCMC.

\section{Joint Quantile-Vector Regression with Regularized-Horseshoe Shrinkage}
\label{sec:joint_qdesn_supp}

This section derives the joint multi-quantile regression introduced in the main
article. The reservoir features remain unchanged, and the model adds a prior
over the quantile-indexed regression-coefficient sequence,
conditional on the fixed non-intercept DESN design
\(\mat Z=[\tilde{\vect x}_1^\top;\ldots;\tilde{\vect x}_T^\top]\in\R^{T\times r}\).
The intercepts are kept separate from the high-dimensional slope shrinkage.

\subsection{Stacked exAL Working Likelihood}

Let \(0<p_1<\cdots<p_Q<1\) be the quantile grid. At level \(p_q\),
\[
Q_{p_q}(y_t\mid\mathcal F_t)=
\alpha_q+\tilde{\vect x}_t^\top\vect\beta_q,
\qquad
\alpha_q\in\R,\quad \vect\beta_q\in\R^r .
\]
For \(Q=1\), set \(\alpha_1=\alpha\), \(\vect\beta_1=\vect\beta\), and
\(\vect\beta_{\mathrm{base}}=\zeros{r}\); assign the regularized-horseshoe
prior directly to \(\Delta_1=\vect\beta_1\), and omit the baseline hierarchy.
The adjacent-difference hierarchy then coincides with the single-quantile
regularized-horseshoe prior.
The exAL augmented likelihood is a quantile-indexed copy of
\eqref{eq:supp_exal_hierarchy}:
\begin{align}
v_{q,t}\mid\sigma_q&\sim\Exp(\text{rate}=1/\sigma_q),&
s_{q,t}&\sim\TN(0,1),\\
y_t\mid\alpha_q,\vect\beta_q,\sigma_q,\gamma_q,v_{q,t},s_{q,t}
&\sim
\Normal\left(
\alpha_q+\tilde{\vect x}_t^\top\vect\beta_q
+\sigma_qD_qs_{q,t}+A_qv_{q,t},\,
\sigma_qB_qv_{q,t}
\right),
\label{eq:supp_joint_exal_hierarchy}
\end{align}
where \(A_q=A_{p_q}(\gamma_q)\), \(B_q=B_{p_q}(\gamma_q)\), and
\(D_q=\lambda_{p_q}(\gamma_q)\). The AL special case fixes the AL constants
and omits \(s_{q,t}\) and \(\gamma_q\).

Define
\[
\vect y_{\mathrm{st}}=\ones{Q}\otimes\vect y,\qquad
\mat Z_{\mathrm{st}}=\Id{Q}\otimes\mat Z,\qquad
\vect\beta_{\mathrm{st}}=(\vect\beta_1^\top,\ldots,\vect\beta_Q^\top)^\top,
\]
and
\[
\vect\alpha_{\mathrm{st}}
=
(\alpha_1\ones{T}^\top,\ldots,\alpha_Q\ones{T}^\top)^\top .
\]
Let
\[
d_{q,t}=\sigma_qD_qs_{q,t}+A_qv_{q,t},\qquad
\omega_{q,t}=\sigma_qB_qv_{q,t},
\]
and stack these into \(\vect d\) and
\(\mat\Omega=\diag(\omega_{q,t})\). Conditional on the latent variables and
likelihood parameters,
\begin{equation}
\vect y_{\mathrm{st}}\mid\vect\beta_{\mathrm{st}},\vect\alpha,\vect d,\mat\Omega
\sim
\Normal(
\vect\alpha_{\mathrm{st}}+\mat Z_{\mathrm{st}}\vect\beta_{\mathrm{st}}+\vect d,\,
\mat\Omega).
\label{eq:supp_joint_stacked_likelihood}
\end{equation}
The joint analyses use the product of the \(Q\) quantile-specific
working-likelihood contributions without tempering.

\subsection{Quantile-Indexed Regularized-Horseshoe Prior}
\label{subsec:supp_joint_qvp_prior}

For a genuine multi-quantile grid, let
\(\vect\beta_{\mathrm{base}}\in\R^r\) denote a baseline slope vector. Adjacent
coefficient differences are
\[
\Delta_1=\vect\beta_1-\vect\beta_{\mathrm{base}},\qquad
\Delta_q=\vect\beta_q-\vect\beta_{q-1},\quad q=2,\ldots,Q.
\]
The baseline slope and adjacent differences use separate RHS hierarchies:
\begin{align}
\beta_{\mathrm{base},j}\mid\lambda^0_j,\tau^0,\zeta_0
&\sim\Normal(0,V^0_j),&
V^0_j
&=
\frac{\zeta_0^2(\tau^0)^2(\lambda^0_j)^2}
{\zeta_0^2+(\tau^0)^2(\lambda^0_j)^2},
\label{eq:supp_joint_baseline_var}\\
\Delta_{q,j}\mid\lambda^\Delta_{q,j},\tau^\Delta_q,\zeta_\Delta
&\sim\Normal(0,V^\Delta_{q,j}),&
V^\Delta_{q,j}
&=
\frac{\zeta_\Delta^2(\tau^\Delta_q)^2(\lambda^\Delta_{q,j})^2}
{\zeta_\Delta^2+(\tau^\Delta_q)^2(\lambda^\Delta_{q,j})^2}.
\label{eq:supp_joint_innovation_var}
\end{align}
The local and global half-Cauchy scales use the inverse-gamma augmentation in
\eqref{eq:supp_hs_aux}; it is applied once to the baseline coefficients and
separately to each adjacent-difference block.

Let \(\mat H\in\R^{Qr\times Qr}\) be the block first-difference matrix
satisfying
\[
\mat H\vect\beta_{\mathrm{st}}=
(\vect\beta_1^\top,
(\vect\beta_2-\vect\beta_1)^\top,\ldots,
(\vect\beta_Q-\vect\beta_{Q-1})^\top)^\top .
\]
Set
\[
\tilde{\vect\beta}_{\mathrm{base}}=
(\vect\beta_{\mathrm{base}}^\top,\zeros{r}^\top,\ldots,\zeros{r}^\top)^\top,
\qquad
\mat D_\Delta=\diag(V^\Delta_{q,j}:q=1,\ldots,Q,\ j=1,\ldots,r).
\]
Then the conditional prior over the quantile-indexed slope path can be written as
\begin{equation}
p(\vect\beta_{\mathrm{st}}\mid\vect\beta_{\mathrm{base}},\mat D_\Delta)
\propto
|\mat D_\Delta|^{-1/2}
\exp\left[
-\frac12
(\mat H\vect\beta_{\mathrm{st}}-\tilde{\vect\beta}_{\mathrm{base}})^\top
\mat D_\Delta^{-1}
(\mat H\vect\beta_{\mathrm{st}}-\tilde{\vect\beta}_{\mathrm{base}})
\right].
\label{eq:supp_joint_qvp_prior}
\end{equation}
This representation is sparse and block banded. It is the computational form
used by both the MCMC and VB updates.

\subsection{Complete-Data Posterior}

Let \(\Theta_{\Delta}\) collect all local, global, auxiliary, and slab scales
for the adjacent-difference RHS hierarchy, and let \(\Theta_0\) collect the
corresponding baseline RHS scales. Up to constants not depending on unknowns,
the complete-data posterior for the multi-quantile case is
\begin{align}
&p(\vect\beta_{\mathrm{st}},\vect\beta_{\mathrm{base}},\vect\alpha,
\{\sigma_q,\gamma_q\}_{q=1}^Q,
\{v_{q,t},s_{q,t}\}_{q,t},
\Theta_0,\Theta_\Delta\mid\vect y,\mat Z)
\nonumber\\
&\quad\propto
\exp\left[
-\frac{1}{2}
(\vect y_{\mathrm{st}}-\vect\alpha_{\mathrm{st}}
-\mat Z_{\mathrm{st}}\vect\beta_{\mathrm{st}}-\vect d)^\top
\mat\Omega^{-1}
(\vect y_{\mathrm{st}}-\vect\alpha_{\mathrm{st}}
-\mat Z_{\mathrm{st}}\vect\beta_{\mathrm{st}}-\vect d)
\right]
\nonumber\\
&\qquad\times
\prod_{q=1}^Q\prod_{t=1}^T
\left\{
(\omega_{q,t})^{-1/2}
p(v_{q,t}\mid\sigma_q)p(s_{q,t})
\right\}
\prod_{q=1}^Q p(\sigma_q)\pi_{\gamma_q}(\gamma_q)\ind\{L_q<\gamma_q<U_q\}
\nonumber\\
&\qquad\times
p(\vect\alpha)\,
p(\vect\beta_{\mathrm{base}}\mid\Theta_0)p(\Theta_0)\,
p(\vect\beta_{\mathrm{st}}\mid\vect\beta_{\mathrm{base}},\mat D_\Delta)p(\Theta_\Delta).
\label{eq:supp_joint_complete_posterior}
\end{align}
The reported analysis uses the untempered complete-data posterior in
\eqref{eq:supp_joint_complete_posterior}. Weighting the quantile-specific
contributions would define a different generalized posterior.
For \(Q=1\), remove
\(\vect\beta_{\mathrm{base}}\), \(\Theta_0\), and the baseline-prior factor
from \eqref{eq:supp_joint_complete_posterior}, set
\(\Delta_1=\vect\beta_1\), and use the adjacent-difference RHS hierarchy as the
ordinary single-level RHS hierarchy.

\subsection{Gaussian Blocks}

Let
\(\vect r_\beta=\vect y_{\mathrm{st}}-\vect\alpha_{\mathrm{st}}-\vect d\).
Combining \eqref{eq:supp_joint_stacked_likelihood} and
\eqref{eq:supp_joint_qvp_prior} gives
\begin{align}
\mat K_\beta
&=
\mat Z_{\mathrm{st}}^\top\mat\Omega^{-1}\mat Z_{\mathrm{st}}
+\mat H^\top\mat D_\Delta^{-1}\mat H,\\
\vect m_\beta
&=
\mat K_\beta^{-1}
\left\{
\mat Z_{\mathrm{st}}^\top\mat\Omega^{-1}\vect r_\beta
+\mat H^\top\mat D_\Delta^{-1}\tilde{\vect\beta}_{\mathrm{base}}
\right\}.
\label{eq:supp_joint_beta_conditional}
\end{align}
Thus
\[
\vect\beta_{\mathrm{st}}\mid\cdots\sim\Normal(\vect m_\beta,\mat K_\beta^{-1}).
\]
Numerical computation uses a sparse factorization of \(\mat K_\beta\); forming
its dense inverse is unnecessary.

The baseline coefficients are conditionally independent by feature under the
diagonal RHS representation. For \(j=1,\ldots,r\),
\begin{align}
K_{0,j}&=(V^0_j)^{-1}+(V^\Delta_{1,j})^{-1},\\
m_{0,j}&=K_{0,j}^{-1}(V^\Delta_{1,j})^{-1}\beta_{1,j},
\end{align}
so
\[
\beta_{\mathrm{base},j}\mid\cdots\sim\Normal(m_{0,j},K_{0,j}^{-1}).
\]
This update uses the fact that only the first innovation
\(\Delta_1=\vect\beta_1-\vect\beta_{\mathrm{base}}\) contains \(\vect\beta_{\mathrm{base}}\).

For unconstrained intercepts with independent priors
\(\alpha_q\sim\Normal(a_{\alpha,q},V_{\alpha,q})\), define
\[
e_{q,t}^{\alpha}=y_t-d_{q,t}-\tilde{\vect x}_t^\top\vect\beta_q .
\]
Then
\begin{align}
K_{\alpha,q}
&=V_{\alpha,q}^{-1}+\sum_{t=1}^T\omega_{q,t}^{-1},\\
m_{\alpha,q}
&=K_{\alpha,q}^{-1}
\left\{
V_{\alpha,q}^{-1}a_{\alpha,q}
+\sum_{t=1}^T\omega_{q,t}^{-1}e_{q,t}^{\alpha}
\right\},
\end{align}
and \(\alpha_q\mid\cdots\sim\Normal(m_{\alpha,q},K_{\alpha,q}^{-1})\). For a
monotone-gap version, write
\(\alpha_q=\alpha_1+\sum_{\ell=2}^q\exp(\eta_\ell)\) for
\(q\ge2\). The conditional log density for
\((\alpha_1,\eta_2,\ldots,\eta_Q)\) is the Gaussian likelihood term above plus
the chosen priors on the transformed coordinates and, if the prior is placed on
the positive gaps rather than on \(\eta\), the Jacobian
\(\sum_{\ell=2}^Q\eta_\ell\). This block is then updated by a
low-dimensional Metropolis--Hastings, slice, or Laplace step.

\subsection{exAL Latent-Variable and Scale Blocks}

Conditional on the joint quantile-vector regression, each exAL latent update is a
quantile-indexed copy of the single-level update. Define
\[
e_{q,t}^{v}
=y_t-\alpha_q-\tilde{\vect x}_t^\top\vect\beta_q-\sigma_qD_qs_{q,t}.
\]
Then
\[
p(v_{q,t}\mid\cdots)
\propto
v_{q,t}^{-1/2}
\exp\left[
-\frac12\left\{
\chi_{q,t}^{v}v_{q,t}^{-1}
+\psi_{q,t}^{v}v_{q,t}
\right\}
\right],
\]
which is a GIG density with
\[
\chi_{q,t}^{v}=\frac{(e_{q,t}^{v})^2}{\sigma_qB_q},
\qquad
\psi_{q,t}^{v}=\frac{A_q^2}{\sigma_qB_q}+\frac{2}{\sigma_q}.
\]
The GIG index is \(1/2\). Terms independent of \(v_{q,t}\) have
been omitted from the kernel.

For the positive-shift latent variable, define
\[
e_{q,t}^{s}
=y_t-\alpha_q-\tilde{\vect x}_t^\top\vect\beta_q-A_qv_{q,t}.
\]
The conditional distribution is positive-truncated Normal with precision and
mean
\[
K_{s,q,t}=1+\frac{\sigma_qD_q^2}{B_qv_{q,t}},
\qquad
m_{s,q,t}=K_{s,q,t}^{-1}
\frac{D_qe_{q,t}^{s}}{B_qv_{q,t}},
\]
that is,
\[
s_{q,t}\mid\cdots\sim\TN(m_{s,q,t},K_{s,q,t}^{-1}).
\]
The AL case omits this update and uses the AL constants in the \(v_{q,t}\)
kernel.

For each quantile level, the non-conjugate scale-asymmetry block has log
kernel
\begin{align}
\ell_q(\sigma_q,\gamma_q)
&=
-\frac{1}{2}\sum_{t=1}^T
\left[
\log\{\sigma_qB_qv_{q,t}\}
+
\frac{
\{y_t-\alpha_q-\tilde{\vect x}_t^\top\vect\beta_q
-\sigma_qD_qs_{q,t}-A_qv_{q,t}\}^2}
{\sigma_qB_qv_{q,t}}
\right]
\nonumber\\
&\quad
-T\log\sigma_q-\sigma_q^{-1}\sum_{t=1}^Tv_{q,t}
+\log p(\sigma_q)+\log\pi_{\gamma_q}(\gamma_q).
\label{eq:supp_joint_sigma_gamma_kernel}
\end{align}
This kernel is evaluated on transformed coordinates
\(\eta_{q,1}=\log\sigma_q\) and
\(\eta_{q,2}=\log\{(\gamma_q-L_q)/(U_q-\gamma_q)\}\), adding the
transformation Jacobian for MCMC proposals or the Laplace--Delta approximation.

\subsection{RHS Scale Updates}

The RHS scale updates are copies of the single-level
Nishimura--Suchard product updates. For a baseline coefficient,
\[
\lambda_{0,j}^2\mid\cdots
\sim
\IG\left(
1,\,
\frac{1}{\nu_{0,j}}+\frac{\beta_{\mathrm{base},j}^2}{2(\tau^0)^2}
\right),
\qquad
\nu_{0,j}\mid\cdots
\sim
\IG\left(1,\ 1+\frac{1}{\lambda_{0,j}^2}\right),
\]
and
\[
(\tau^0)^2\mid\cdots
\sim
\IG\left(
\frac{r+1}{2},\,
\frac{1}{\xi_0}
+\frac12\sum_{j=1}^r\frac{\beta_{\mathrm{base},j}^2}{(\lambda^0_j)^2}
\right),
\qquad
\xi_0\mid\cdots
\sim
\IG\left(1,\frac{1}{(\tau^0)^2}+\frac{1}{(\tau_0^0)^2}\right).
\]
For each adjacent-difference block \(q\), replace \(\beta_{\mathrm{base},j}\),
\(\lambda^0_j\), and \(\tau^0\) by \(\Delta_{q,j}\),
\(\lambda^\Delta_{q,j}\), and \(\tau^\Delta_q\). If the slab scale is random
and shared across adjacent-difference blocks, then
\[
\zeta_\Delta^2\mid\cdots
\sim
\IG\left(
a_{\zeta_\Delta}+\frac{Qr}{2},\,
b_{\zeta_\Delta}
+\frac12\sum_{q=1}^Q\sum_{j=1}^r\Delta_{q,j}^2
\right),
\]
with the analogous \(r/2\) update for a random baseline slab scale
\(\zeta_0^2\). If slab scales are fixed, these updates are omitted.

\subsection{MCMC Algorithm and VB--LD Pointer}

\begin{suppalgorithm}{MCMC sampler for the joint quantile-vector Q--DESN posterior}
\label{alg:joint_qdesn_mcmc}
\emph{Input:} fixed non-intercept design \(\mat Z\), response \(\vect y\),
quantile grid \(p_1<\cdots<p_Q\), priors, and MCMC
controls. For \(Q=1\), use the collapsed convention
\(\vect\beta_{\mathrm{base}}\equiv\zeros{r}\) and
\(\Delta_1=\vect\beta_1\), and skip every baseline update below.
\begin{enumerate}[label=(\arabic*)]
\item Initialize \(\vect\alpha\), \(\vect\beta_{\mathrm{st}}\), likelihood
parameters, exAL latent variables, and adjacent-difference RHS scales. If \(Q>1\),
also initialize \(\vect\beta_{\mathrm{base}}\) and baseline RHS scales.
\item Repeat:
  \begin{enumerate}[label=(\alph*)]
  \item draw each \(v_{q,t}\mid\cdot\) from its quantile-indexed GIG full
  conditional; for AL use the AL constants and omit the \(s\)-latent block;
  \item for exAL, draw each \(s_{q,t}\mid\cdot\) from its
  positive-truncated Normal full conditional;
  \item compute the sparse precision and mean in
  \eqref{eq:supp_joint_beta_conditional}, then draw
  \(\vect\beta_{\mathrm{st}}\mid\cdot\sim
  \Normal(\vect m_\beta,\mat K_\beta^{-1})\);
  \item if \(Q>1\), draw \(\vect\beta_{\mathrm{base}}\) featurewise from the
  Gaussian baseline conditionals;
  \item draw unconstrained intercepts from their Gaussian conditionals, or
  update the monotone-gap transformed block when exact intercept ordering is
  imposed;
  \item draw adjacent-difference RHS scales, and if \(Q>1\) draw baseline RHS scales,
  from their inverse-gamma full conditionals;
  \item update each exAL \((\sigma_q,\gamma_q)\) block on transformed
  coordinates targeting \eqref{eq:supp_joint_sigma_gamma_kernel}; for AL,
  use the corresponding \(\sigma_q\) full conditional and omit \(\gamma_q\);
  \item store retained draws and crossing diagnostics after burn-in.
  \end{enumerate}
\end{enumerate}
\end{suppalgorithm}

For the joint VB--LD approximation, the same block structure is used as a
mean-field computational approximation, with Wang--Blei Laplace--Delta updates
only for the non-conjugate exAL scale-asymmetry factors
\citep{wang2013nonconjugatevb}. The generic fixed-design factorization and
ELBO decomposition are given in Sections~\ref{sec:cavi}--\ref{sec:elbo}; the
joint computation applies those same factors to the stacked Gaussian block, with
the required covariance blocks retained for the coordinate updates.

\subsection{Crossing Diagnostics}

For adjacent levels \(q-1\) and \(q\), define the fitted gap at time \(t\) by
\[
g_{q,t}=
(\alpha_q-\alpha_{q-1})
+\tilde{\vect x}_t^\top(\vect\beta_q-\vect\beta_{q-1}).
\]
The crossing count before rearrangement is
\[
C_{\mathrm{raw}}=\sum_{q=2}^Q\sum_{t=1}^T\ind\{g_{q,t}<0\}.
\]
If fitted quantile curves are projected or rearranged, the corresponding
post-rearrangement crossing count should be computed after that stated monotone
rearrangement. A zero crossing count after rearrangement does not indicate how
strongly the fitted curves departed from monotonicity before rearrangement. The
maximum and mean absolute projection adjustments quantify this discrepancy. On a bounded feature domain
\(\tilde{\vect x}_t\in[-1,1]^r\), the sufficient condition
\[
\alpha_q-\alpha_{q-1}\ge
\|\vect\beta_q-\vect\beta_{q-1}\|_1
\]
is a diagnostic reference for adjacent-level separation. Enforcing this
inequality requires explicit constrained estimation.

\section{Multi-Step Quantile Forecasting and Monotone Rearrangement}
\label{sec:forecast_synthesis}

Posterior or variational parameter draws are combined with recursively updated
reservoir states to obtain multi-step quantile forecasts. The analyses use
equal-weight isotonic projection unless a specific evaluation design states
another choice. A
joint multi-quantile posterior model is specified separately. For fitted ordinates
\(\widehat{\vect q}_t=(\widehat q_{1,t},\ldots,\widehat q_{K,t})^\top\) at
levels \(p_1<\cdots<p_K\), the projection is
\begin{equation}
\vect q_t^*
=
\argmin_{q_1\le\cdots\le q_K}
\sum_{k=1}^K w_k(\widehat q_{k,t}-q_k)^2,
\qquad w_k>0.
\label{eq:supp_monotone_projection}
\end{equation}
The projection may be applied draw by draw when joint posterior or variational
quantile-path draws are available, or to posterior summary ordinates for
independently estimated quantiles. Draw-level projection preserves more
posterior dependence information; summary-level projection orders a displayed
grid.

For a single-level fit at level \(p_0\), write a posterior draw or a draw from
the variational approximation as
\(\theta_{p_0}^{(m)}
=(\vect\beta_{p_0}^{(m)},\sigma_{p_0}^{(m)},\gamma_{p_0}^{(m)})\). Forecasting
proceeds by propagating the deterministic reservoir state forward from the
forecast origin. At horizon \(h\), the current lag vector is built from
observed values for one-step forecasts and from previously simulated values for
free-running multi-step forecasts. Conditional on the draw and the resulting
\(p_0\)-indexed design row, the working-likelihood draw is
\[
y_{T+h}^{(m,p_0)}
\sim
\exAL_{p_0}
\left(\vect x_{T+h}^{(m,p_0)\top}\vect\beta_{p_0}^{(m)},
\sigma_{p_0}^{(m)},\gamma_{p_0}^{(m)}\right),
\]
with the \(s\)-latent and \(\gamma\) block omitted for AL. Repetition over
parameter draws yields response-level simulations when the working likelihood
is sampled and conditional-quantile summaries when the regression location is
evaluated directly. The reported scoring criterion is defined for the
corresponding forecast summary. Both summaries condition on the fixed reservoir
realization and the stated working likelihood. To report several
independently fitted levels, the same recursion is run separately for each
fitted \(p_k\).

For a joint quantile-vector regression, a posterior or variational draw returns a
raw quantile vector over \(p_1,\ldots,p_Q\). If recursive lag inputs require a
future response path, that path is generated by a stated propagation rule.
For draw \(m\) at horizon \(h\), write
\[
\mu_{j,T+h}^{(m)}
=\alpha_j^{(m)}
+\tilde{\vect x}_{T+h}^{(m)\top}\vect\beta_j^{(m)},
\qquad j=1,\ldots,Q,
\]
and let
\(\vect q_{T+h}^{(m)}
=(\mu_{1,T+h}^{(m)},\ldots,\mu_{Q,T+h}^{(m)})^\top\). If exact monotone output
is required, apply the stated monotone rearrangement, such as isotonic projection
or rearrangement
\citep{BarlowBrunk1972IsotonicDual,chernozhukov2010},
and set \(\vect q_{T+h}^{*,(m)}=\mathcal M(\vect q_{T+h}^{(m)})\). The ordered
values define a draw-specific quantile curve on the fitted interval
\([p_1,p_Q]\) by linear interpolation.

For independently estimated quantile levels \(p_1<\cdots<p_K\), monotone
rearrangement is applied to the fitted conditional-quantile summaries and may
use isotonic projection \eqref{eq:supp_monotone_projection} or monotone
rearrangement \citep{BarlowBrunk1972IsotonicDual,chernozhukov2010}. Linear
interpolation defines a quantile function over \([p_1,p_K]\), and reported
summaries are restricted to this fitted range. The reported joint study
uses a sequential conditional evaluation. Coefficients and reservoir weights
remain fixed throughout the held-out period, while a realized response becomes
available before it enters the lag vector for subsequent evaluation times.
Joint and independently estimated quantile grids are scored on the same 990
origin--horizon rows for each simulation setting. A free-running response
trajectory over \((0,1)\) would additionally require a tail extension and a
rule for propagating one simulated history across quantile levels.

Forecast scores use the following convention. Let
\(0<p_1<\cdots<p_K<1\), with \(K\geq2\), and write
\(\mathcal P_K=\{p_1,\ldots,p_K\}\) for the evaluation grid. The evaluation
levels may differ from the fitted levels when an estimated quantile function
is available at other probabilities. Let
\(q_{T,h,k}^*\) denote the reported conditional \(p_k\)-quantile at origin
\(T\) and horizon \(h\), after monotone rearrangement when it is applied.
Here \(K\) counts evaluation levels, while \(m\) indexes posterior or
variational draws. With check loss
\(\rho_p(e)=e\{p-\ind(e<0)\}\), the CRPS of a predictive distribution \(F\)
with finite first moment has the integrated representation
\citep{GneitingRaftery2007}:
\begin{equation}
\operatorname{CRPS}(F,y)
=2\int_0^1\rho_p\!\left\{y-F^{-1}(p)\right\}\,dp.
\label{eq:supp-crps-integrated-check}
\end{equation}
For quantiles evaluated on \(\mathcal P_K\), the reported finite-grid
approximation over \([p_1,p_K]\) is
\begin{equation}
\aCRPS_{\mathcal P_K}(y;\vect q_{T,h}^*)
=2\sum_{k=1}^K\omega_k
\rho_{p_k}\!\left(y-q_{T,h,k}^*\right),
\label{eq:supp-acrps}
\end{equation}
where
\begin{equation}
\begin{aligned}
\omega_1&=\frac{p_2-p_1}{2},
&\omega_k&=\frac{p_{k+1}-p_{k-1}}{2},\quad 2\le k\le K-1,
&\omega_K&=\frac{p_K-p_{K-1}}{2}.
\end{aligned}
\label{eq:supp-acrps-weights}
\end{equation}
Thus \(\sum_k\omega_k=p_K-p_1\). The joint simulation and GloFAS analyses
use seven common evaluation levels with endpoints 0.05 and 0.95, so their
weights sum to 0.90. For fitted quantile regressions, the evaluation levels
coincide with the fitted grid. Their interior levels differ between studies;
comparisons of \(\aCRPS\) are therefore made within each study.

The joint simulation also evaluates the quantile grid with respect to the
known conditional response distribution. Let \(\mathcal G\) denote its
origin--horizon evaluation set, let \(F_{0,r}\) be the true conditional
distribution for evaluation pair \(r\), and let
\(\vect q_r^{*,(b)}=(q_{r,1}^{*,(b)},\ldots,q_{r,K}^{*,(b)})^\top\) be the
monotonically rearranged quantile vector from posterior draw \(b\). The
data-generating-process (DGP)-integrated finite-grid quantile score is
\begin{equation}
S_{\mathrm{DGP}}^{(b)}
=\frac{2}{|\mathcal G|}
\sum_{r\in\mathcal G}\sum_{k=1}^K\omega_k
\E_{Y_r\sim F_{0,r}}
\left[
\rho_{p_k}\!\left\{Y_r-q_{r,k}^{*,(b)}\right\}
\right].
\label{eq:supp-dgp-integrated-score}
\end{equation}
For the seven-level grid, the weights are
\[
(0.025,\ 0.10,\ 0.20,\ 0.25,\ 0.20,\ 0.10,\ 0.025),
\]
which sum to 0.90 and retain the scale of integrated check loss over
\([0.05,0.95]\). The expectation in
Eq.~\eqref{eq:supp-dgp-integrated-score} is evaluated analytically or by
deterministic numerical integration under the known simulation mechanism.
It averages over the conditional response distribution at the fixed held-out
design points and conditions on the simulated fitting series and selected
specification. Its posterior interval therefore excludes repeated-simulation
and model-selection uncertainty.

For a joint fit, each score draw uses the quantile vector from one joint
posterior draw and therefore retains its cross-quantile dependence. For
independently estimated levels, the marginal draws are combined with a
pre-specified seeded product coupling that allocates each chain evenly across
the paired draws. Joint-minus-independent contrasts use a separate
pre-specified within-chain permutation to pair the resulting scalar score
draws. The resulting posterior contrast intervals are conditional on this
pairing, which defines the coupling between separately fitted models.
Posterior means and equal-tailed 95\% intervals of
\(S_{\mathrm{DGP}}^{(b)}\) form the primary comparison. The score evaluated
at the posterior-mean quantile curve is reported separately as a sensitivity
summary. GloFAS applies monotone rearrangement to its variational
posterior-mean quantile estimates and uses realized outcomes in
Eq.~\eqref{eq:supp-acrps}.

Response-level posterior predictive draws determine an empirical predictive
quantile function that can be evaluated on any prespecified grid; integrating
its check loss over \((0,1)\) gives the empirical predictive CRPS. Posterior
draws of a fitted \(p\)-level quantile provide credible summaries for that
conditional quantile. The single-level interval analysis below computes its
criteria for each retained quantile draw.

\clearpage
\section{Supplementary Simulation and Application Tables}
\label{sec:supp_validation_tables}

This section provides numerical values corresponding to the single-quantile
MCMC figures, analogous variational summaries, and additional comparisons of
joint and independently estimated quantiles.

\subsection{Single-Quantile MCMC Posterior Metric Tables}

The following tables give posterior means and equal-tailed 95\% credible
intervals for
the criteria displayed in the MCMC figures. Q--DESN and exQ--DESN entries use
fitting-sample preprocessing, and ridge rows are omitted. Within each model and
simulation setting, the specification minimizing each criterion was selected
separately. Consequently, criteria within a displayed model row can correspond
to different fitted specifications. The intervals condition on the
simulated data, evaluation grid, and reservoir realization.

\clearpage
\begin{table}[!ht]
\centering
\scriptsize
\setstretch{1}
\setlength{\tabcolsep}{4pt}
\begin{tabular}{@{}clccc@{}}
\toprule
Target & Model & Fit RMSE & Forecast MAE & Forecast check loss \\
\midrule
$p=0.05$ & DQLM & \shortstack{2.602\\{\scriptsize [1.977, 3.252]}} & \shortstack{3.935\\{\scriptsize [3.830, 4.040]}} & \shortstack{1.113\\{\scriptsize [1.096, 1.131]}} \\
 & exDQLM & \shortstack{\textbf{2.153}\\{\scriptsize [1.368, 3.003]}} & \shortstack{\textbf{2.611}\\{\scriptsize [2.489, 2.733]}} & \shortstack{1.123\\{\scriptsize [1.081, 1.167]}} \\
 & Q--DESN AL--RHS & \shortstack{2.765\\{\scriptsize [2.070, 3.468]}} & \shortstack{7.173\textsuperscript{\(\dagger\)}\\{\scriptsize [2.718, 12.23]}} & \shortstack{1.235\textsuperscript{\(\dagger\)}\\{\scriptsize [1.082, 1.454]}} \\
 & Q--DESN exAL--RHS & \shortstack{2.302\\{\scriptsize [1.529, 3.092]}} & \shortstack{3.654\\{\scriptsize [1.246, 7.917]}} & \shortstack{\textbf{1.112}\\{\scriptsize [1.052, 1.253]}} \\
\addlinespace[2pt]
$p=0.25$ & DQLM & \shortstack{2.213\\{\scriptsize [1.531, 2.960]}} & \shortstack{3.156\\{\scriptsize [3.019, 3.294]}} & \shortstack{3.437\\{\scriptsize [3.370, 3.505]}} \\
 & exDQLM & \shortstack{1.832\\{\scriptsize [1.171, 2.570]}} & \shortstack{\textbf{2.441}\\{\scriptsize [2.330, 2.557]}} & \shortstack{\textbf{3.378}\\{\scriptsize [3.308, 3.449]}} \\
 & Q--DESN AL--RHS & \shortstack{2.256\\{\scriptsize [1.450, 3.139]}} & \shortstack{3.155\\{\scriptsize [1.419, 6.687]}} & \shortstack{3.400\\{\scriptsize [3.243, 3.841]}} \\
 & Q--DESN exAL--RHS & \shortstack{\textbf{1.664}\\{\scriptsize [0.9861, 2.385]}} & \shortstack{3.029\\{\scriptsize [0.9567, 6.550]}} & \shortstack{3.393\\{\scriptsize [3.229, 3.814]}} \\
\addlinespace[2pt]
$p=0.50$ & DQLM & \shortstack{1.866\\{\scriptsize [1.255, 2.573]}} & \shortstack{2.718\\{\scriptsize [2.592, 2.845]}} & \shortstack{4.227\\{\scriptsize [4.140, 4.315]}} \\
 & exDQLM & \shortstack{1.863\\{\scriptsize [1.260, 2.543]}} & \shortstack{2.722\\{\scriptsize [2.598, 2.850]}} & \shortstack{4.228\\{\scriptsize [4.141, 4.315]}} \\
 & Q--DESN AL--RHS & \shortstack{1.470\\{\scriptsize [0.8016, 2.207]}} & \shortstack{2.753\\{\scriptsize [1.067, 6.308]}} & \shortstack{4.191\\{\scriptsize [3.960, 4.875]}} \\
 & Q--DESN exAL--RHS & \shortstack{\textbf{1.415}\\{\scriptsize [0.7952, 2.108]}} & \shortstack{\textbf{2.694}\\{\scriptsize [1.027, 6.117]}} & \shortstack{\textbf{4.184}\\{\scriptsize [3.962, 4.821]}} \\
\bottomrule
\end{tabular}
\caption{Posterior metric intervals for the Gaussian single-quantile simulation family. Entries are posterior means with equal-tailed 95\% credible intervals; lower is better, and boldface marks the lowest unrounded posterior mean by target and criterion. A dagger marks a diagnostic caution recorded for the source analysis.}
\label{tab:simulation-500obs-mcmc-intervals-normal}
\end{table}

\clearpage
\begin{table}[!ht]
\centering
\scriptsize
\setstretch{1}
\setlength{\tabcolsep}{4pt}
\begin{tabular}{@{}clccc@{}}
\toprule
Target & Model & Fit RMSE & Forecast MAE & Forecast check loss \\
\midrule
$p=0.05$ & DQLM & \shortstack{\textbf{4.967}\\{\scriptsize [3.787, 6.175]}} & \shortstack{9.817\\{\scriptsize [9.638, 10.00]}} & \shortstack{1.974\\{\scriptsize [1.945, 2.003]}} \\
 & exDQLM & \shortstack{7.250\\{\scriptsize [5.823, 8.610]}} & \shortstack{4.702\textsuperscript{\(\dagger\)}\\{\scriptsize [4.517, 4.891]}} & \shortstack{1.935\\{\scriptsize [1.878, 1.993]}} \\
 & Q--DESN AL--RHS & \shortstack{5.324\\{\scriptsize [3.839, 6.647]}} & \shortstack{7.652\\{\scriptsize [4.731, 10.76]}} & \shortstack{2.252\\{\scriptsize [1.903, 2.966]}} \\
 & Q--DESN exAL--RHS & \shortstack{6.593\\{\scriptsize [5.160, 7.967]}} & \shortstack{\textbf{3.980}\\{\scriptsize [2.220, 9.085]}} & \shortstack{\textbf{1.912}\\{\scriptsize [1.840, 2.172]}} \\
\addlinespace[2pt]
$p=0.25$ & DQLM & \shortstack{2.530\\{\scriptsize [1.848, 3.247]}} & \shortstack{4.225\\{\scriptsize [4.059, 4.397]}} & \shortstack{4.647\\{\scriptsize [4.571, 4.723]}} \\
 & exDQLM & \shortstack{2.067\\{\scriptsize [1.387, 2.860]}} & \shortstack{3.015\\{\scriptsize [2.883, 3.152]}} & \shortstack{4.521\\{\scriptsize [4.438, 4.604]}} \\
 & Q--DESN AL--RHS & \shortstack{2.275\\{\scriptsize [1.454, 3.106]}} & \shortstack{2.971\\{\scriptsize [1.110, 6.860]}} & \shortstack{4.518\\{\scriptsize [4.355, 5.046]}} \\
 & Q--DESN exAL--RHS & \shortstack{\textbf{1.758}\\{\scriptsize [0.9837, 2.568]}} & \shortstack{\textbf{2.213}\\{\scriptsize [0.7827, 5.293]}} & \shortstack{\textbf{4.445}\\{\scriptsize [4.337, 4.773]}} \\
\addlinespace[2pt]
$p=0.50$ & DQLM & \shortstack{1.784\\{\scriptsize [1.252, 2.374]}} & \shortstack{3.104\\{\scriptsize [2.964, 3.248]}} & \shortstack{5.335\\{\scriptsize [5.239, 5.430]}} \\
 & exDQLM & \shortstack{1.781\\{\scriptsize [1.264, 2.373]}} & \shortstack{3.111\\{\scriptsize [2.970, 3.252]}} & \shortstack{5.337\\{\scriptsize [5.241, 5.433]}} \\
 & Q--DESN AL--RHS & \shortstack{\textbf{1.324}\\{\scriptsize [0.7016, 2.000]}} & \shortstack{\textbf{1.819}\\{\scriptsize [0.6747, 4.363]}} & \shortstack{\textbf{5.104}\\{\scriptsize [4.995, 5.434]}} \\
 & Q--DESN exAL--RHS & \shortstack{1.337\\{\scriptsize [0.7227, 2.002]}} & \shortstack{1.886\\{\scriptsize [0.7423, 4.371]}} & \shortstack{5.112\\{\scriptsize [4.996, 5.447]}} \\
\bottomrule
\end{tabular}
\caption{Posterior metric intervals for the Laplace single-quantile simulation family. Entries are posterior means with equal-tailed 95\% credible intervals; lower is better, and boldface marks the lowest unrounded posterior mean by target and criterion. A dagger marks a diagnostic caution recorded for the source analysis.}
\label{tab:simulation-500obs-mcmc-intervals-laplace}
\end{table}

\clearpage
\begin{table}[!ht]
\centering
\scriptsize
\setstretch{1}
\setlength{\tabcolsep}{4pt}
\begin{tabular}{@{}clccc@{}}
\toprule
Target & Model & Fit RMSE & Forecast MAE & Forecast check loss \\
\midrule
$p=0.05$ & DQLM & \shortstack{\textbf{2.997}\\{\scriptsize [2.132, 3.897]}} & \shortstack{\textbf{3.478}\\{\scriptsize [3.347, 3.606]}} & \shortstack{\textbf{1.521}\\{\scriptsize [1.495, 1.549]}} \\
 & exDQLM & \shortstack{3.732\\{\scriptsize [2.479, 4.977]}} & \shortstack{4.849\\{\scriptsize [4.655, 5.039]}} & \shortstack{1.628\\{\scriptsize [1.564, 1.693]}} \\
 & Q--DESN AL--RHS & \shortstack{3.186\\{\scriptsize [2.130, 4.339]}} & \shortstack{3.810\\{\scriptsize [1.997, 7.136]}} & \shortstack{1.594\\{\scriptsize [1.509, 1.822]}} \\
 & Q--DESN exAL--RHS & \shortstack{3.558\\{\scriptsize [2.291, 4.817]}} & \shortstack{3.770\\{\scriptsize [1.306, 8.629]}} & \shortstack{1.579\\{\scriptsize [1.488, 1.868]}} \\
\addlinespace[2pt]
$p=0.25$ & DQLM & \shortstack{2.252\\{\scriptsize [1.511, 3.146]}} & \shortstack{3.866\\{\scriptsize [3.699, 4.038]}} & \shortstack{4.681\\{\scriptsize [4.592, 4.769]}} \\
 & exDQLM & \shortstack{\textbf{1.797}\\{\scriptsize [1.092, 2.649]}} & \shortstack{3.295\textsuperscript{\(\dagger\)}\\{\scriptsize [3.119, 3.473]}} & \shortstack{4.656\\{\scriptsize [4.564, 4.748]}} \\
 & Q--DESN AL--RHS & \shortstack{1.892\\{\scriptsize [1.014, 2.892]}} & \shortstack{\textbf{3.066}\\{\scriptsize [1.039, 7.451]}} & \shortstack{\textbf{4.631}\\{\scriptsize [4.467, 5.251]}} \\
 & Q--DESN exAL--RHS & \shortstack{1.923\\{\scriptsize [1.066, 2.871]}} & \shortstack{3.422\\{\scriptsize [1.208, 8.390]}} & \shortstack{4.642\textsuperscript{\(\dagger\)}\\{\scriptsize [4.467, 5.244]}} \\
\addlinespace[2pt]
$p=0.50$ & DQLM & \shortstack{1.831\\{\scriptsize [1.269, 2.537]}} & \shortstack{3.773\\{\scriptsize [3.606, 3.941]}} & \shortstack{5.826\\{\scriptsize [5.712, 5.940]}} \\
 & exDQLM & \shortstack{1.831\\{\scriptsize [1.260, 2.528]}} & \shortstack{3.777\\{\scriptsize [3.611, 3.949]}} & \shortstack{5.826\\{\scriptsize [5.710, 5.941]}} \\
 & Q--DESN AL--RHS & \shortstack{\textbf{1.274}\\{\scriptsize [0.6121, 2.094]}} & \shortstack{\textbf{2.293}\\{\scriptsize [0.7000, 5.832]}} & \shortstack{\textbf{5.607}\\{\scriptsize [5.421, 6.178]}} \\
 & Q--DESN exAL--RHS & \shortstack{1.326\\{\scriptsize [0.6313, 2.218]}} & \shortstack{2.737\\{\scriptsize [0.8724, 6.996]}} & \shortstack{5.677\\{\scriptsize [5.430, 6.473]}} \\
\bottomrule
\end{tabular}
\caption{Posterior metric intervals for the Gaussian mixture single-quantile simulation family. Entries are posterior means with equal-tailed 95\% credible intervals; lower is better, and boldface marks the lowest unrounded posterior mean by target and criterion. A dagger marks a diagnostic caution recorded for the source analysis.}
\label{tab:simulation-500obs-mcmc-intervals-gausmix}
\end{table}

\clearpage

\clearpage
\subsection{Single-Quantile VB Posterior Metric Intervals}

The following figures and tables report approximate posterior means and
equal-tailed 95\% variational intervals for the same draw-wise fitting-sample
root-mean-square error (RMSE), forecast mean absolute error (MAE), and forecast
check-loss criteria.

\begin{figure}[!htbp]
\centering
\includegraphics[width=0.98\textwidth]{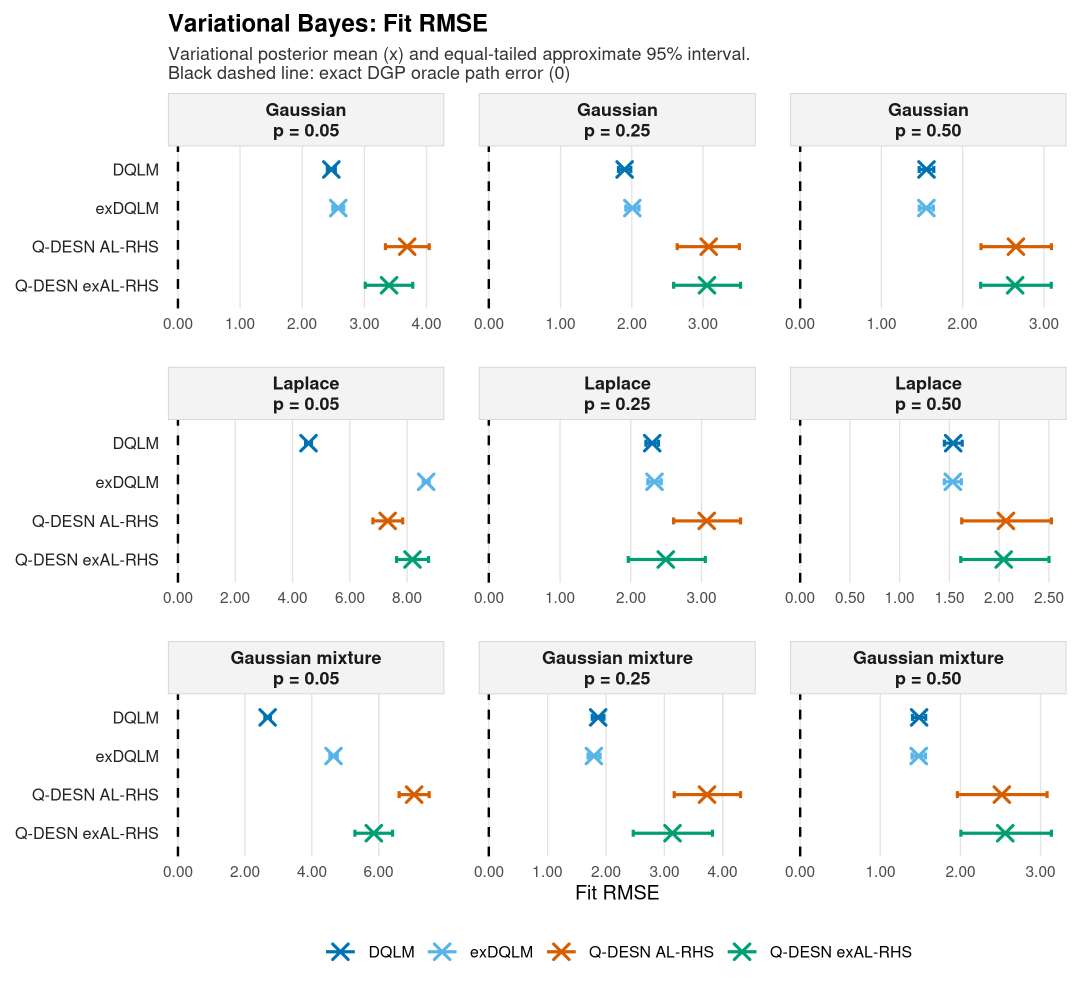}
\caption{Variational Bayes posterior uncertainty for fit RMSE in the single-quantile simulation study. Horizontal segments show equal-tailed approximate 95\% variational posterior intervals; crosses mark posterior means. Each panel uses its own horizontal scale, and lower values are better. The black dashed line marks the exact DGP oracle value of zero for this conditional-quantile path-error criterion. Intervals condition on the simulated data, evaluation design, and case-specific model specification.}
\label{fig:simulation-500obs-vb-fit-rmse-intervals}
\end{figure}

\begin{figure}[!htbp]
\centering
\includegraphics[width=0.98\textwidth]{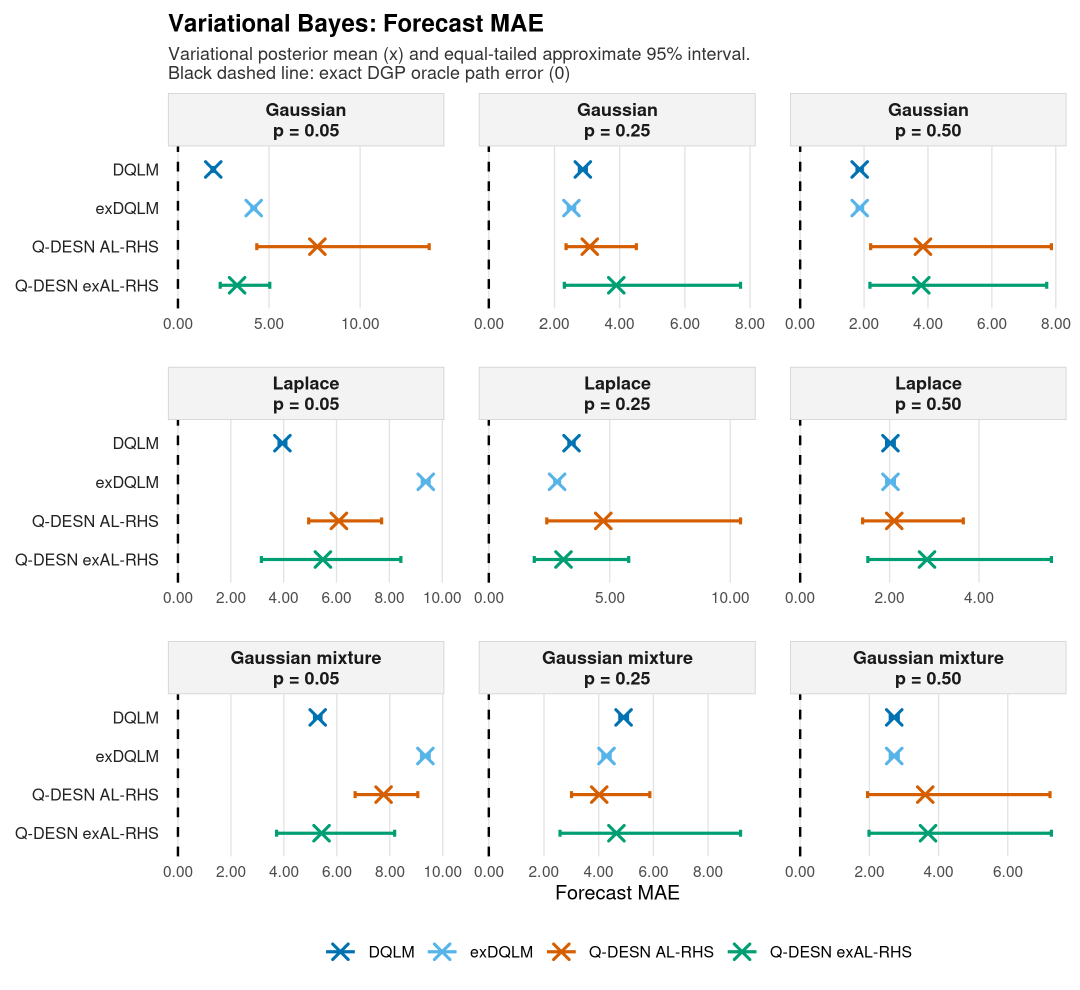}
\caption{Variational Bayes posterior uncertainty for forecast MAE in the single-quantile simulation study. Horizontal segments show equal-tailed approximate 95\% variational posterior intervals; crosses mark posterior means. Each panel uses its own horizontal scale, and lower values are better. The black dashed line marks the exact DGP oracle value of zero for this conditional-quantile path-error criterion. Intervals condition on the simulated data, evaluation design, and case-specific model specification.}
\label{fig:simulation-500obs-vb-forecast-mae-intervals}
\end{figure}

\begin{figure}[!htbp]
\centering
\includegraphics[width=0.98\textwidth]{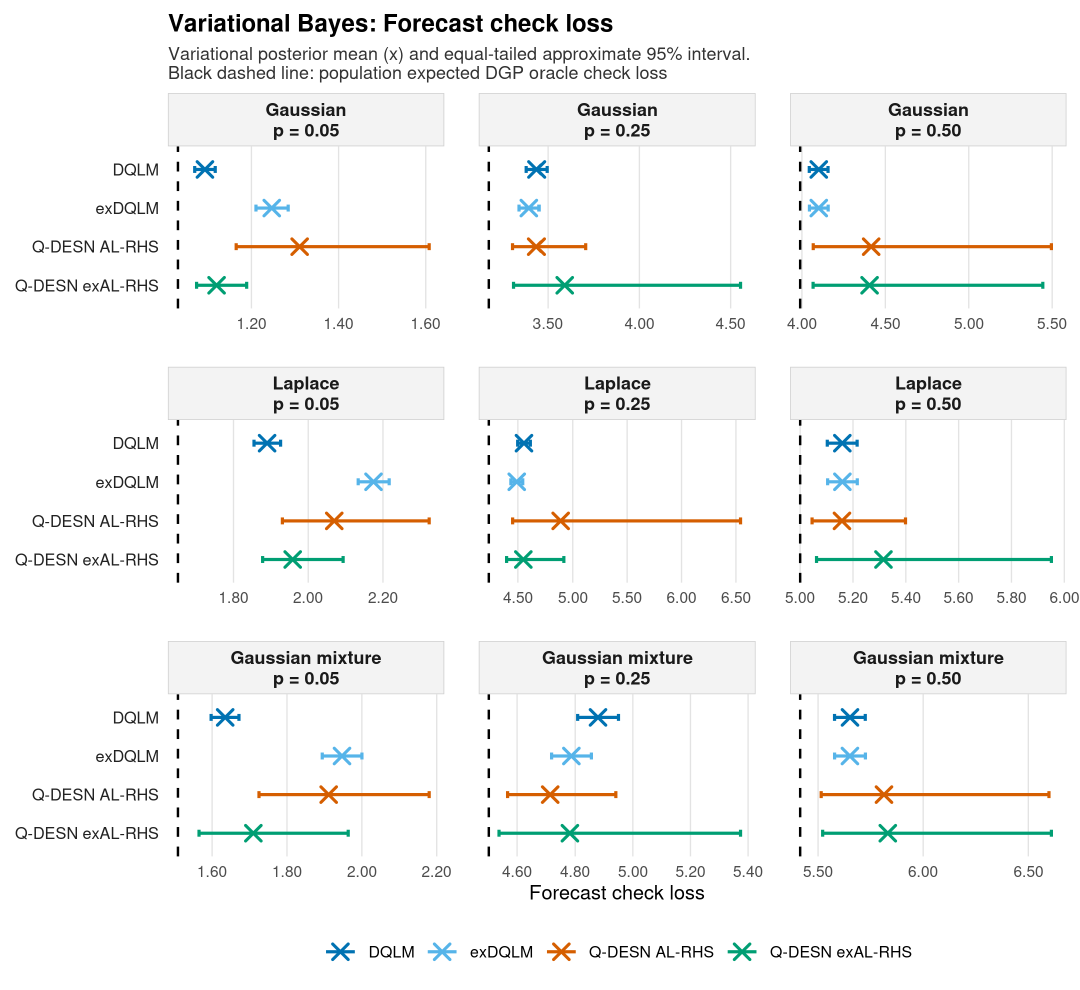}
\caption{Variational Bayes posterior uncertainty for forecast check loss in the single-quantile simulation study. Horizontal segments show equal-tailed approximate 95\% variational posterior intervals; crosses mark posterior means. Each panel uses its own horizontal scale, and lower values are better. The black dashed line marks the population expected check loss at the true conditional quantile. Because the intervals condition on one simulated series, finite-sample check-loss summaries may cross this population reference. Intervals condition on the simulated data, evaluation design, and case-specific model specification.}
\label{fig:simulation-500obs-vb-forecast-check-loss-intervals}
\end{figure}

\clearpage
\begin{table}[!ht]
\centering
\scriptsize
\setstretch{1}
\setlength{\tabcolsep}{4pt}
\begin{tabular}{@{}clccc@{}}
\toprule
Target & Model & Fit RMSE & Forecast MAE & Forecast check loss \\
\midrule
$p=0.05$ & DQLM & \shortstack{\textbf{2.468}\\{\scriptsize [2.404, 2.533]}} & \shortstack{\textbf{1.938}\\{\scriptsize [1.864, 2.013]}} & \shortstack{\textbf{1.093}\\{\scriptsize [1.070, 1.117]}} \\
 & exDQLM & \shortstack{2.580\\{\scriptsize [2.491, 2.668]}} & \shortstack{4.159\\{\scriptsize [4.051, 4.271]}} & \shortstack{1.247\\{\scriptsize [1.210, 1.284]}} \\
 & Q--DESN AL--RHS & \shortstack{3.689\\{\scriptsize [3.339, 4.043]}} & \shortstack{7.655\\{\scriptsize [4.329, 13.79]}} & \shortstack{1.311\\{\scriptsize [1.165, 1.608]}} \\
 & Q--DESN exAL--RHS & \shortstack{3.397\\{\scriptsize [3.014, 3.778]}} & \shortstack{3.247\\{\scriptsize [2.320, 5.041]}} & \shortstack{1.120\\{\scriptsize [1.074, 1.189]}} \\
\addlinespace[2pt]
$p=0.25$ & DQLM & \shortstack{\textbf{1.899}\\{\scriptsize [1.813, 1.986]}} & \shortstack{2.869\\{\scriptsize [2.764, 2.977]}} & \shortstack{3.439\\{\scriptsize [3.383, 3.495]}} \\
 & exDQLM & \shortstack{2.008\\{\scriptsize [1.916, 2.102]}} & \shortstack{\textbf{2.520}\\{\scriptsize [2.423, 2.619]}} & \shortstack{\textbf{3.397}\\{\scriptsize [3.343, 3.452]}} \\
 & Q--DESN AL--RHS & \shortstack{3.080\\{\scriptsize [2.638, 3.510]}} & \shortstack{3.083\\{\scriptsize [2.355, 4.511]}} & \shortstack{3.437\\{\scriptsize [3.308, 3.706]}} \\
 & Q--DESN exAL--RHS & \shortstack{3.055\\{\scriptsize [2.587, 3.527]}} & \shortstack{3.895\\{\scriptsize [2.303, 7.708]}} & \shortstack{3.591\\{\scriptsize [3.313, 4.554]}} \\
\addlinespace[2pt]
$p=0.50$ & DQLM & \shortstack{1.555\\{\scriptsize [1.465, 1.645]}} & \shortstack{\textbf{1.863}\\{\scriptsize [1.780, 1.944]}} & \shortstack{\textbf{4.101}\\{\scriptsize [4.044, 4.157]}} \\
 & exDQLM & \shortstack{\textbf{1.554}\\{\scriptsize [1.467, 1.639]}} & \shortstack{1.863\\{\scriptsize [1.781, 1.948]}} & \shortstack{4.101\\{\scriptsize [4.045, 4.157]}} \\
 & Q--DESN AL--RHS & \shortstack{2.656\\{\scriptsize [2.226, 3.093]}} & \shortstack{3.843\\{\scriptsize [2.206, 7.863]}} & \shortstack{4.415\\{\scriptsize [4.068, 5.495]}} \\
 & Q--DESN exAL--RHS & \shortstack{2.646\\{\scriptsize [2.223, 3.091]}} & \shortstack{3.791\\{\scriptsize [2.184, 7.716]}} & \shortstack{4.406\\{\scriptsize [4.067, 5.444]}} \\
\bottomrule
\end{tabular}
\caption{Approximate posterior metric intervals for the Gaussian single-quantile simulation family. Entries are variational posterior means with equal-tailed approximate 95\% intervals; lower is better, and boldface marks the lowest unrounded posterior mean by target and criterion.}
\label{tab:simulation-500obs-vb-intervals-normal}
\end{table}

\clearpage
\begin{table}[!ht]
\centering
\scriptsize
\setstretch{1}
\setlength{\tabcolsep}{4pt}
\begin{tabular}{@{}clccc@{}}
\toprule
Target & Model & Fit RMSE & Forecast MAE & Forecast check loss \\
\midrule
$p=0.05$ & DQLM & \shortstack{\textbf{4.559}\\{\scriptsize [4.458, 4.661]}} & \shortstack{\textbf{3.953}\\{\scriptsize [3.824, 4.085]}} & \shortstack{\textbf{1.890}\\{\scriptsize [1.855, 1.926]}} \\
 & exDQLM & \shortstack{8.667\\{\scriptsize [8.564, 8.772]}} & \shortstack{9.378\\{\scriptsize [9.246, 9.506]}} & \shortstack{2.175\\{\scriptsize [2.134, 2.217]}} \\
 & Q--DESN AL--RHS & \shortstack{7.321\\{\scriptsize [6.806, 7.847]}} & \shortstack{6.091\\{\scriptsize [4.946, 7.706]}} & \shortstack{2.070\\{\scriptsize [1.931, 2.324]}} \\
 & Q--DESN exAL--RHS & \shortstack{8.189\\{\scriptsize [7.634, 8.750]}} & \shortstack{5.485\\{\scriptsize [3.158, 8.435]}} & \shortstack{1.958\\{\scriptsize [1.878, 2.093]}} \\
\addlinespace[2pt]
$p=0.25$ & DQLM & \shortstack{\textbf{2.303}\\{\scriptsize [2.215, 2.392]}} & \shortstack{3.413\\{\scriptsize [3.303, 3.522]}} & \shortstack{4.556\\{\scriptsize [4.498, 4.615]}} \\
 & exDQLM & \shortstack{2.335\\{\scriptsize [2.241, 2.431]}} & \shortstack{\textbf{2.814}\\{\scriptsize [2.716, 2.916]}} & \shortstack{\textbf{4.490}\\{\scriptsize [4.436, 4.545]}} \\
 & Q--DESN AL--RHS & \shortstack{3.074\\{\scriptsize [2.606, 3.552]}} & \shortstack{4.731\\{\scriptsize [2.385, 10.43]}} & \shortstack{4.889\\{\scriptsize [4.452, 6.542]}} \\
 & Q--DESN exAL--RHS & \shortstack{2.495\\{\scriptsize [1.965, 3.054]}} & \shortstack{3.079\\{\scriptsize [1.887, 5.786]}} & \shortstack{4.550\\{\scriptsize [4.396, 4.918]}} \\
\addlinespace[2pt]
$p=0.50$ & DQLM & \shortstack{1.538\\{\scriptsize [1.450, 1.629]}} & \shortstack{\textbf{2.019}\\{\scriptsize [1.932, 2.105]}} & \shortstack{5.160\\{\scriptsize [5.103, 5.216]}} \\
 & exDQLM & \shortstack{\textbf{1.534}\\{\scriptsize [1.449, 1.620]}} & \shortstack{2.019\\{\scriptsize [1.934, 2.107]}} & \shortstack{5.160\\{\scriptsize [5.104, 5.216]}} \\
 & Q--DESN AL--RHS & \shortstack{2.068\\{\scriptsize [1.624, 2.525]}} & \shortstack{2.104\\{\scriptsize [1.397, 3.648]}} & \shortstack{\textbf{5.158}\\{\scriptsize [5.045, 5.399]}} \\
 & Q--DESN exAL--RHS & \shortstack{2.046\\{\scriptsize [1.614, 2.501]}} & \shortstack{2.836\\{\scriptsize [1.514, 5.618]}} & \shortstack{5.315\\{\scriptsize [5.062, 5.951]}} \\
\bottomrule
\end{tabular}
\caption{Approximate posterior metric intervals for the Laplace single-quantile simulation family. Entries are variational posterior means with equal-tailed approximate 95\% intervals; lower is better, and boldface marks the lowest unrounded posterior mean by target and criterion.}
\label{tab:simulation-500obs-vb-intervals-laplace}
\end{table}

\clearpage
\begin{table}[!ht]
\centering
\scriptsize
\setstretch{1}
\setlength{\tabcolsep}{4pt}
\begin{tabular}{@{}clccc@{}}
\toprule
Target & Model & Fit RMSE & Forecast MAE & Forecast check loss \\
\midrule
$p=0.05$ & DQLM & \shortstack{\textbf{2.680}\\{\scriptsize [2.597, 2.764]}} & \shortstack{\textbf{5.278}\\{\scriptsize [5.158, 5.402]}} & \shortstack{\textbf{1.635}\\{\scriptsize [1.597, 1.672]}} \\
 & exDQLM & \shortstack{4.648\\{\scriptsize [4.541, 4.758]}} & \shortstack{9.345\\{\scriptsize [9.198, 9.489]}} & \shortstack{1.947\\{\scriptsize [1.894, 2.000]}} \\
 & Q--DESN AL--RHS & \shortstack{7.058\\{\scriptsize [6.608, 7.507]}} & \shortstack{7.766\\{\scriptsize [6.691, 9.058]}} & \shortstack{1.911\\{\scriptsize [1.725, 2.180]}} \\
 & Q--DESN exAL--RHS & \shortstack{5.853\\{\scriptsize [5.288, 6.412]}} & \shortstack{5.424\\{\scriptsize [3.725, 8.184]}} & \shortstack{1.710\\{\scriptsize [1.565, 1.964]}} \\
\addlinespace[2pt]
$p=0.25$ & DQLM & \shortstack{1.866\\{\scriptsize [1.765, 1.963]}} & \shortstack{4.909\\{\scriptsize [4.771, 5.054]}} & \shortstack{4.879\\{\scriptsize [4.808, 4.950]}} \\
 & exDQLM & \shortstack{\textbf{1.793}\\{\scriptsize [1.693, 1.898]}} & \shortstack{4.286\\{\scriptsize [4.153, 4.419]}} & \shortstack{4.787\\{\scriptsize [4.718, 4.856]}} \\
 & Q--DESN AL--RHS & \shortstack{3.732\\{\scriptsize [3.169, 4.306]}} & \shortstack{\textbf{4.019}\\{\scriptsize [3.006, 5.869]}} & \shortstack{\textbf{4.713}\\{\scriptsize [4.567, 4.941]}} \\
 & Q--DESN exAL--RHS & \shortstack{3.140\\{\scriptsize [2.467, 3.824]}} & \shortstack{4.648\\{\scriptsize [2.586, 9.186]}} & \shortstack{4.782\\{\scriptsize [4.537, 5.374]}} \\
\addlinespace[2pt]
$p=0.50$ & DQLM & \shortstack{1.485\\{\scriptsize [1.402, 1.567]}} & \shortstack{2.719\\{\scriptsize [2.604, 2.835]}} & \shortstack{5.652\\{\scriptsize [5.578, 5.725]}} \\
 & exDQLM & \shortstack{\textbf{1.480}\\{\scriptsize [1.396, 1.564]}} & \shortstack{\textbf{2.718}\\{\scriptsize [2.604, 2.835]}} & \shortstack{\textbf{5.652}\\{\scriptsize [5.579, 5.725]}} \\
 & Q--DESN AL--RHS & \shortstack{2.517\\{\scriptsize [1.962, 3.081]}} & \shortstack{3.615\\{\scriptsize [1.947, 7.221]}} & \shortstack{5.814\\{\scriptsize [5.515, 6.598]}} \\
 & Q--DESN exAL--RHS & \shortstack{2.560\\{\scriptsize [2.006, 3.135]}} & \shortstack{3.695\\{\scriptsize [1.990, 7.261]}} & \shortstack{5.832\\{\scriptsize [5.522, 6.610]}} \\
\bottomrule
\end{tabular}
\caption{Approximate posterior metric intervals for the Gaussian mixture single-quantile simulation family. Entries are variational posterior means with equal-tailed approximate 95\% intervals; lower is better, and boldface marks the lowest unrounded posterior mean by target and criterion.}
\label{tab:simulation-500obs-vb-intervals-gausmix}
\end{table}

\clearpage

\clearpage
\subsection{Single-Quantile Computation and Diagnostics}
\label{subsec:supp-independent-exal-mcmc-details}

The independent Q--DESN exAL--RHS MCMC analysis uses the mixture representation
for \(v\) described earlier in the supplement and an analysis-specific
scale-collapsed transition. Conditional on the remaining unknowns, the scale is
integrated from the conditional density used to update the asymmetry parameter
on a logit transformation of its bounded support, and the scale is then drawn
from its full conditional distribution. This transition is used for the
independent simulation analysis; Algorithm~\ref{alg:qdesn_mcmc} gives the generic joint or
conditionally split update. The resulting draws determine the fitted and
forecast quantile paths used in the reported criteria.

The exDQLM baseline uses version 1.1.1 of the CRAN package \pkg{exdqlm}
\citep{exdqlmRPackage}. For unrestricted exAL
fits, its MCMC sampler updates the asymmetry parameter from its conditional
distribution with the scale analytically integrated out, then draws the scale
from its full conditional distribution. Its variational calculation uses the
structured factorization \(q(\gamma)q(\sigma\mid\gamma)\). These model-specific
computations produce the
quantile-path draws evaluated by the common criteria above.
For retained pairs \((\sigma_b,\gamma_b)\),
\(b=1,\ldots,B_{\mathrm{post}}\), let
\[
m_{\varepsilon,b}=\sigma_b\{A(\gamma_b)+\lambda(\gamma_b)\sqrt{2/\pi}\},\qquad
V_{\varepsilon,b}=\sigma_b^2\{B(\gamma_b)+A(\gamma_b)^2
  +\lambda(\gamma_b)^2(1-2/\pi)\}.
\]
The rolling-origin filter uses
\(\bar m_\varepsilon=B_{\mathrm{post}}^{-1}\sum_b m_{\varepsilon,b}\) and
\(\bar V_\varepsilon=B_{\mathrm{post}}^{-1}
\sum_b(V_{\varepsilon,b}+m_{\varepsilon,b}^2)-\bar m_\varepsilon^2\).
The parameter posterior from the initial fit is retained, while observations
available at successive origins update the state filter before forecasts are
propagated over the requested horizons.

The MCMC metric distributions generally pool 12,000 retained evaluations from
three independently initialized chains. The Q--DESN AL--RHS Gaussian
\(p=0.05\) forecast MAE and check-loss summaries use 3,000 retained evaluations
from three chains. Diagnostics assess between-chain agreement for each
draw-wise criterion.
Across 54 fitted configurations and three criteria, 158 of the 162 diagnostic
comparisons satisfy the specified criteria and four receive warnings. Five of
the 108 displayed MCMC summaries are marked by daggers because their
corresponding fits warrant additional caution. These include the exDQLM
Gaussian-mixture \(p=0.25\) forecast MAE and Laplace \(p=0.05\) forecast MAE
summaries.

\clearpage
\subsection{Single-Quantile Five-Chain Point-Path Sensitivity}
\label{subsec:supp-independent-five-chain-sensitivity}

The primary single-quantile MCMC tables report draw-wise posterior summaries of
the evaluation criteria. The analysis below instead evaluates each criterion
on a posterior point estimate of the quantile path and provides a separate
sensitivity analysis across independently initialized chains.
For the Gaussian \(p=0.25\) comparison, a five-chain sensitivity analysis uses
five independently seeded chains for each model specification. For each chain,
the posterior point path is computed first; the reported sensitivity path is the
coordinatewise median of those five paths. All criteria are then recomputed
from that path. This construction summarizes chain-specific point estimates.

One common specification is reported for each working likelihood. Relative to
the corresponding criterion-specific summary, the \(\AL\)--\(\RHS\)
specification reduced fit RMSE by 8.4\% and forecast MAE by 1.6\%, while its
forecast check loss increased by 0.4\%. The \(\exAL\)--\(\RHS\) specification
reduced fit RMSE by 18.3\%, forecast MAE by 8.7\%, and forecast check loss by
0.4\%. Within the Gaussian \(p=0.25\) comparison, these results indicate
limited sensitivity to MCMC initialization for the selected specifications.

\begin{table}[!htbp]
\centering
\small
\setlength{\tabcolsep}{3pt}
\begin{tabular}{@{}llrrr@{}}
\toprule
Model & Comparison summary & Fit RMSE & Forecast MAE & Forecast check loss \\
\midrule
Q--DESN \(\AL\)--\(\RHS\)
  & Reported criterion-specific summary & 2.176 & 2.361 & \textbf{3.297} \\
  & Five-chain common specification & \textbf{1.993} & \textbf{2.323} & 3.309 \\
\addlinespace
Q--DESN \(\exAL\)--\(\RHS\)
  & Reported criterion-specific summary & 1.710 & 2.709 & 3.333 \\
  & Five-chain common specification & \textbf{1.396} & \textbf{2.473} & \textbf{3.319} \\
\bottomrule
\end{tabular}
\caption{Five-chain MCMC sensitivity analysis for the Gaussian simulation
family at \(p=0.25\) with 500 fitting observations. The criterion-specific rows
reproduce the corresponding summaries in the main article's Gaussian MCMC
table and are interpreted separately by criterion. Each five-chain row uses one
pre-specified model specification and reports criteria recomputed from
the coordinatewise median of five chain-specific posterior point paths. This
aggregation summarizes posterior point paths across independent chains.
Forecast criteria use the 1000-observation held-out period, with origins spaced
30 observations apart and horizons up to 30 steps ahead. Boldface
marks the lower value within each model and criterion. This sensitivity analysis
uses a point-path estimator distinct from the draw-wise posterior summaries.}
\label{tab:supp-independent-mcmc-five-chain-sensitivity}
\end{table}

\clearpage
\subsection{Joint Multi-Quantile MCMC Evaluation Details}

The following tables give the MCMC allocation,
joint-minus-independent DGP-integrated score contrasts, quantile-crossing
counts, and forecast quantile-path recovery diagnostics. All 16 paired score
intervals include zero. MCMC diagnostics for the posterior score meet the
pre-specified rank-normalized \(\widehat R\) and effective-sample-size
thresholds in 21 of 32 comparisons; the remaining 11 comparisons are
interpreted cautiously.

For the independent \(\exAL\) analyses, the scale-collapsed update yields maximum
rank-normalized \(\widehat R\) values of approximately 1.020 for \(\gamma\)
and 1.016 for \(\sigma\), with minimum bulk effective sample sizes of
approximately 410 and 472, respectively. The corresponding joint-fit values
are approximately 1.015 and 1.012, with minimum bulk effective sample sizes
of 849 and 959. All 16 \(\exAL\) comparisons use this scale-collapsed update
separately at each quantile level. The weaker remaining diagnostics occur
mainly in intercept and coefficient coordinates. Crossings in
Table~\ref{tab:supp-joint-qdesn-crossings} refer to the posterior-mean
quantile grids. The table also reports draw-level crossing rates, for which
10 of the 32 scenario--model comparisons meet the pre-specified diagnostic
criterion and 22 warrant review.

\begin{table}[!htbp]
\centering
\small
\begin{tabular}{@{}>{\raggedright\arraybackslash}p{0.23\textwidth}>{\raggedright\arraybackslash}p{0.67\textwidth}@{}}
\toprule
Item & Value
\\
\midrule
Evaluation components & Variational approximations provide initial values; MCMC draws determine the reported posterior score distributions and paired contrasts. \\
Simulation settings & Eight data-generating mechanisms with known conditional quantile paths, spanning Gaussian, Laplace, Gaussian-mixture, asymmetric-Laplace, and Student-\(t\) innovations with several dynamic structures. \\
Model comparison & Joint and independently estimated quantile regressions under \(\AL\) (Q--DESN) and \(\exAL\) (exQDESN), all with the regularized horseshoe prior. \\
Quantile grid & 0.05, 0.10, 0.25, 0.50, 0.75, 0.90, and 0.95. \\
Fitting sample & 500 observations after the pre-specified DESN washout. \\
Forecast evaluation & Sequential conditional forecasts on 990 aligned origin--horizon rows per comparison. Coefficients and reservoir weights remain fixed, while newly observed responses enter subsequent lag vectors. \\
MCMC effort & Every comparison uses eight chains. Fourteen \(\AL\) comparisons use 8,000 iterations, 2,000 burn-in iterations, and thinning by 4; two use 24,000, 4,000, and 4. Five \(\exAL\) comparisons use 24,000, 4,000, and 4; eleven use 48,000, 8,000, and 8. \\
Posterior score summary & Each reported score distribution and paired contrast uses 8,000 chain-balanced score draws. Entries are posterior means with equal-tailed 95\% credible intervals. \\
Contrast construction & Joint-minus-independent posterior contrasts use a pre-specified within-chain pairing of scalar score draws and are conditional on that pairing. \\
Quantile ordering & Scores are computed after monotone rearrangement; crossings in the posterior-mean forecast quantile grids are reported before and after rearrangement. \\
Repeated-simulation comparison & The repeated-simulation comparison contains 1600 independent VB fits. \\
\bottomrule
\end{tabular}
\caption{Design of the joint multi-quantile evaluation. Variational methods initialize the chains, while MCMC supplies the reported DGP-integrated finite-grid score summaries.}
\label{tab:joint-qdesn-article-validation-mcmc-balanced-protocol}
\end{table}

\begin{table}[!htbp]
\centering
\small
\begin{tabular}{@{}>{\raggedright\arraybackslash}p{0.39\textwidth}rr@{}}
\toprule
Simulation setting & \(\AL\) & \(\exAL\) \\
\midrule
Asymmetric-Laplace tail & 0.0003 [-0.0035, 0.0049] & -0.0004 [-0.0047, 0.0038] \\
Gaussian-mixture innovations & 0.0000 [-0.0038, 0.0038] & -0.0002 [-0.0055, 0.0050] \\
Laplace innovations & -0.0008 [-0.0039, 0.0027] & -0.0011 [-0.0065, 0.0032] \\
Nonlinear reservoir dynamics & -0.0012 [-0.0059, 0.0033] & -0.0009 [-0.0073, 0.0055] \\
Gaussian innovations & -0.0008 [-0.0051, 0.0035] & -0.0001 [-0.0066, 0.0068] \\
Persistent heavy tails & 0.0009 [-0.0042, 0.0060] & 0.0014 [-0.0047, 0.0077] \\
Regime shift & 0.0058 [-0.0095, 0.0285] & 0.0079 [-0.0090, 0.0329] \\
Student-\(t\) location--scale & -0.0006 [-0.0042, 0.0033] & 0.0006 [-0.0047, 0.0076] \\
\bottomrule
\end{tabular}
\caption{Joint-minus-independent posterior contrasts in the data-generating-process-integrated finite-grid quantile score. Entries are means with equal-tailed 95\% intervals from 8,000 score draws under the pre-specified within-chain pairing; negative values favor joint estimation. The intervals are conditional on this pairing, and every interval includes zero.}
\label{tab:supp-joint-qdesn-dgp-score-contrasts}
\end{table}

\begin{table}[!htbp]
\centering
\small
\begin{tabular}{@{}>{\raggedright\arraybackslash}p{0.43\textwidth}rrr@{}}
\toprule
Model & \shortstack{Before\\rearrangement} & \shortstack{After\\rearrangement} & \shortstack{Mean draw-level\\crossing rate (\%)} \\
\midrule
Joint Q--DESN \(\AL\)--\(\RHS\) & 1 & 0 & 0.35 \\
Independent Q--DESN \(\AL\)--\(\RHS\) & 25 & 0 & 1.42 \\
Joint exQDESN \(\exAL\)--\(\RHS\) & 0 & 0 & 1.50 \\
Independent exQDESN \(\exAL\)--\(\RHS\) & 0 & 0 & 3.37 \\
\bottomrule
\end{tabular}
\caption{Adjacent-level crossings in the posterior-mean forecast quantile grids, summed over the eight simulation settings. Counts before rearrangement describe departures from monotonicity in the fitted quantiles; monotone rearrangement produces ordered grids in all 32 scenario--model comparisons. Within a setting, the draw-level rate is the number of adjacent-level crossings divided by the six adjacent quantile pairs across 8,000 draws and 990 forecast rows; the final column averages this rate over the eight settings for each model.}
\label{tab:supp-joint-qdesn-crossings}
\end{table}

\begin{table}[!htbp]
\centering
\scriptsize
\resizebox{\textwidth}{!}{%
\begin{tabular}{@{}>{\raggedright\arraybackslash}p{0.23\textwidth}rrrr@{}}
\toprule
Simulation setting & \shortstack{Joint Q--DESN\\\(\AL\)--\(\RHS\)} & \shortstack{Independent Q--DESN\\\(\AL\)--\(\RHS\)} & \shortstack{Joint exQDESN\\\(\exAL\)--\(\RHS\)} & \shortstack{Independent exQDESN\\\(\exAL\)--\(\RHS\)} \\
\midrule
Asymmetric-Laplace tail & 0.087 (0.136) & 0.086 (0.141) & 0.094 (0.150) & 0.089 (0.146) \\
Gaussian-mixture innovations & 0.095 (0.126) & 0.098 (0.131) & 0.106 (0.132) & 0.094 (0.119) \\
Laplace innovations & 0.085 (0.117) & 0.094 (0.133) & 0.111 (0.147) & 0.105 (0.144) \\
Nonlinear reservoir dynamics & 0.104 (0.141) & 0.116 (0.153) & 0.155 (0.200) & 0.149 (0.194) \\
Gaussian innovations & 0.083 (0.110) & 0.082 (0.109) & 0.114 (0.133) & 0.100 (0.117) \\
Persistent heavy tails & 0.130 (0.169) & 0.125 (0.166) & 0.128 (0.167) & 0.108 (0.141) \\
Regime shift & 0.151 (0.170) & 0.101 (0.124) & 0.180 (0.217) & 0.103 (0.127) \\
Student-\(t\) location--scale & 0.072 (0.095) & 0.072 (0.098) & 0.116 (0.146) & 0.095 (0.121) \\
\bottomrule
\end{tabular}
}%
\caption{Forecast quantile-path recovery in the joint multi-quantile simulation. Entries are mean absolute error with root-mean-square error in parentheses for the monotonically rearranged posterior-mean forecast quantile paths, computed relative to the known conditional quantiles. These diagnostics complement the data-generating-process-integrated score comparison in the main article.}
\label{tab:supp-joint-qdesn-oracle-recovery}
\end{table}

\begin{table}[!htbp]
\centering
\scriptsize
\begin{tabular}{@{}>{\raggedright\arraybackslash}p{0.30\textwidth}rrr@{}}
\toprule
Simulation setting & Joint: AL-exAL & Independent: AL-exAL & Replicates
\\
\midrule
Asymmetric Laplace Tail & -0.054 (100\%) & -0.042 (94\%) & 50 \\
Gaussian-Mixture Benchmark & -0.047 (96\%) & -0.045 (96\%) & 50 \\
Laplace Benchmark & -0.045 (96\%) & -0.037 (92\%) & 50 \\
Nonlinear Reservoir-Feature Mechanism & -0.047 (94\%) & -0.044 (94\%) & 50 \\
Gaussian Benchmark & -0.031 (94\%) & -0.027 (92\%) & 50 \\
Persistent Heavy Tails & -0.035 (88\%) & -0.031 (82\%) & 50 \\
Regime Shift & -0.037 (70\%) & -0.041 (96\%) & 50 \\
Student-t Location-Scale & -0.043 (86\%) & -0.037 (84\%) & 50 \\
\bottomrule
\end{tabular}
\caption{Replicated VB comparison of AL and exAL forecast quantile-path MAE over 50 independent data-generating realizations for each simulation setting. Entries are the median paired difference AL minus exAL, with the percentage of replicates favoring AL in parentheses.}
\label{tab:joint-qdesn-article-validation-phase153-replication-summary}
\end{table}

The final table reports a separate 50-replicate variational sensitivity
analysis. Its paired \(\AL\)-minus-\(\exAL\) forecast-MAE differences assess
recovery of the known quantile paths; the MCMC DGP-integrated score remains
the ranking criterion in the main comparison.

\clearpage
\subsection{PriceFM Application Diagnostic Breakdowns}
\label{subsec:supp_pricefm_diagnostics}

The main article reports the aggregate PriceFM comparison over the same region
and fold combinations. The tables below summarize results by fold, predictor
set, and forecast-horizon group. All three summaries use the same retrospective
information set described in the main article.

\begin{table}[!htbp]
\centering
\caption{PriceFM retrospective comparison by fold. Average quantile loss (AQL)
is computed on the original electricity-price scale using the quantile levels
reported in the PriceFM paper. The last two
columns report Q--DESN AQL minus PriceFM AQL; negative values indicate lower
Q--DESN AQL. Near ties retain the pre-specified classification used in the
full-panel comparison. No paired uncertainty interval is reported for
these fold-level contrasts.}
\label{tab:supp-pricefm-full-fold-summary}
\begingroup
\TableStyle
\scriptsize
\setlength{\tabcolsep}{3pt}
\begin{tabular}{@{}lrrrrrrrr@{}}
\toprule
Comparison set & $n$ & \shortstack{Q--DESN\\lower} & \shortstack{Near\\ties} & \shortstack{PriceFM\\lower} & \shortstack{Mean\\Q--DESN AQL} & \shortstack{Mean\\PriceFM AQL} & \shortstack{Mean\\$\Delta$} & \shortstack{Median\\$\Delta$} \\
\midrule
Overall & 114 & 66 & 12 & 36 & 6.826 & 7.039 & -0.213 & -0.083 \\
Fold 1 & 38 & 22 & 6 & 10 & 7.047 & 7.297 & -0.250 & -0.073 \\
Fold 2 & 38 & 30 & 2 & 6 & 6.429 & 6.826 & -0.397 & -0.295 \\
Fold 3 & 38 & 14 & 4 & 20 & 7.001 & 6.993 & 0.008 & 0.281 \\
\bottomrule
\end{tabular}
\endgroup

\end{table}

\begin{table}[!htbp]
\centering
\caption{PriceFM retrospective comparison by predictor set. Own-region rows and rows with
neighborhood summaries both use retrospectively observed own-region load,
solar, and wind lead covariates in the retrospective comparison. Rows with
neighborhood summaries additionally include chosen neighboring-region
summaries or neighboring-region lead/lag features.}
\label{tab:supp-pricefm-full-input-set-summary}
\begingroup
\TableStyle
\scriptsize
\setlength{\tabcolsep}{3pt}
\begin{tabular}{@{}lrrrrrr@{}}
\toprule
Input set & $n$ & \shortstack{Q--DESN\\lower} & \shortstack{Near\\ties} & \shortstack{PriceFM\\lower} & \shortstack{Mean\\$\Delta$} & \shortstack{Median\\$\Delta$} \\
\midrule
Neighboring-region predictors & 56 & 34 & 5 & 17 & -0.366 & -0.113 \\
Own-region predictors & 58 & 32 & 7 & 19 & -0.064 & -0.046 \\
\bottomrule
\end{tabular}
\endgroup

\end{table}

\begin{table}[!htbp]
\centering
\caption{PriceFM retrospective comparison by forecast-horizon group.
Entries denote scored region--fold and forecast-horizon combinations for the
subset with horizon-specific results. Negative \(\Delta\) values indicate
lower Q--DESN AQL; positive values indicate lower PriceFM AQL.}
\label{tab:supp-pricefm-full-horizon-diagnostic-summary}
\begingroup
\TableStyle
\scriptsize
\setlength{\tabcolsep}{3pt}
\begin{tabular}{@{}lrrrr@{}}
\toprule
Horizon block & $n$ & \shortstack{Q--DESN\\lower} & \shortstack{Mean\\$\Delta$} & \shortstack{Median\\$\Delta$} \\
\midrule
1-24 & 72 & 22 & 0.421 & 0.400 \\
25-48 & 72 & 45 & -0.683 & -0.382 \\
49-72 & 72 & 34 & -0.196 & 0.032 \\
73-96 & 72 & 42 & -0.152 & -0.114 \\
\bottomrule
\end{tabular}
\endgroup

\end{table}

\clearpage
\section{GloFAS Forecast-Discrepancy Model with DESN Features}
\label{sec:glofas_discrepancy}

\noindent\textbf{Model overview.}
The streamflow application combines forecasts from the Global Flood Awareness
System (GloFAS) with a reference gauge record
\citep{AlfieriEtAl2013GloFAS,HarriganEtAl2023GloFASForecasts}. The reference
quantile and the GloFAS discrepancy receive separate coefficient vectors. The
historical reservoir states are fixed after preprocessing and washout. At
issued forecast horizons, however, the reference path is unobserved at the
forecast origin, so future reservoir states may depend on latent future
reference values. This section derives the fixed-design distributions given a
proposed future reference path and the corresponding latent-path sampler. The
full latent-path posterior also
contains the missing future reference values and the induced future reservoir
states.

\subsection{Observed Quantities and Augmented Design Matrix}
\label{subsec:glofas_aug_design}

\noindent\textbf{Observed and future quantities.}
Fix a quantile level \(p_0\). Let \(y_t\) denote transformed reference
streamflow, \(g_t^{\mathrm{ret}}\) the retrospective GloFAS product, and
\(g_{T+h,m}^{\mathrm{ens}}\) GloFAS ensemble member \(m\) at forecast origin
\(T\) and horizon \(h\). Write
\(\vect y_F=(y_{T+1},\ldots,y_{T+H})^\top\) for the future reference path. Let
\(\mathcal I_{\mathrm{ref}}\) index reference
observations and let \(\mathcal I_{\mathrm{glo}}\) index both retrospective
and issued-ensemble GloFAS values. For \(i\in\mathcal I_c\),
\(c\in\mathcal C=\{\mathrm{ref},\mathrm{glo}\}\), write the observed value as
\(z_i^c\). The reference feature vector
\(\vect x_i^{\mathrm{ref}}\) is generated from the reference-process history.
The discrepancy feature vector \(\vect x_i^{\mathrm{disc}}\) is generated from
the GloFAS-minus-reference discrepancy path. Both vectors are fixed after
conditioning on the relevant historical or latent future path.

\noindent\textbf{Augmented design.}
Define
\[
\vect\theta_{p_0}
=
\begin{bmatrix}
\vect\beta_{p_0}\\
\vect\alpha_{p_0}
\end{bmatrix},
\qquad
\vect h_t^{\mathrm{ref}}=
\begin{bmatrix}
\vect x_t^{\mathrm{ref}}\\
\vect 0
\end{bmatrix},
\qquad
\vect h_t^{\mathrm{glo}}=
\begin{bmatrix}
\vect x_t^{\mathrm{ref}}\\
\vect x_t^{\mathrm{disc}}
\end{bmatrix}.
\]
The same definition applies to any observation indexed by \(i\). The
source-specific quantile locations are
\[
q^{\mathrm{ref}}_{p_0,i}
=\vect h_i^{\mathrm{ref}\,\top}\vect\theta_{p_0}
=\vect x_i^{\mathrm{ref}\,\top}\vect\beta_{p_0},
\qquad
q^{\mathrm{glo}}_{p_0,i}
=\vect h_i^{\mathrm{glo}\,\top}\vect\theta_{p_0}
=\vect x_i^{\mathrm{ref}\,\top}\vect\beta_{p_0}
+\vect x_i^{\mathrm{disc}\,\top}\vect\alpha_{p_0}.
\]
Thus \(\vect\beta_{p_0}\) represents the reference-process quantile regression and
\(\vect\alpha_{p_0}\) represents an additive GloFAS discrepancy component at the same
level. In the historical rows, \(\vect x_i^{\mathrm{disc}}\) is driven by
\(g_i^{\mathrm{ret}}-y_i\). In the reported retrospective forecast-period
specification, the
future discrepancy predictor uses the last available historical retrospective
discrepancy on the transformed scale as a persistence baseline and then adds
the estimated discrepancy regression. The recursive path
\(\hat q^{\mathrm{ens}}_{p_0,T+h}-y_{T+h}\), where
\(\hat q^{\mathrm{ens}}_{p_0,T+h}\) is the empirical GloFAS ensemble quantile
at the fitted level, is an alternative specification for future sensitivity
analysis.

\noindent\textbf{Forecast summaries.}
Stacking the reference and GloFAS observations conditional on the specified
reservoir-state construction gives a vector \(\vect z\), a design matrix \(\mat H\), and
labels \(c_i\). With observed historical data only, \(\mat H\) is fixed.
Forecast-period summaries are variational quantile estimates followed by
monotone rearrangement.

\noindent\textbf{Forecast-time decomposition.}
The forecast-time decomposition is additive around the
persistence baseline. For horizons covered by an issued GloFAS ensemble at
origin \(T\), the ensemble provides the GloFAS forecasts and the unadjusted
GloFAS benchmark. The Q--DESN reference quantile is
\[
q^{\mathrm{ref},(s)}_{p_0,T+h}
=\vect x_{T+h}^{\mathrm{ref},(s)\top}\vect\beta_{p_0}^{(s)},
\]
and the associated forecast-system quantile is
\[
q^{\mathrm{glo},(s)}_{p_0,T+h}
=q^{\mathrm{ref},(s)}_{p_0,T+h}
+d_T^{\mathrm{ret}}
+\vect x_{T+h}^{\mathrm{disc},(s)\top}\vect\alpha_{p_0}^{(s)}.
\]
Here \(d_T^{\mathrm{ret}}\) is the last historical retrospective discrepancy.
Point summaries are taken after this quantile construction. Observed held-out
reference flows are reserved for evaluation and excluded from every
forecast-period lagged-response design row. The analysis remains retrospective because the
precipitation and soil-moisture predictors include blended contributions from
weather observations realized after that origin.

\noindent\textbf{Reservoir specification.}
Separate DESN feature maps are used for the reference process and the GloFAS
discrepancy. Both maps have depth
\(D=\GlofasApplicationCurrentReferenceReservoirDepth{}\), layer widths
\((\GlofasApplicationCurrentReferenceReservoirSize{})\), reducer widths
\((\GlofasApplicationCurrentReferenceReducerSize{})\), memory
\(\GlofasApplicationCurrentReferenceReservoirMemory{}\), washout
\(\GlofasApplicationCurrentReferenceReservoirWashout{}\), spectral radii
\((\GlofasApplicationCurrentReferenceReservoirRho{})\), recurrent sparsity
\((\GlofasApplicationCurrentReferenceReservoirPiW{})\), and input inclusion
probabilities \((\GlofasApplicationCurrentReferenceReservoirPiIn{})\). The
reference map uses leak rates
\((\GlofasApplicationCurrentReferenceReservoirAlpha{})\), global and intercept
input scales
\(\GlofasApplicationCurrentReferenceReservoirWinScaleGlobal{}/\GlofasApplicationCurrentReferenceReservoirWinScaleBias{}\),
and random-number seed \(\GlofasApplicationCurrentReferenceReservoirSeed{}\).
The discrepancy map uses leak rates
\((\GlofasApplicationCurrentDiscrepancyReservoirAlpha{})\), global and intercept
input scales
\(\GlofasApplicationCurrentDiscrepancyReservoirWinScaleGlobal{}/\GlofasApplicationCurrentDiscrepancyReservoirWinScaleBias{}\),
and random-number seed \(\GlofasApplicationCurrentDiscrepancyReservoirSeed{}\).

\subsection{AL and exAL Working Likelihoods}
\label{subsec:glofas_likelihoods}

For exAL, each source \(c\in\mathcal C\) has scale and asymmetry parameters
\((\sigma_c,\gamma_c)\). Conditional on latent variables
\(\{v_i^c,s_i^c:i\in\mathcal I_c\}\),
\begin{align}
v_i^c\mid\sigma_c&\sim\Exp(\text{rate}=1/\sigma_c),&
s_i^c&\sim\TN(0,1),\\
z_i^c\mid\vect\theta_{p_0},\sigma_c,\gamma_c,v_i^c,s_i^c
&\sim
\Normal\left(
\vect h_i^{c\top}\vect\theta_{p_0}
+\sigma_cD_{\gamma_c}s_i^c+A_{\gamma_c}v_i^c,
\sigma_cB_{\gamma_c}v_i^c
\right).
\label{eq:supp_glofas_exal_hierarchy}
\end{align}
The marginal \(p_0\)-quantile of \(z_i^c\mid\vect h_i^c\) is
\(\vect h_i^{c\top}\vect\theta_{p_0}\). The AL working likelihood is obtained
by omitting \(s_i^c\) and \(\gamma_c\), replacing \(A_{\gamma_c}\) and
\(B_{\gamma_c}\) with \(A_0\) and \(B_0\), and using
\begin{equation}
z_i^c\mid\vect\theta_{p_0},\sigma_c,v_i^c
\sim
\Normal\left(
\vect h_i^{c\top}\vect\theta_{p_0}+A_0v_i^c,
\sigma_cB_0v_i^c
\right).
\label{eq:supp_glofas_al_hierarchy}
\end{equation}

The GloFAS observations \(\mathcal I_{\mathrm{glo}}\) may contain both retrospective values and
issued ensemble members. Ensemble members at a common origin and horizon enter
through a working-likelihood factorization that treats them as conditionally
independent and exchangeable given the forecast-system quantile path and the
GloFAS likelihood parameters. This factorization is motivated by the
operational ensemble construction, in which perturbed initial states and model
perturbations generate distinct forecast alternatives
\citep{ECMWF2026ENSGeneration}; it is a working-likelihood approximation for
the regression analysis.

\subsection{Default Regularized-Horseshoe Coefficient Prior}
\label{subsec:glofas_rhs_prior}

For the application derivation, the regularized horseshoe is the default
coefficient prior. Let \(\mathcal J_\beta\) and \(\mathcal J_\alpha\) denote the
non-intercept entries of \(\vect\beta_{p_0}\) and \(\vect\alpha_{p_0}\). The
reference and discrepancy intercepts are assigned weak Gaussian priors. For
block \(b\in\{\beta,\alpha\}\) and \(j\in\mathcal J_b\), write the corresponding
coefficient as \(\theta_{b,j}\). The product representation uses
\begin{equation}
\Normal(\theta_{b,j}\mid0,\tau_b^2\lambda_{b,j}^2)
\Normal(0\mid\theta_{b,j},\zeta_b^2).
\label{eq:supp_glofas_rhs_product}
\end{equation}
Conditional on the scales, this factor is proportional in \(\theta_{b,j}\) to a
Gaussian density with variance
\begin{equation}
V_{b,j}=\left(\zeta_b^{-2}+\tau_b^{-2}\lambda_{b,j}^{-2}\right)^{-1}.
\label{eq:supp_glofas_rhs_cond}
\end{equation}
The auxiliary representation for
\((\lambda_{b,j}^2,\nu_{b,j},\tau_b^2,\xi_b,\zeta_b^2)\) is the same as
\eqref{eq:supp_hs_aux}--\eqref{eq:supp_zeta_prior}, with
\(\beta_j\) replaced by \(\theta_{b,j}\) and \(r\) replaced by
\(|\mathcal J_b|\). The two regularized-horseshoe hierarchies are independent a priori, although
the Gaussian coefficient update couples \(\vect\beta_{p_0}\) and
\(\vect\alpha_{p_0}\) through the augmented likelihood. The source-specific
exAL scale and asymmetry parameters are the only non-conjugate components introduced by
the working likelihood.

\subsection{Complete-Data Posterior}
\label{subsec:glofas_posterior}

Let
\[
\Theta_b
=\{(\lambda_{b,j}^2,\nu_{b,j}):j\in\mathcal J_b\},\tau_b^2,\xi_b,\zeta_b^2,
\qquad b\in\{\beta,\alpha\},
\]
and let \(\vect v^c=(v_i^c:i\in\mathcal I_c)\) and
\(\vect s^c=(s_i^c:i\in\mathcal I_c)\). Under exAL,
\begin{align}
&p(\vect z,\{\vect v^c,\vect s^c,\sigma_c,\gamma_c\}_{c\in\mathcal C},
\vect\theta_{p_0},\Theta_\beta,\Theta_\alpha\mid\mat H)
\nonumber\\
&\quad\propto
\prod_{c\in\mathcal C}
\left[
\left\{
\prod_{i\in\mathcal I_c}
p(z_i^c\mid\vect\theta_{p_0},\sigma_c,\gamma_c,v_i^c,s_i^c,\vect h_i^c)
p(v_i^c\mid\sigma_c)p(s_i^c)
\right\}
p(\sigma_c)\pi_{\gamma_c}(\gamma_c)\ind\{L<\gamma_c<U\}
\right]
\nonumber\\
&\qquad\times
p(\vect\theta_{\mathrm{int}})
\prod_{b\in\{\beta,\alpha\}}
\Biggl[
\prod_{j\in\mathcal J_b}
\left\{
\Normal(\theta_{b,j}\mid0,\tau_b^2\lambda_{b,j}^2)
\Normal(0\mid\theta_{b,j},\zeta_b^2)
p(\lambda_{b,j}^2\mid\nu_{b,j})p(\nu_{b,j})
\right\}
\nonumber\\
&\qquad\qquad\times
p(\tau_b^2\mid\xi_b)p(\xi_b)p(\zeta_b^2)
\Biggr].
\label{eq:supp_glofas_joint_exal_rhs}
\end{align}
Here \(p(\vect\theta_{\mathrm{int}})\) denotes the weak Gaussian prior for the
two intercepts. Under AL,
\begin{align}
&p(\vect z,\{\vect v^c,\sigma_c\}_{c\in\mathcal C},
\vect\theta_{p_0},\Theta_\beta,\Theta_\alpha\mid\mat H)
\nonumber\\
&\quad\propto
\prod_{c\in\mathcal C}
\left[
\left\{
\prod_{i\in\mathcal I_c}
p(z_i^c\mid\vect\theta_{p_0},\sigma_c,v_i^c,\vect h_i^c)
p(v_i^c\mid\sigma_c)
\right\}
p(\sigma_c)
\right]
\nonumber\\
&\qquad\times
p(\vect\theta_{\mathrm{int}})
\prod_{b\in\{\beta,\alpha\}}
\Biggl[
\prod_{j\in\mathcal J_b}
\left\{
\Normal(\theta_{b,j}\mid0,\tau_b^2\lambda_{b,j}^2)
\Normal(0\mid\theta_{b,j},\zeta_b^2)
p(\lambda_{b,j}^2\mid\nu_{b,j})p(\nu_{b,j})
\right\}
\nonumber\\
&\qquad\qquad\times
p(\tau_b^2\mid\xi_b)p(\xi_b)p(\zeta_b^2)
\Biggr].
\label{eq:supp_glofas_joint_al_rhs}
\end{align}
The fixed-design posterior is the corresponding joint density viewed as a
function of unknown quantities conditional on \((\vect z,\mat H)\). In the
latent-path ensemble-likelihood model, \(\mat H\) should be read as
\(\mat H(\vect y_F)\) for forecast-period rows, and the posterior is
augmented by the missing future reference path.
Forecast-period output inputs are strictly lagged: the design row for
\(T+h\) may depend on \(y_{T+h-1},y_{T+h-2},\ldots\), but not on
\(y_{T+h}\) itself.

\subsection{Full Conditional Posteriors}
\label{subsec:glofas_conditionals}

The full conditionals follow from the augmented Gaussian representation.
Define, for exAL,
\[
z_i^{c\star}
=z_i^c-\sigma_cD_{\gamma_c}s_i^c-A_{\gamma_c}v_i^c,\qquad
w_i^c=(\sigma_cB_{\gamma_c}v_i^c)^{-1}.
\]
For AL, use
\[
z_i^{c\star}=z_i^c-A_0v_i^c,\qquad
w_i^c=(\sigma_cB_0v_i^c)^{-1}.
\]
Stacking \(z_i^{c\star}\) and \(w_i^c\) over sources gives
\(\vect z^\star\) and \(\mat\Omega=\diag(w_i^c)\). Let
\(\mat P_\theta\) denote the conditional prior precision implied by the
weak Gaussian intercept priors and the RHS scales, with diagonal entries
\(\tau^{-2}\lambda_j^{-2}+\zeta^{-2}\) for \(j\in\mathcal J\). Let
\(\vect b_\theta\) collect the prior means, with zero entries for RHS-shrunk
coefficients. Then
\begin{align}
\mat\Sigma_{\theta\mid\cdot}
&=(\mat H^\top\mat\Omega\mat H+\mat P_\theta)^{-1},\\
\vect m_{\theta\mid\cdot}
&=\mat\Sigma_{\theta\mid\cdot}
(\mat H^\top\mat\Omega\vect z^\star+\mat P_\theta\vect b_\theta),\\
\vect\theta_{p_0}\mid\cdot
&\sim
\Normal(\vect m_{\theta\mid\cdot},\mat\Sigma_{\theta\mid\cdot}).
\label{eq:supp_glofas_theta_mcmc}
\end{align}

For exAL latent variables, define
\[
\delta_i^c
=z_i^c-\vect h_i^{c\top}\vect\theta_{p_0}
-\sigma_cD_{\gamma_c}s_i^c,
\qquad
\chi_i^c=\frac{(\delta_i^c)^2}{\sigma_cB_{\gamma_c}},
\qquad
\psi_i^c=\frac{A_{\gamma_c}^2}{\sigma_cB_{\gamma_c}}+\frac{2}{\sigma_c}.
\]
Then
\begin{equation}
v_i^c\mid\cdot\sim\GIG(1/2,\chi_i^c,\psi_i^c).
\label{eq:supp_glofas_v_mcmc}
\end{equation}
For \(s_i^c\), set
\(\mu_{0i}^c=z_i^c-\vect h_i^{c\top}\vect\theta_{p_0}
-A_{\gamma_c}v_i^c\). Then
\begin{align}
a_{s,i}^c&=1+\frac{\sigma_cD_{\gamma_c}^2}{B_{\gamma_c}v_i^c},&
m_{s,i}^c&=
\frac{D_{\gamma_c}\mu_{0i}^c}
{B_{\gamma_c}v_i^c+\sigma_cD_{\gamma_c}^2},\\
s_i^c\mid\cdot&\sim\TN(m_{s,i}^c,(a_{s,i}^c)^{-1}).
\label{eq:supp_glofas_s_mcmc}
\end{align}
For AL,
\begin{equation}
v_i^c\mid\cdot
\sim
\GIG\left(
1/2,
\frac{(z_i^c-\vect h_i^{c\top}\vect\theta_{p_0})^2}{\sigma_cB_0},
\frac{A_0^2}{\sigma_cB_0}+\frac{2}{\sigma_c}
\right).
\label{eq:supp_glofas_al_v_mcmc}
\end{equation}
If \(\sigma_c\sim\IG(a_{\sigma,c},b_{\sigma,c})\), then under AL
\begin{align}
\sigma_c\mid\cdot
&\sim
\IG\left(
a_{\sigma,c}+\frac{3n_c}{2},
b_{\sigma,c}+\sum_{i\in\mathcal I_c}v_i^c
+\frac12\sum_{i\in\mathcal I_c}
\frac{(z_i^c-\vect h_i^{c\top}\vect\theta_{p_0}-A_0v_i^c)^2}
{B_0v_i^c}
\right),
\label{eq:supp_glofas_al_sigma_mcmc}
\end{align}
where \(n_c=|\mathcal I_c|\).

For exAL, update \((\sigma_c,\gamma_c)\) source by source. The joint
log-kernel is
\begin{align}
\ell_c(\sigma_c,\gamma_c)
&=
-\frac12\sum_{i\in\mathcal I_c}\log(\sigma_cB_{\gamma_c}v_i^c)
-\frac12\sum_{i\in\mathcal I_c}
\frac{
\{z_i^c-\vect h_i^{c\top}\vect\theta_{p_0}
-\sigma_cD_{\gamma_c}s_i^c-A_{\gamma_c}v_i^c\}^2}
{\sigma_cB_{\gamma_c}v_i^c}
\nonumber\\
&\quad
-n_c\log\sigma_c-\sum_{i\in\mathcal I_c}\frac{v_i^c}{\sigma_c}
-(a_{\sigma,c}+1)\log\sigma_c-\frac{b_{\sigma,c}}{\sigma_c}
+\log\pi_{\gamma_c}(\gamma_c),
\label{eq:supp_glofas_sigmagam_kernel}
\end{align}
for \(L<\gamma_c<U\), with value \(-\infty\) outside this interval. This
kernel is the target for the source-specific scale and asymmetry update.

For each block \(b\in\{\beta,\alpha\}\) and \(j\in\mathcal J_b\), the RHS
scale updates are
\begin{align}
\lambda_{b,j}^2\mid\cdot
&\sim
\IG\left(1,\frac{1}{\nu_{b,j}}+\frac{\theta_{b,j}^2}{2\tau_b^2}\right),\\
\nu_{b,j}\mid\cdot
&\sim
\IG\left(1,1+\frac{1}{\lambda_{b,j}^2}\right),\\
\tau_b^2\mid\cdot
&\sim
\IG\left(\frac{|\mathcal J_b|+1}{2},
\frac{1}{\xi_b}+\frac12\sum_{j\in\mathcal J_b}\frac{\theta_{b,j}^2}{\lambda_{b,j}^2}\right),\\
\xi_b\mid\cdot
&\sim
\IG\left(1,\frac{1}{\tau_{0,b}^2}+\frac{1}{\tau_b^2}\right),\\
\zeta_b^2\mid\cdot
&\sim
\IG\left(a_{\zeta,b}+\frac{|\mathcal J_b|}{2},
b_{\zeta,b}+\frac12\sum_{j\in\mathcal J_b}\theta_{b,j}^2\right),
\label{eq:supp_glofas_rhs_mcmc}
\end{align}
with the final line omitted when the slab scale is fixed.

\subsection{MCMC Sampler}
\label{subsec:glofas_mcmc}

For exAL, use unconstrained coordinates
\[
\eta_{\sigma,c}=\log\sigma_c,\qquad
\eta_{\gamma,c}=\logit\{(\gamma_c-L)/(U-L)\}.
\]
The slice sampler targets
\begin{equation}
\ell_c\{T_c(\vect\eta_c)\}+\log|J_{T_c}(\vect\eta_c)|,
\qquad
\vect\eta_c=(\eta_{\gamma,c},\eta_{\sigma,c})^\top,
\label{eq:supp_glofas_slice_target}
\end{equation}
where \(T_c\) maps unconstrained coordinates to
\((\gamma_c,\sigma_c)\). This bivariate slice update may be replaced by a
Laplace-informed proposal or by one-dimensional conditional slice updates, but
the target density is \eqref{eq:supp_glofas_slice_target}.

\begin{suppalgorithm}{MCMC sampler for the GloFAS forecast-discrepancy model}
\label{alg:glofas_mcmc}
\emph{Input:} stacked observations \(\vect z\), source labels \(c_i\), target
quantile \(p_0\), likelihood choice AL or exAL, RHS hyperparameters, MCMC
controls, and either a fixed augmented design \(\mat H\) or a mapping from a
proposed future reference path to its reservoir states and design rows.
\begin{enumerate}[label=(\arabic*)]
\item Initialize
\(\vect\theta_{p_0}=(\vect\beta_{p_0}^\top,\vect\alpha_{p_0}^\top)^\top\),
the separate \(\beta\)- and \(\alpha\)-block RHS scales, source scales
\(\{\sigma_c\}\), latent variables \(\{v_i^c\}\), and, for exAL,
\(\{\gamma_c,s_i^c\}\). For the latent-path model, also initialize
\(\vect y_F\).
\item For each MCMC iteration:
  \begin{enumerate}[label=(\alph*)]
  \item for the latent-path model, recompute the forecast-period rows of
  \(\mat H(\vect y_F)\);
  \item draw each \(v_i^c\mid\cdot\) from the GIG full conditional in
  \eqref{eq:supp_glofas_v_mcmc} for exAL or
  \eqref{eq:supp_glofas_al_v_mcmc} for AL;
  \item for exAL, draw each \(s_i^c\mid\cdot\) from
  \eqref{eq:supp_glofas_s_mcmc};
  \item draw
  \(\vect\theta_{p_0}\mid\cdot\sim
  \Normal(\vect m_{\theta\mid\cdot},\mat\Sigma_{\theta\mid\cdot})\)
  using \eqref{eq:supp_glofas_theta_mcmc};
  \item for each block \(b\in\{\beta,\alpha\}\), draw the RHS local, global,
  auxiliary, and slab scales from \eqref{eq:supp_glofas_rhs_mcmc};
  \item for AL, draw each \(\sigma_c\mid\cdot\) from
  \eqref{eq:supp_glofas_al_sigma_mcmc};
  \item for exAL, update each \((\sigma_c,\gamma_c)\) by slice sampling on
  the transformed target \eqref{eq:supp_glofas_slice_target};
  \item for the latent-path model, update \(\vect y_F\) using a
  Metropolis--Hastings, slice-sampling, or componentwise transition targeting
  the complete-data posterior
  \eqref{eq:supp_glofas_joint_exal_rhs} or
  \eqref{eq:supp_glofas_joint_al_rhs} with \(\mat H=\mat H(\vect y_F)\),
  recomputing only forecast-period state rows after a proposal;
  \item store retained draws and diagnostics after burn-in, with thinning only
  if used.
  \end{enumerate}
\end{enumerate}
\end{suppalgorithm}

\subsection{Variational Bayes and CAVI Updates}
\label{subsec:glofas_vb}

\noindent\textbf{Mean-field family.}
For exAL, use the mean-field family
\begin{equation}
q=
q_\theta(\vect\theta_{p_0})
q_{\Theta_\beta}(\Theta_\beta)
q_{\Theta_\alpha}(\Theta_\alpha)
\prod_{c\in\mathcal C}
\left[
q_{\sigma_c,\gamma_c}(\sigma_c,\gamma_c)
\prod_{i\in\mathcal I_c}q_{v_i^c}(v_i^c)q_{s_i^c}(s_i^c)
\right]
\label{eq:supp_glofas_vb_family}
\end{equation}
where each \(q_{\Theta_b}\) factorizes over the auxiliary inverse-gamma RHS
components used in Section~\ref{subsec:glofas_rhs_prior}.
For AL, replace \(q_{\sigma_c,\gamma_c}\) by \(q_{\sigma_c}\) and omit
\(q_{s_i^c}\). The coefficient factor is Gaussian. Let
\[
\bar w_i^c=\E\{(\sigma_cB_{\gamma_c}v_i^c)^{-1}\},
\qquad
\bar m_i^c=
\E\left[
\frac{z_i^c-\sigma_cD_{\gamma_c}s_i^c-A_{\gamma_c}v_i^c}
{\sigma_cB_{\gamma_c}v_i^c}
\right]
\]
for exAL, with \(A_{\gamma_c},B_{\gamma_c},D_{\gamma_c}\) replaced by
\(A_0,B_0,0\) for AL. Stacking \(\bar w_i^c\) and \(\bar m_i^c\) gives
\(\bar{\mat\Omega}\) and \(\bar{\vect m}\). Let
\(\bar{\mat P}_\theta\) replace each block-specific RHS precision by
\(\E(\tau_b^{-2})\E(\lambda_{b,j}^{-2})+\E(\zeta_b^{-2})\). Then
\begin{align}
q_\theta(\vect\theta_{p_0})&=\Normal(\vect m_\theta,\mat\Sigma_\theta),\\
\mat\Sigma_\theta&=(\mat H^\top\bar{\mat\Omega}\mat H+\bar{\mat P}_\theta)^{-1},\\
\vect m_\theta&=\mat\Sigma_\theta(\mat H^\top\bar{\vect m}
+\bar{\mat P}_\theta\vect b_\theta).
\label{eq:supp_glofas_vb_theta}
\end{align}

\noindent\textbf{Regression residual moments.}
For each contribution, define
\[
\bar r_i^c=z_i^c-\vect h_i^{c\top}\vect m_\theta,\qquad
\overline{(r_i^c)^2}
=(\bar r_i^c)^2+\vect h_i^{c\top}\mat\Sigma_\theta\vect h_i^c .
\]

\noindent\textbf{AL coordinate factors.}
For AL,
\begin{align}
q(v_i^c)
&=
\GIG\left(
1/2,
B_0^{-1}\E(\sigma_c^{-1})\overline{(r_i^c)^2},
\E(\sigma_c^{-1})\left(\frac{A_0^2}{B_0}+2\right)
\right),
\label{eq:supp_glofas_vb_al_v}\\
q(\sigma_c)
&=
\IG(a_{\sigma,c}^\star,b_{\sigma,c}^\star),\\
a_{\sigma,c}^\star
&=a_{\sigma,c}+\frac{3n_c}{2},\\
b_{\sigma,c}^\star
&=
b_{\sigma,c}+\sum_{i\in\mathcal I_c}\E(v_i^c)
+\frac{1}{2B_0}\sum_{i\in\mathcal I_c}
\left[
\overline{(r_i^c)^2}\E\{(v_i^c)^{-1}\}
-2A_0\bar r_i^c+A_0^2\E(v_i^c)
\right].
\label{eq:supp_glofas_vb_al_sigma}
\end{align}

\noindent\textbf{exAL coordinate factors.}
For exAL, define source-specific branchwise smooth expectations
\[
M_{1c}=\E\{(\sigma_cB_{\gamma_c})^{-1}\},\quad
M_{2c}=\E(D_{\gamma_c}/B_{\gamma_c}),\quad
M_{3c}=\E(\sigma_cD_{\gamma_c}^2/B_{\gamma_c}),\quad
M_{4c}=\E(A_{\gamma_c}D_{\gamma_c}/B_{\gamma_c}),
\]
and
\[
M_{5c}=\E\{A_{\gamma_c}^2/(\sigma_cB_{\gamma_c})+2/\sigma_c\}.
\]
These expectations involve the source-specific exAL scale and asymmetry factor
and are evaluated by Laplace--Delta within the active \(\gamma_c\) branch. The
AL point \(\gamma_c=0\) is handled as the nested AL case rather than by a
generic two-sided Hessian. The latent factors are
\begin{align}
q(v_i^c)
&=
\GIG(1/2,\bar\chi_i^c,\bar\psi_i^c),\\
\bar\chi_i^c
&=
M_{1c}\overline{(r_i^c)^2}
-2M_{2c}\bar r_i^c\E(s_i^c)
+M_{3c}\E\{(s_i^c)^2\},\\
\bar\psi_i^c&=M_{5c},
\label{eq:supp_glofas_vb_exal_v}\\
q(s_i^c)&=\TN(m_{s,i}^{c,\mathrm{VB}},V_{s,i}^{c,\mathrm{VB}}),\\
V_{s,i}^{c,\mathrm{VB}}
&=\left[1+M_{3c}\E\{(v_i^c)^{-1}\}\right]^{-1},\\
m_{s,i}^{c,\mathrm{VB}}
&=
V_{s,i}^{c,\mathrm{VB}}
\left[
M_{2c}\bar r_i^c\E\{(v_i^c)^{-1}\}-M_{4c}
\right].
\label{eq:supp_glofas_vb_exal_s}
\end{align}

\noindent\textbf{RHS coordinate factors.}
The RHS CAVI updates are applied independently to each block
\(b\in\{\beta,\alpha\}\):
\begin{align}
q(\lambda_{b,j}^2)
&=\IG\left(1,\ \E(\nu_{b,j}^{-1})
+\frac12\overline{\theta_{b,j}^2}\E(\tau_b^{-2})\right),\\
q(\nu_{b,j})
&=\IG\left(1,\ 1+\E(\lambda_{b,j}^{-2})\right),\\
q(\tau_b^2)
&=\IG\left(\frac{|\mathcal J_b|+1}{2},
\E(\xi_b^{-1})
+\frac12\sum_{j\in\mathcal J_b}\overline{\theta_{b,j}^2}
\E(\lambda_{b,j}^{-2})\right),\\
q(\xi_b)
&=\IG\left(1,\ \tau_{0,b}^{-2}+\E(\tau_b^{-2})\right),\\
q(\zeta_b^2)
&=\IG\left(a_{\zeta,b}+\frac{|\mathcal J_b|}{2},
b_{\zeta,b}+\frac12\sum_{j\in\mathcal J_b}\overline{\theta_{b,j}^2}\right),
\label{eq:supp_glofas_vb_rhs}
\end{align}
where
\(\overline{\theta_{b,j}^2}=m_{\theta,bj}^2+(\mat\Sigma_\theta)_{bj,bj}\).

\noindent\textbf{Source scale--asymmetry factor.}
For exAL, the source-specific Laplace--Delta target is
\begin{align}
\ell_{c,\mathrm{LD}}(\vect\eta_c)
&=
\E_{q_{-(\sigma_c,\gamma_c)}}\{
\log p(\vect z^c\mid\vect\theta_{p_0},\sigma_c,\gamma_c,
\vect v^c,\vect s^c,\{\vect h_i^c:i\in\mathcal I_c\})
\nonumber\\
&\qquad
+\log p(\vect v^c\mid\sigma_c)
+\log p(\sigma_c)+\log\pi_{\gamma_c}(\gamma_c)
\}
+\log|J_{T_c}(\vect\eta_c)|.
\label{eq:supp_glofas_ld_target}
\end{align}
At the mode \(\hat{\vect\eta}_c\), with Hessian
\(\mat K_c=\nabla^2\ell_{c,\mathrm{LD}}(\hat{\vect\eta}_c)\), set
\begin{equation}
q_{\sigma_c,\gamma_c}(\vect\eta_c)
\approx
\Normal\{\hat{\vect\eta}_c,(-\mat K_c)^{-1}\}.
\label{eq:supp_glofas_ld_gaussian}
\end{equation}
For any branchwise smooth function \(h(\sigma_c,\gamma_c)\),
\begin{equation}
\E\{h(\sigma_c,\gamma_c)\}
\approx
h\{T_c(\hat{\vect\eta}_c)\}
+\frac12\tr\left[
\nabla^2_\eta h\{T_c(\hat{\vect\eta}_c)\}(-\mat K_c)^{-1}
\right].
\label{eq:supp_glofas_delta}
\end{equation}

\noindent\textbf{Future-path factor.}
For the latent-path ensemble-likelihood model, the variational family is
augmented by a factor \(q_F(\vect y_F)\) for the future reference path. Write
\(z_i^c(\vect y_F)=z_{i0}^c+\vect u_i^\top\vect y_F\), where
\(\vect u_i=\vect0\) for observed rows and identifies the appropriate future
reference value for forecast-period reference rows. The forecast-period
fixed-design row is \(\vect h_i^c(\vect y_F)\), because output lags in the
reservoir input may depend on \(\vect y_F\). For a current coefficient factor
\(q_\theta=\Normal(\vect m_\theta,\mat\Sigma_\theta)\), define
\[
e_i^c(\vect y_F)
=z_i^c(\vect y_F)-\vect h_i^c(\vect y_F)^\top\vect m_\theta,
\qquad
R_i^c(\vect y_F)
=\{e_i^c(\vect y_F)\}^2+
\vect h_i^c(\vect y_F)^\top
\mat\Sigma_\theta\vect h_i^c(\vect y_F).
\]

\noindent\textbf{Future-path likelihood target.}
For AL, the future-path Laplace target is, up to constants not depending on
\(\vect y_F\),
\begin{equation}
\ell_F^{\mathrm{AL}}(\vect y_F)
=
-\frac{1}{2B_0}\sum_{i\in\mathcal I_F}
\E(\sigma_{c_i}^{-1})
\left[
\E\{(v_i^{c_i})^{-1}\}R_i^{c_i}(\vect y_F)
-2A_0e_i^{c_i}(\vect y_F)
\right],
\label{eq:supp_glofas_latent_path_al_ld_target}
\end{equation}
where \(\mathcal I_F\) contains rows whose response or fixed-design row depends on
the future path. Terms independent of \(\vect y_F\), including the
\(A_0^2\E(v_i^{c_i})\) part of the Gaussian likelihood and the exponential
latent-prior terms, are handled in the complete-data ELBO and omitted from the
future-path mode or Hessian. For exAL, the corresponding target is
\begin{align}
\ell_F^{\mathrm{exAL}}(\vect y_F)
&=
-\frac12\sum_{i\in\mathcal I_F}
\left[
M_{1,c_i}\E\{(v_i^{c_i})^{-1}\}R_i^{c_i}(\vect y_F)
-2M_{2,c_i}\E(s_i^{c_i})\E\{(v_i^{c_i})^{-1}\}
e_i^{c_i}(\vect y_F)
-2M_{6,c_i}e_i^{c_i}(\vect y_F)
\right],
\label{eq:supp_glofas_latent_path_exal_ld_target}
\end{align}
with
\[
M_{6,c}=\E\{A_{\gamma_c}/(\sigma_cB_{\gamma_c})\}.
\]
The remaining exAL terms in the complete-data likelihood and latent-prior
density are independent of \(\vect y_F\) and are included in the ELBO
decomposition below.

\noindent\textbf{Future-path Gaussian approximation.}
At the mode \(\hat{\vect y}_F\), set
\[
\mat K_F=-\nabla^2_{\vect y_F}\ell_F(\hat{\vect y}_F),
\qquad
q_F(\vect y_F)\approx
\Normal(\hat{\vect y}_F,\mat K_F^{-1}),
\]
with any Hessian regularization reported with the numerical diagnostics.
Delta-method
moments of the nonlinear state map are then used in the remaining coordinate
updates.

\noindent\textbf{Delta moments.}
If \(\mat J_i\) is the Jacobian of \(\vect h_i^c(\vect y_F)\) at
\(\hat{\vect y}_F\), and \(\mat V_F=\mat K_F^{-1}\), then the default
first-order Delta moments are
\[
\vect a_i^c
\approx
\vect h_i^c(\hat{\vect y}_F),
\qquad
\mat S_i^c
\approx
\vect a_i^c\vect a_i^{c\top}
+\mat J_i\mat V_F\mat J_i^\top .
\]
Also,
\[
\bar z_i^c=z_{i0}^c+\vect u_i^\top\hat{\vect y}_F,
\qquad
\overline{(z_i^c)^2}
=(\bar z_i^c)^2+\vect u_i^\top\mat V_F\vect u_i,
\qquad
\vect b_i^c
\approx
\vect a_i^c\bar z_i^c+\mat J_i\mat V_F\vect u_i .
\]
Second-order mean corrections can be added for chosen low-dimensional
summaries. The first-order construction above avoids featurewise Hessian
calculations while retaining the mean and covariance terms needed for the
Gaussian future-path approximation.

\noindent\textbf{Entropy and objective.}
Replacing fixed-design products by
\((\vect a_i^c,\mat S_i^c,\vect b_i^c,\bar z_i^c,
\overline{(z_i^c)^2})\) gives the latent-path VB--LD updates. The entropy
contribution of the future-path factor is
\[
H(q_F)=\frac12\log |2\pi e\,\mat V_F|.
\]
The resulting monitored objective is a Laplace--Delta approximation to the
ELBO of the nonlinear latent-path model.

\begin{suppalgorithm}{Variational Bayes updates for the GloFAS forecast-discrepancy model}
\label{alg:glofas_vb}
\emph{Input:} stacked observations \(\vect z\), source labels \(c_i\), target
quantile \(p_0\), likelihood choice AL or exAL, block-specific RHS hyperparameters,
convergence controls, and either a fixed augmented design \(\mat H\) or a
mapping from a proposed future reference path to its reservoir states and
design rows.
\begin{enumerate}[label=(\arabic*)]
\item Initialize \(q_\theta\), latent factors, source scale factors, and RHS
scale factors. For the latent-path model, also initialize \(q_F\).
\item Repeat until convergence:
  \begin{enumerate}[label=(\alph*)]
  \item for the latent-path model, set \(q_F\) by the Laplace target
  \eqref{eq:supp_glofas_latent_path_al_ld_target} for AL or
  \eqref{eq:supp_glofas_latent_path_exal_ld_target} for exAL, then recompute
  the Delta-method design moments;
  \item set \(q(v_i^c)\) by \eqref{eq:supp_glofas_vb_al_v} for AL or
  \eqref{eq:supp_glofas_vb_exal_v} for exAL;
  \item for exAL, set \(q(s_i^c)\) by
  \eqref{eq:supp_glofas_vb_exal_s};
  \item set \(q_\theta\) by \eqref{eq:supp_glofas_vb_theta};
  \item set the beta and alpha RHS factors by \eqref{eq:supp_glofas_vb_rhs};
  \item for AL, set each \(q(\sigma_c)\) by
  \eqref{eq:supp_glofas_vb_al_sigma};
  \item for exAL, set each \(q_{\sigma_c,\gamma_c}\) by
  \eqref{eq:supp_glofas_ld_target}--\eqref{eq:supp_glofas_ld_gaussian};
  \item recompute branchwise smooth expectations by \eqref{eq:supp_glofas_delta} and
  monitor the ELBO, or the Laplace--Delta ELBO approximation when \(q_F\) is
  present.
  \end{enumerate}
\item Return variational factors, fitted reference and discrepancy summaries,
forecast summaries, ELBO history, and convergence diagnostics.
\end{enumerate}
\end{suppalgorithm}

\subsection{ELBO}
\label{subsec:glofas_elbo}

\noindent\textbf{Application ELBO.}
For exAL, the application ELBO is
\begin{align}
\elbo(q)
&=
\sum_{c\in\mathcal C}
\left[
\E_q\{\log p(\vect z^c\mid\vect\theta_{p_0},\sigma_c,\gamma_c,
\vect v^c,\vect s^c,\mat H)\}
+\E_q\{\log p(\vect v^c\mid\sigma_c)\}
+\E_q\{\log p(\vect s^c)\}
\right.
\nonumber\\
&\qquad\left.
+\E_q\{\log p(\sigma_c)\}
+\E_q\{\log\pi_{\gamma_c}(\gamma_c)\}
\right]
+\E_q\{\log p(\vect\theta_{p_0},\Theta_\beta,\Theta_\alpha)\}
+H(q).
\label{eq:supp_glofas_elbo_decomp}
\end{align}

\noindent\textbf{Likelihood terms.}
For AL, omit the \(s\)-latent and \(\gamma_c\) prior terms. The exAL
likelihood term for source \(c\) is
\begin{align}
\elbo_{\mathrm{like},c}
&=
-\frac{n_c}{2}\log(2\pi)
-\frac12\sum_{i\in\mathcal I_c}
\left[
\E(\log\sigma_c)+\E(\log B_{\gamma_c})+\E(\log v_i^c)
\right]
\nonumber\\
&\quad
-\frac12\sum_{i\in\mathcal I_c}
\E\left[
\frac{
\{z_i^c-\vect h_i^{c\top}\vect\theta_{p_0}
-\sigma_cD_{\gamma_c}s_i^c-A_{\gamma_c}v_i^c\}^2}
{\sigma_cB_{\gamma_c}v_i^c}
\right].
\label{eq:supp_glofas_elbo_like}
\end{align}
The corresponding AL likelihood term is
\begin{align}
\elbo_{\mathrm{like},c}^{\mathrm{AL}}
&=
-\frac{n_c}{2}\log(2\pi)
-\frac12\sum_{i\in\mathcal I_c}
\left[
\E(\log\sigma_c)+\log B_0+\E(\log v_i^c)
\right]
\nonumber\\
&\quad
-\frac12\sum_{i\in\mathcal I_c}
\E\left[
\frac{
\{z_i^c-\vect h_i^{c\top}\vect\theta_{p_0}-A_0v_i^c\}^2}
{\sigma_cB_0v_i^c}
\right].
\label{eq:supp_glofas_elbo_like_al}
\end{align}

\noindent\textbf{Latent-path objective.}
For the latent-path model, these likelihood terms are evaluated with the
Laplace--Delta future-path moments above, and \(H(q)\) includes
\(H(q_F)\). Thus the monitored quantity is an ELBO approximation for the
nonlinear latent-path model; it reduces to the fixed-design ELBO when
\(\vect y_F\) is absent or held fixed.

\noindent\textbf{Prior and entropy terms.}
The latent and source-parameter prior terms are
\begin{align}
\elbo_{v,c}
&=
-n_c\E(\log\sigma_c)-\E(\sigma_c^{-1})
\sum_{i\in\mathcal I_c}\E(v_i^c),\\
\elbo_{s,c}
&=
n_c\log2-\frac{n_c}{2}\log(2\pi)
-\frac12\sum_{i\in\mathcal I_c}\E\{(s_i^c)^2\},\\
\elbo_{\sigma,c}
&=
a_{\sigma,c}\log b_{\sigma,c}-\log\Gamma(a_{\sigma,c})
-(a_{\sigma,c}+1)\E(\log\sigma_c)-b_{\sigma,c}\E(\sigma_c^{-1}),\\
\elbo_{\gamma,c}
&=\E\{\log\pi_{\gamma_c}(\gamma_c)\}.
\label{eq:supp_glofas_elbo_source_terms}
\end{align}
The RHS coefficient-prior and shrinkage-scale terms are the corresponding
blockwise RHS ELBO terms, with \(\vect\beta\) replaced separately by the
non-intercept entries of \(\vect\beta_{p_0}\) and \(\vect\alpha_{p_0}\);
the decomposition follows the generic regularized-horseshoe representation in
Sections~\ref{sec:cavi}--\ref{sec:elbo}.
Entropy terms are summed over the Gaussian coefficient factor, GIG latent
factors, truncated-normal latent factors, inverse-gamma RHS factors, and the
source-specific Laplace--Delta factors.
Terms involving smooth nonlinear functions of \((\sigma_c,\gamma_c)\) are
evaluated with the same Laplace--Delta approximation used in
\eqref{eq:supp_glofas_delta}.

\subsection{Limiting Cases and Computational Checks}
\label{subsec:glofas_checks}

The derivation has several useful reductions. If the GloFAS source is removed
and \(\vect\alpha_{p_0}\) is omitted, the model reduces to the ordinary
Q--DESN regression in Sections~\ref{sec:model_joint}--\ref{sec:elbo}. If
\(\gamma_c=0\) and the \(s_i^c\) variables are omitted, the exAL updates reduce
to the AL updates with constants \(A_0\) and \(B_0\). The coefficient update
remains Gaussian because the augmented likelihood and the RHS prior are
conditionally Gaussian. The RHS hierarchy contributes only inverse-gamma scale
updates under the product representation; the source-specific scale and
asymmetry block is the non-conjugate component. Finally, the MCMC
sampler and the variational approximation are written for the same augmented
working model when the likelihood family, source labels, and reservoir-state
specification are matched. The ELBO in \eqref{eq:supp_glofas_elbo_decomp} is therefore an
approximate objective for the corresponding variational factorization. Direct
checks of posterior marginals or interval properties require separate
comparison with posterior simulation.

\bibliographystyle{apalike}
\bibliography{refs}